\documentclass{aa} 
\usepackage{natbib}
\usepackage{amsmath} 
\usepackage{booktabs}
\bibpunct{(}{)}{;}{a}{}{,}

\usepackage{graphicx}
\usepackage{txfonts}
\usepackage{hyperref}
\hypersetup{
  colorlinks   = true,    
  urlcolor     = blue,    
  linkcolor    = blue,    
  citecolor    = blue,    
  breaklinks   = true
}

\usepackage[utf8]{inputenc} 
\usepackage[T1]{fontenc}

\usepackage{multirow}

\usepackage{footnote}
\makesavenoteenv{tabular} 
\makesavenoteenv{table}

\usepackage{subfigure}
\usepackage[font=small,labelfont=bf,labelsep=period]{caption}

\usepackage{breakurl}

\usepackage[export]{adjustbox}

\def\HII{H~\textsc{ii} }
\def\HI{H~\textsc{i} }
\def\OI{[O~\textsc{i}] }
\def\OIII{[O~\textsc{iii}] }
\def\CI{[C~\textsc{i}] }
\def\CII{[C~\textsc{ii}] }
\def\NII{[N~\textsc{ii}] }

\begin{document} 

\title{Radiative and mechanical feedback into the molecular gas in the Large Magellanic Cloud. II. 30 Doradus\thanks{\textit{Herschel} is an ESA 
space observatory with science instruments provided by European-led Principal Investigator consortia and with important participation from NASA.}}

\author{M.-Y. Lee\inst{1,2,3}\and
        S. C. Madden\inst{2}\and
        F. Le Petit\inst{4}\and 
        A. Gusdorf\inst{5,6}\and  
        P. Lesaffre\inst{5,6}\and 
        R. Wu\inst{4}\and 
        V. Lebouteiller\inst{2}\and
        F. Galliano\inst{2}\and
        M. Chevance\inst{7}}

\institute{Max-Planck-Institut f\"ur Radioastronomie, Auf dem H\"ugel 69, 53121 Bonn, Germany\label{1} \\ \email{mlee@mpifr-bonn.mpg.de} \and
           AIM, CEA, CNRS, Universit\'e Paris-Saclay, Universit\'e Paris Diderot, Sorbonne Paris Cit\'e, 91191 Gif-sur-Yvette, France\label{2} \and 
           Korea Astronomy and Space Science Institute, 776 Daedeokdae-ro, 34055 Daejeon, Republic of Korea\label{3} \and
           LERMA, Observatoire de Paris, PSL Research University, CNRS, Sorbonne Universit\'e, 92190 Meudon, France\label{4} \and
           Laboratoire de Physique de l'ENS, ENS, Universit\'e PSL, CNRS, Sorbonne Universit\'e, Universit\'e de Paris, Paris, France\label{5} \and
           Observatoire de Paris, Universit\'e PSL, Sorbonne Universit\'e, LERMA, 75014 Paris, France\label{6} \and
           Astronomisches Rechen-Institut, Zentrum f\"{u}r Astronomie der Universit\"{a}t Heidelberg, M\"{o}nchhofstra\ss e 12-14, 69120 Heidelberg, Germany\label{7}
          }

\date{Received; accepted}

\abstract{With an aim of probing the physical conditions and excitation mechanisms of warm molecular gas in individual star-forming regions, 
we performed \textit{Herschel} SPIRE Fourier Transform Spectrometer (FTS) observations of 30 Doradus in the Large Magellanic Cloud (LMC). 
In our FTS observations, important far-infrared (FIR) cooling lines in the interstellar medium (ISM),
including CO $J$=4--3 to $J$=13--12, \CI 370 $\mu$m, and \NII 205 $\mu$m, were clearly detected. 
In combination with ground-based CO $J$=1--0 and $J$=3--2 data, 
we then constructed CO spectral line energy distributions (SLEDs) on $\sim$10 pc scales over a $\sim$60 pc $\times$ 60 pc area 
and found that the shape of the observed CO SLEDs considerably changes across 30 Doradus, e.g., 
the peak transition $J_{\rm p}$ varies from $J$=6--5 to $J$=10--9, 
while the slope characterized by the high-to-intermediate $J$ ratio $\alpha$ ranges from $\sim$0.4 to $\sim$1.8. 
To examine the source(s) of these variations in CO transitions, we analyzed the CO observations,   
along with \CII 158 $\mu$m, \CI 370 $\mu$m, \OI 145 $\mu$m, H$_{2}$ 0--0 S(3), and FIR luminosity data, 
using state-of-the-art models of photodissociation regions (PDRs) and shocks. 
Our detailed modeling showed that the observed CO emission likely originates from 
highly-compressed (thermal pressure $P/k_{\rm B}$ $\sim$ 10$^{7}$--10$^{9}$ K cm$^{-3}$) clumps on $\sim$0.7--2 pc scales, 
which could be produced by either ultraviolet (UV) photons (UV radiation field $G_{\rm UV}$ $\sim$ 10$^{3}$--10$^{5}$ Mathis fields)
or low-velocity C-type shocks (pre-shock medium density $n_{\rm pre}$ $\sim$ 10$^{4}$--10$^{6}$ cm$^{-3}$ and 
shock velocity $\varv_{\rm s}$ $\sim$ 5--10 km s$^{-1}$).
Considering the stellar content in 30 Doradus, however, we tentatively excluded the stellar origin of CO excitation 
and concluded that low-velocity shocks driven by $\sim$kpc scale processes 
(e.g., interaction between the Milky Way and the Magellanic Clouds) are likely the dominant source of heating for CO.
The shocked CO-bright medium was then found to be warm (temperature $T$ $\sim$ 100--500 K) 
and surrounded by a UV-regulated low pressure component ($P/k_{\rm B}$ $\sim$ a few (10$^{4}$--10$^{5}$) K cm$^{-3}$)
that is bright in \CII 158 $\mu$m, \CI 370 $\mu$m, \OI 145 $\mu$m, and FIR dust continuum emission.} 


   \maketitle
%

\section{Introduction} 
\label{s:intro} 

As a nascent fuel for star formation, molecular gas plays an important role in the evolution of galaxies (e.g., \citealt{Kennicutt12}). 
The rotational transitions of carbon monoxide (CO)\footnote{In this paper, we focus on $^{12}$CO and refer to it as CO.} 
have been the most widely used tracer of such a key component of the interstellar medium (ISM) 
and in particular enable the examination of the physical conditions of molecular gas in diverse environments
(e.g., kinetic temperature $T_{\rm k}$ $\sim$ 10--1000 K and density $n$ $\sim$ 10$^{3}$--10$^{8}$ cm$^{-3}$)
thanks to their large range of critical densities. 

\begin{figure*}
\centering
\includegraphics[scale=0.23]{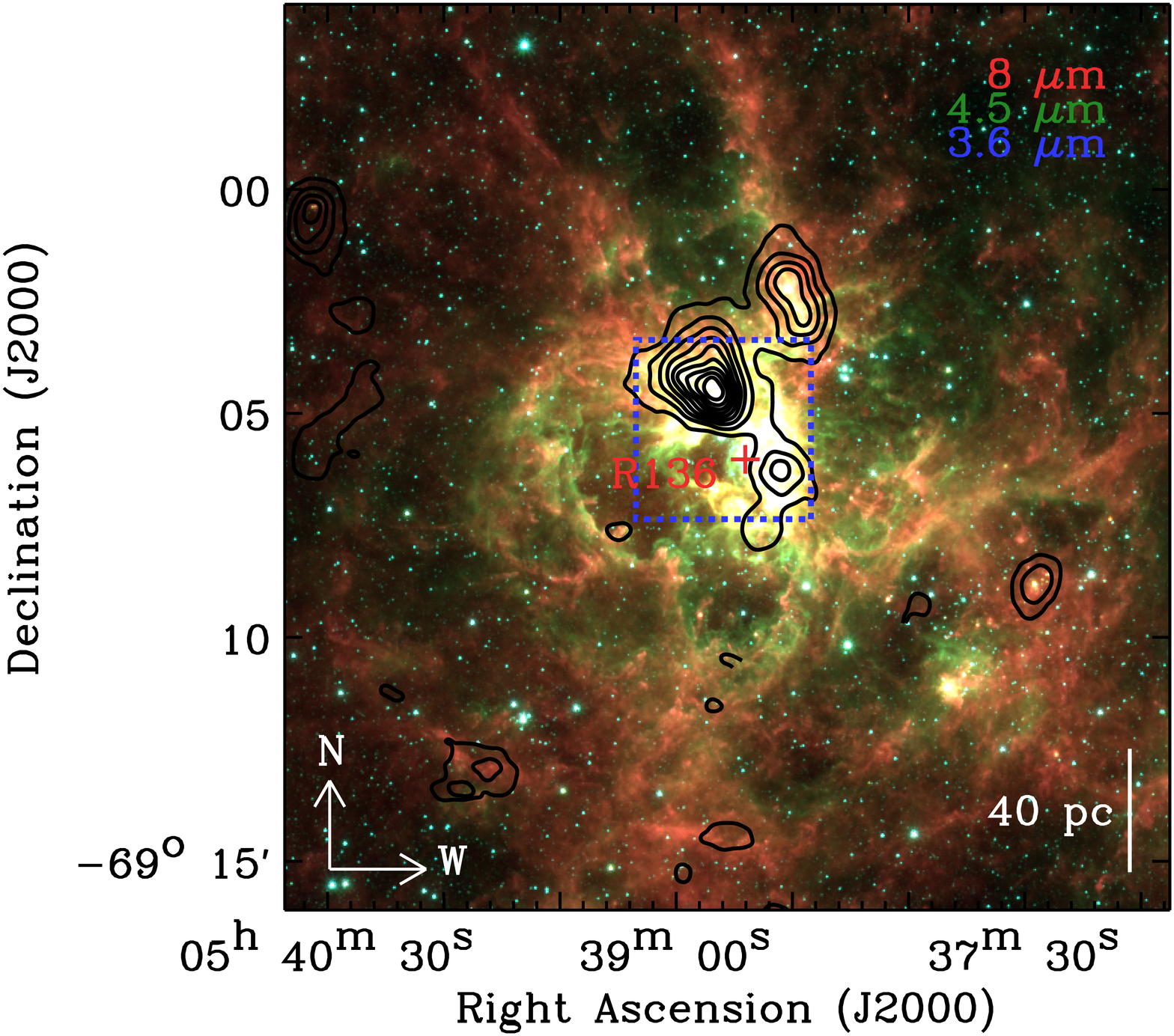}\hspace{0.8cm} 
\includegraphics[scale=0.2]{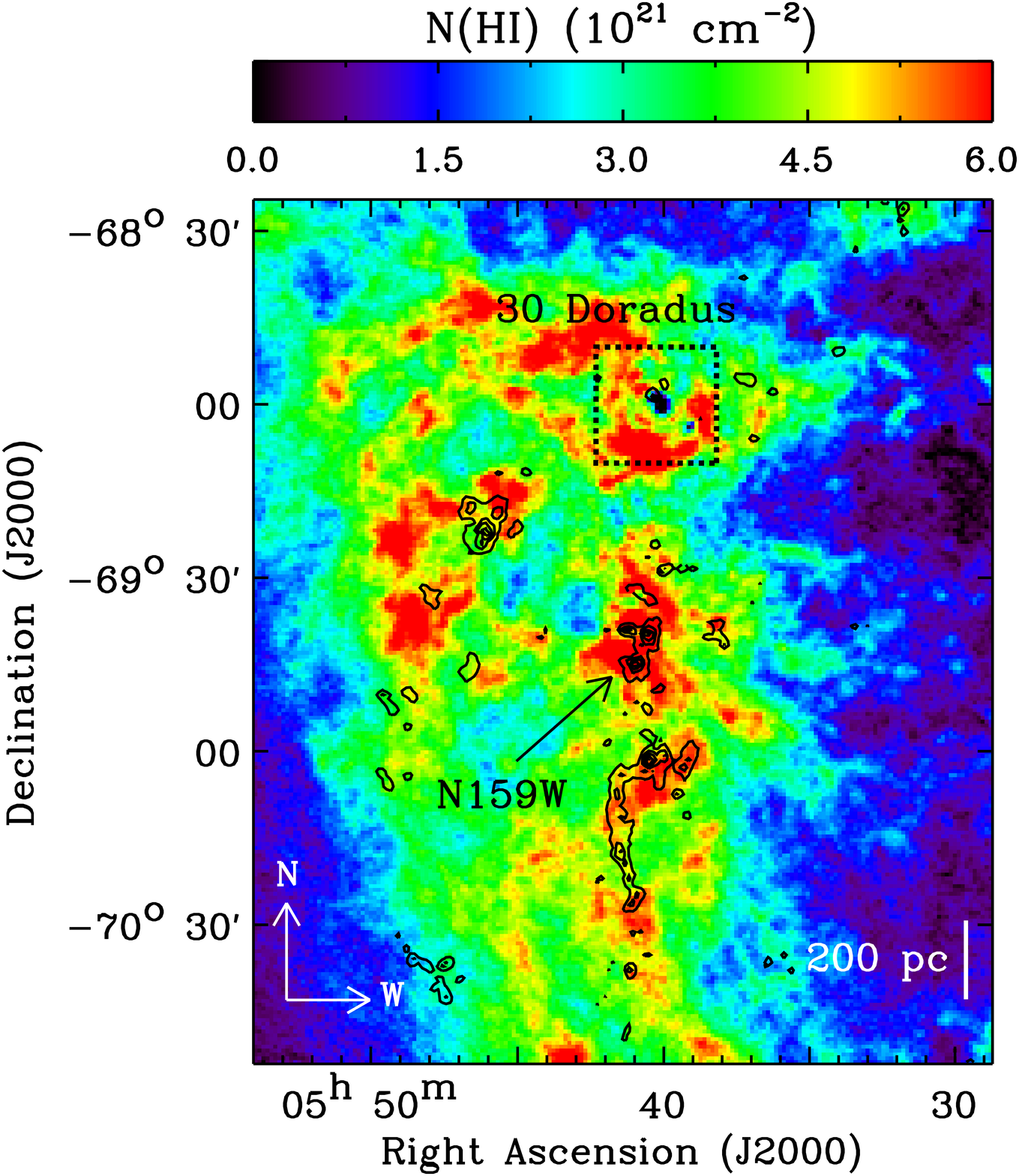}
\caption{\label{f:intro} \textit{Left}: Three-color composite image of 30 Doradus 
(\textit{Spitzer} 3.6 $\mu$m, 4.5 $\mu$m, and 8 $\mu$m in blue, green, and red; \citealt{Meixner06}). 
The central star cluster R136 is marked as the red cross, 
and the FTS coverage is outlined in blue. 
Additionally, the CO(1--0) integrated intensity from the MAGMA survey (\citealt{Wong11}; Sect. \ref{s:ground_based_CO}) 
is overlaid as the black contours with levels ranging from 10\% to 90\% of the peak (16.6 K km s$^{-1}$) in 10\% steps. 
\textit{Right}: \HI column density image from \cite{Kim98}.  
The MAGMA CO(1--0) integrated intensity is shown as the black contours (10\% to 90\% of the peak value, 39.5 K km s$^{-1}$, in 20\% steps), 
and the coverage of the \textit{left} image is indicated as the black dashed box. 
This large \HI structure, where 30 Doradus and N159W are located, corresponds to the southeastern \HI overdensity region of the LMC 
(e.g., \citealt{D'onghia16}).}
\end{figure*}

The diagnostic power of CO rotational transitions has been explored to a greater extent 
since the advent of the ESA \textit{Herschel Space Observatory} (\citealt{Pilbratt10}).  
The three detectors on board \textit{Herschel}, 
PACS (Photodetector Array Camera and Spectrometer; \citealt{Poglitsch10}), 
SPIRE (Spectral and Photometric Imaging Receiver; \citealt{Griffin10}), 
and HIFI (Heterodyne Instrument for the Far Infrared; \citealt{deGraauw10}), 
provided access to a wavelength window of $\sim$50--670 $\mu$m 
and enabled the studies of CO spectral line energy distributions (CO SLEDs) up to the upper energy level $J_{\rm u}$ = 50
for Galactic and extragalactic sources including 
photodissociation regions (PDRs; e.g., \citealt{Habart10}; \citealt{Kohler14}; \citealt{Stock15}; \citealt{Joblin18}; \citealt{RWu18}), 
protostars (e.g., \citealt{Larson15}), 
infrared (IR) dark clouds (e.g., \citealt{Pon16}), 
IR bright galaxies (e.g., \citealt{Rangwala11}; \citealt{Kamenetzky12}; \citealt{Meijerink13}; 
\citealt{Pellegrini13}; \citealt{Papadopoulos14}; \citealt{Rosenberg14a}; \citealt{Schirm14}; \citealt{Mashian15}; \citealt{RWu15}), 
and Seyfert galaxies (e.g., \citealt{vanderWerf10}; \citealt{Hailey-Dunsheath12}; \citealt{Israel14}). 
These studies have revealed the ubiquitous presence of warm molecular gas ($T_{\rm k}$ $\gtrsim$ 100 K) 
and have proposed various radiative (e.g., ultraviolet (UV) photons, X-rays, and cosmic-rays) 
and mechanical (e.g., shocks) heating sources for CO excitation. 
As the dominant contributor to the total CO luminosity of galaxies ($\sim$70\%; e.g., \citealt{Kamenetzky17}), 
the warm CO is an important phase of the molecular medium.  
Understanding its physical conditions and excitation mechanisms would hence be critical to
fully assess different molecular reservoirs and their roles in the evolution of galaxies.  

While previous \textit{Herschel}-based studies have considered various types of objects, 
they have primarily focused on either small-scale Galactic ($\sim$1 pc or smaller) or large-scale extragalactic ($\sim$1 kpc or larger) sources.  
As recently pointed out by \cite{Indriolo17}, CO SLEDs are affected not only by heating sources, but also by a beam dilution effect, 
suggesting that it is important to examine a wide range of physical scales to comprehensively understand the nature of warm molecular gas in galaxies. 
To bridge the gap in the previous studies and provide insights into the excitation mechanisms of 
warm CO on intermediate scales ($\sim$10--100 pc), 
we then conducted \textit{Herschel} SPIRE Fourier Transform Spectrometer (FTS) observations 
of several star-forming regions in the Large Magellanic Cloud
(LMC) (distance of $\sim$50 kpc and metallicity of $\sim$1/2 Z$_{\odot}$; e.g., \citealt{Russell92}; \citealt{Schaefer08}). 
The first of our LMC studies was \cite{Lee16}, 
where we analyzed \textit{Herschel} observations of the N159W star-forming region (e.g., Fig. \ref{f:intro} of \citealt{Lee16}) 
along with complementary ground-based CO data at $\sim$10 pc resolution.
Specifically, we examined CO transitions from $J$=1--0 to $J$=13--12 ($J$=2--1 not included) over a $\sim$40 pc $\times$ 40 pc area 
by using the non-LTE (Local Thermodynamic Equilibrium) radiative transfer model RADEX (\citealt{vanderTak07})
and found that the CO-emitting gas in N159W is warm ($T_{\rm k}$ $\sim$ 150--750 K) and moderately dense ($n \sim$ a few 10$^{3}$ cm$^{-3}$).
To investigate the origin of this warm molecular gas, we evaluated the impact of several radiative and mechanical heating sources
and concluded that low-velocity C-type shocks ($\sim$10 km s$^{-1}$) provide sufficient energy for CO heating, 
while UV photons regulate fine-structure lines \CII 158 $\mu$m, \CI 370 $\mu$m, and \OI 145 $\mu$m. 
High energy photons and particles including X-rays and cosmic-rays were found not to be significant for CO heating.  

\begin{table*}[t]
\small
\begin{center}
\caption{\label{t:table1} Spectral line and dust continuum data in our study.}
\begin{tabular}{l c c c c c c c} \toprule
Transition & Rest wavelength & $E_{\rm u}^{a}$ & \textit{FWHM}$^{b}$ & $\sigma_{\rm s,med}^{c,e,f}$ & $\sigma_{\rm f,med}^{d,e,f}$ & Reference$^{g}$ \\ 
& ($\mu$m) & (K) & ($''$) & (10$^{-11}$ W m$^{-2}$ sr$^{-1}$) & (10$^{-11}$ W m$^{-2}$ sr$^{-1}$) & \\ \midrule
$^{12}$CO $J$=1--0 & 2600.8 & 6 & 45 & 0.1 & 0.1 & (1) \\ 
$^{12}$CO $J$=3--2 & 867.0 & 33 & 22 & 2.2 & 5.1 & (2) \\
$^{12}$CO $J$=4--3 & 650.3 & 55 & 42 & 24.0 & 26.5 & (3) \\ 
$^{12}$CO $J$=5--4 & 520.2 & 83 & 34 & 15.9 & 20.7 & (3) \\ 
$^{12}$CO $J$=6--5 & 433.6 & 116 & 29 & 9.2 & 16.7 & (3) \\ 
$^{12}$CO $J$=7--6 & 371.7 & 155 & 33 & 7.4 & 18.4 & (3) \\ 
$^{12}$CO $J$=8--7 & 325.2 & 199 & 33 & 18.7 & 25.3 & (3) \\ 
$^{12}$CO $J$=9--8 & 289.1 & 249 & 19 & 25.2 & 30.2 & (3) \\
$^{12}$CO $J$=10--9 & 260.2 & 304 & 18 & 27.6 & 30.6 & (3) \\ 
$^{12}$CO $J$=11--10 & 236.6 & 365 & 17 & 28.7 & 32.2 & (3) \\
$^{12}$CO $J$=12--11 & 216.9 & 431 & 17 & 26.6 & 30.3 & (3) \\ 
$^{12}$CO $J$=13--12 & 200.3 & 503 & 17 & 39.8 & 43.0 & (3) \\
\CI $^{3}P_{1}$--$^{3}P_{0}$ & 609.1 & 24 & 38 & 28.3 & 29.2 & (3) \\ 
\CI $^{3}P_{2}$--$^{3}P_{1}$ & 370.4 & 62 & 33 & 7.0 & 7.9 & (3) \\ 
\CII $^{2}P_{3/2}$--$^{2}P_{1/2}$ & 157.7 & 91 & 12 & 294.6 & 8092.0 & (4) \\ 
\OI $^{3}P_{0}$--$^{3}P_{1}$ & 145.5 & 327 & 12 & 110.2 & 788.0 & (4) \\
\NII $^{3}P_{1}$--$^{3}P_{0}$ & 205.2 & 70 & 17 & 43.6 & 99.0 & (3) \\ 
H$_{2}$ 0--0 S(3) & 9.7 & 2504 & 6 & -- & 1036.0 & (3,5) \\ 
FIR & 60--200 & -- & 42 & -- & 2.9 $\times$ 10$^{6}$ & (3,4) \\
\bottomrule
\end{tabular}
\end{center} 
{$^{(a)}$ Upper level energy. 
$^{(b)}$ Angular resolution of the original data. 
$^{(c)}$ Median $\sigma_{\rm s}$ (statistical 1$\sigma$ uncertainty) on 42$''$ scales. 
$^{(d)}$ Median $\sigma_{\rm f}$ (final 1$\sigma$ uncertainty; statistical and calibration errors added in quadrature) on 42$''$ scales. 
$^{(e)}$ CO(1--0) is exceptionally on 45$''$ scales. 
$^{(f)}$ All pixels are considered.
$^{(g)}$ (1) \cite{Wong11}; (2) \cite{Minamidani08}; (3) This work; (4) \cite{Chevance16}; (5) \cite{Indebetouw09}}
\end{table*}

In this paper, we extend our previous analyses to 30 Doradus (or 30 Dor), the most extreme starburst in the Local Universe. 
30 Doradus harbors more than 1000 massive young stars, e.g., OB-type and Wolf-Rayet (W-R) stars (\citealt{Doran13}), 
and emits $\sim$500 times more ionizing photons than the Orion Nebula (\citealt{Pellegrini10}), producing a giant \HII region. 
In particular, 30 Doradus is primarily powered by the central super star cluster R136, 
which has an extremely high stellar surface density ($\gtrsim$ 10$^{7}$ M$_{\odot}$ pc$^{-3}$; e.g., \citealt{Selman13})
along with the most massive stars known ($\gtrsim$ 150 M$_{\odot}$; e.g., \citealt{Crowther10}). 
R136 is surrounded by vast cavities and bubble-like structures,  
which were likely created by strong stellar winds and supernova explosions (SNe) 
with a total energy of $\sim$10$^{52}$ erg (e.g., \citealt{Chu94}; \citealt{Townsley06a}). 
All in all, these extraordinary star formation activities make 30 Doradus an ideal laboratory 
for examining the impact of radiative and mechanical feedback into the surrounding ISM. 
In Fig. \ref{f:intro}, we show 30 Doradus and its surrounding environment 
(\HI overdensity region where 30 Doradus and N159W are located) in several tracers of gas and dust. 

This paper is organized as follows. 
In Sect. \ref{s:30Dor}, we provide a summary on recent studies of 30 Doradus that are most relevant to our work. 
In Sects. \ref{s:data} and \ref{s:results}, we present the multi-wavelength datasets used in our study 
and describe the spatial distribution of CO and \CI emission, as well as the observed CO SLEDs. 
In Sects. \ref{s:excitation_sources} and \ref{s:discussions}, we then employ state-of-the-art theoretical models of PDRs and shocks 
to examine the physical conditions and excitation mechanisms of CO in 30 Doradus. 
Finally, we summarize the results from our study in Sect. \ref{s:summary}. 




\section{Characteristics of 30 Doradus}
\label{s:30Dor} 

As noted in Sect. \ref{s:intro}, 30 Doradus is one of the best-studied star-forming regions. 
In this section, we summarize recent studies on 30 Doradus that are most relevant to our work. 

\subsection{Stellar content}
\label{s:30dor_stellar_content} 

When it comes to stellar feedback, massive young stars are considered to be a major player; 
their abundant UV photons create \HII regions and PDRs, 
while their powerful stellar winds sweep up the surrounding ISM into shells and bubbles.
The latest view on the massive young stars in 30 Doradus has been offered from the VLT-FLAMES Tarantula Survey (\citealt{Evans11}), 
and we focus here on \cite{Doran13} where the first systematic census of hot luminous stars was presented. 
In \cite{Doran13}, 1145 candidate hot luminous stars were identified based on \textit{UBV} band photometry, 
and $\sim$500 of these stars were spectroscopically confirmed (including 469 OB-type stars and 25 W-R stars). 
The total ionizing and stellar wind luminosities were then estimated to be $\sim$10$^{52}$ photons s$^{-1}$ 
and $\sim$2 $\times$ 10$^{39}$ erg s$^{-1}$ respectively, 
and $\sim$75\% of these luminosities were found to come from the inner 20 pc of 30 Doradus. 
This implies that stellar feedback is highly concentrated in the central cluster R136, 
where one third of the total W-R stars reside along with a majority of the most massive O-type stars. 
As for the age of stellar population, \cite{Doran13} showed that the ionizing stars in 30 Doradus span multiple ages: 
mostly 2--5 Myr with an extension beyond 8 Myr.  

\subsection{Properties of the neutral gas} 
\label{s:30dor_pdr} 

The impact of UV photons on the neutral gas in 30 Doradus was recently studied in detail by \cite{Chevance16}. 
The authors focused on \textit{Herschel} PACS observations of traditional PDR tracers, 
including \CII 158 $\mu$m and \OI 63 $\mu$m and 145 $\mu$m, 
and found that \CII and \OI mostly arise from the neutral medium (PDRs), while \OI 63 $\mu$m is optically thick. 
The observed \CII 158 $\mu$m and \OI 145 $\mu$m were then combined with an image of IR luminosity 
to estimate the thermal pressure of $\sim$(1--20) $\times$ 10$^{5}$ K cm$^{-3}$ 
and the UV radiation of $\sim$(1--300) $\times$ 10$^{2}$ Mathis fields (\citealt{Mathis83}) 
via Meudon PDR modeling  (\citealt{LePetit06}) on 12$''$ scales ($\sim$3 pc). 
In addition, the three-dimensional structure of PDRs was inferred 
based on a comparison between the stellar UV radiation field and the incident UV radiation field determined from PDR modeling:
PDR clouds in 30 Doradus are located at a distance of $\sim$20--80 pc from the central cluster R136. 

As for the molecular ISM in 30 Doradus, \cite{Indebetouw13} provided the sharpest view ever ($\sim$2$''$ or $\sim$0.5 pc scales) 
based on ALMA CO(2--1), $^{13}$CO(2--1), and C$^{18}$O(2--1) observations of the 30Dor-10 cloud (\citealt{Johansson98}). 
The main findings from their study include: 
(1) CO emission mostly arises from dense clumps and filaments on $\sim$0.3--1 pc scales; 
(2) Interclump CO emission is minor, suggesting that there is considerable photodissociation of CO molecules 
by UV photons penetrating between the dense clumps; 
(3) The mass of CO clumps does not change very much with distance from R136. 
More excited CO lines in 30 Doradus (up to $J$=6--5) were recently analyzed by \cite{Okada19}, 
and we discuss this work in detail in Appendix \ref{s:appendix5}. 

\begin{figure*}
\centering
\includegraphics[scale=0.4]{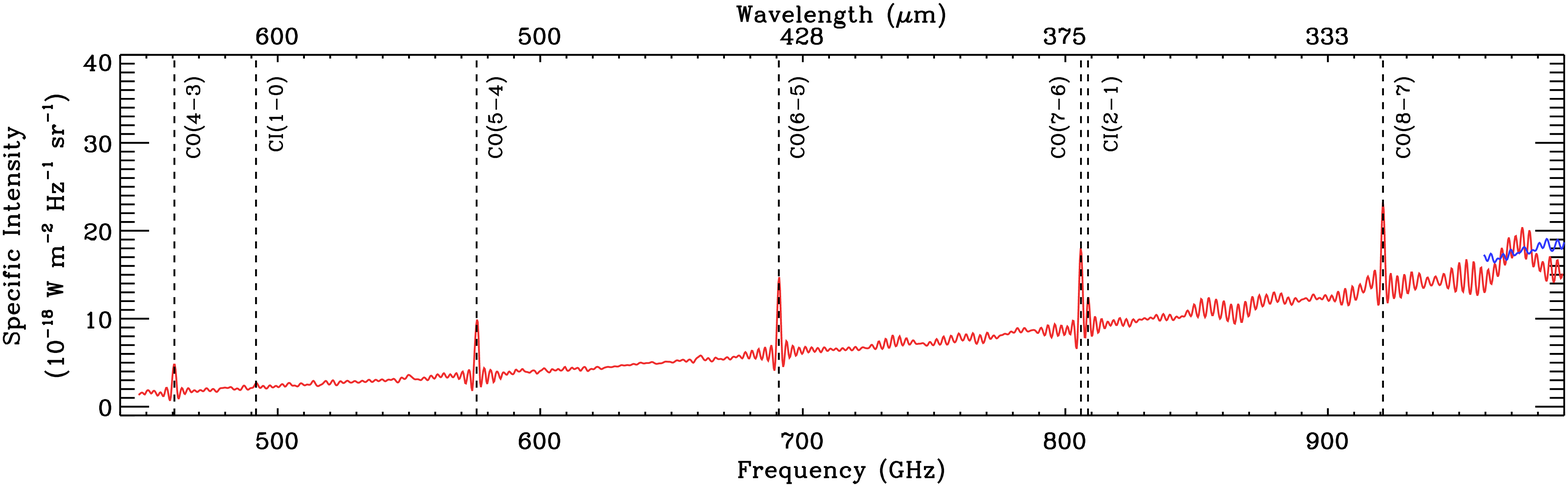}
\includegraphics[scale=0.4]{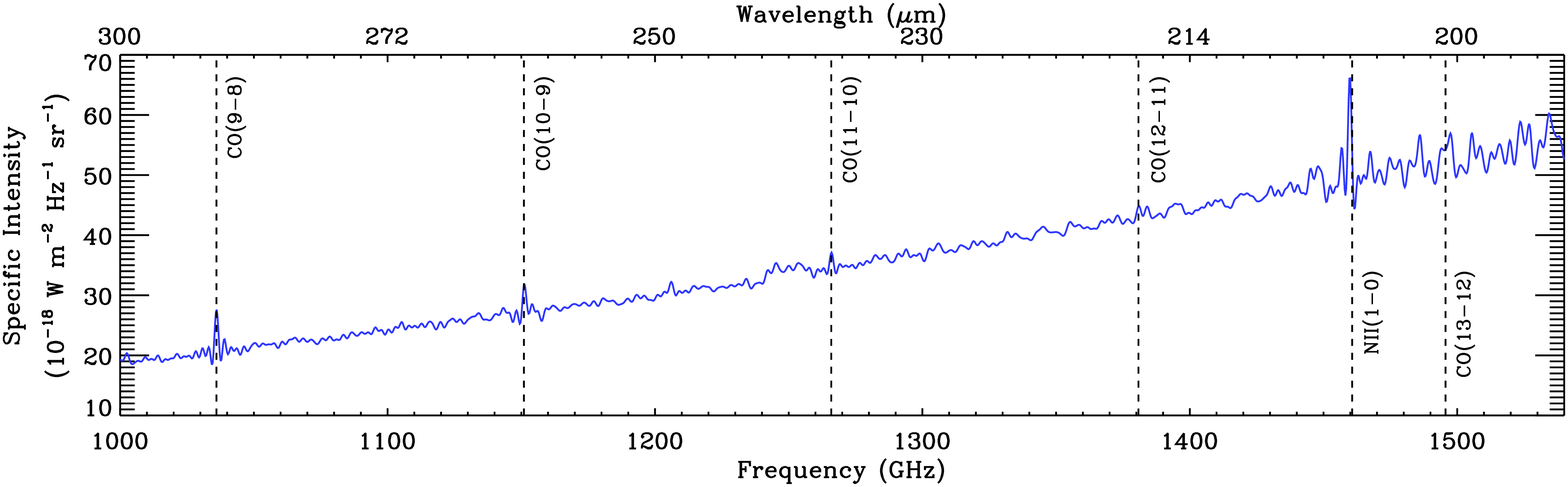}
\caption{\label{f:FTS_lines} Point source calibrated spectra from two central detectors, SLWC3 (red) and SSWD4 (blue). 
These spectra are from the first jiggle position of the Obs. ID = 1342219550, 
and the locations of the two detectors are shown as the yellow and orange crosses in Fig. \ref{f:CO7_6}. 
Additionally, the spectral lines observed with the SPIRE FTS are indicated as the black dashed lines.
Note that no further data processing (e.g., baseline subtraction and smoothing) was done for the spectral lines here, 
which are at their original angular resolutions (e.g., Table \ref{t:table1}).} 
\end{figure*}

\begin{figure}
\centering
\includegraphics[scale=0.22]{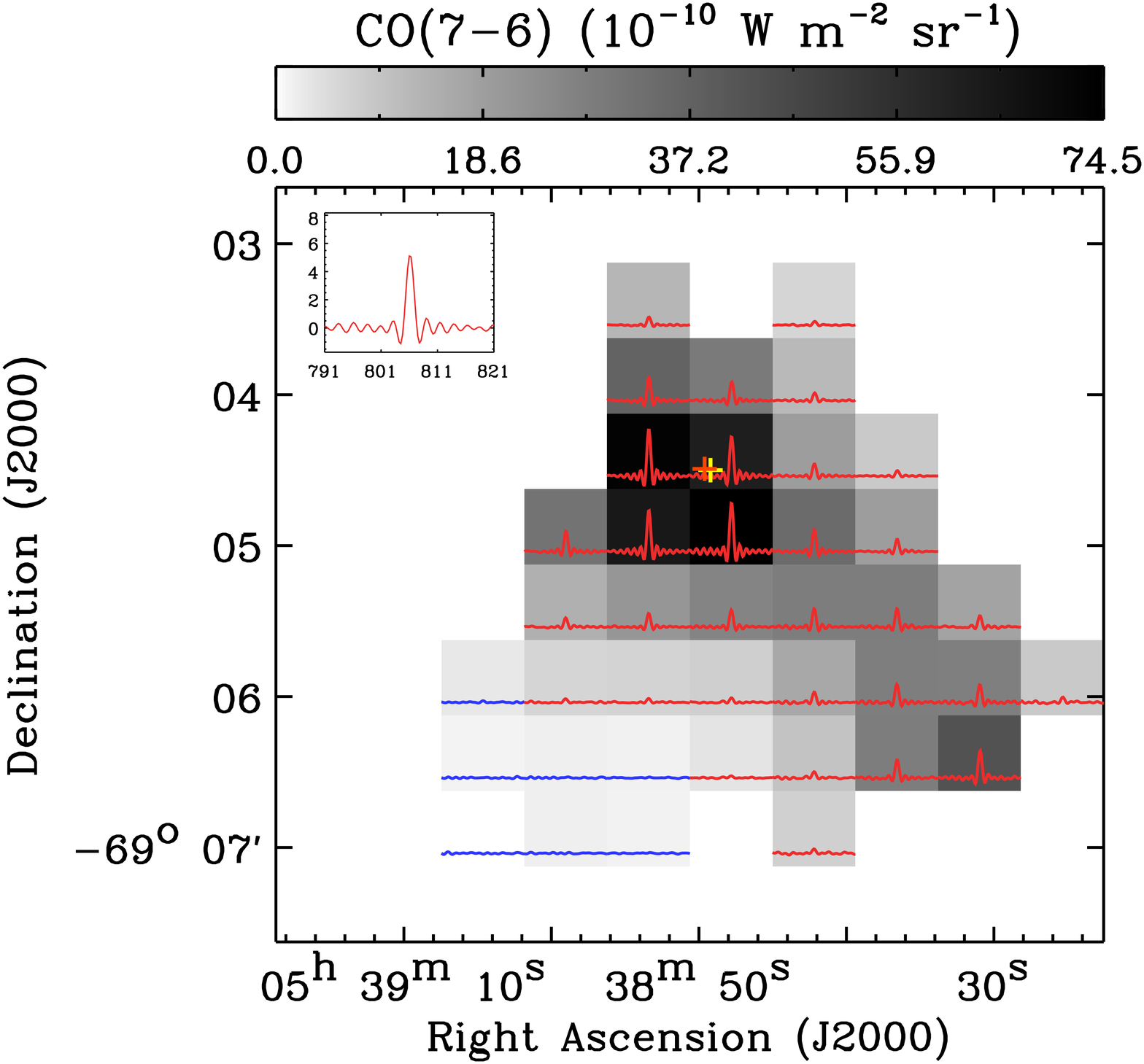} 
\caption{\label{f:CO7_6} CO(7--6) integrated intensity image (FWHM = 42$''$; pixel size = 30$''$). 
In our FTS observations, CO(7--6) is one of the brightest and most sensitive transitions 
(Table \ref{t:table1}; Sect. \ref{s:results}).   
Over the grayscale image, the spectrum of each pixel is overlaid, 
whose $x$- and $y$-axis ranges (in GHz and 10$^{-18}$ W m$^{-2}$ Hz$^{-1}$ sr$^{-1}$) are indicated in the \textit{top left} corner with an example spectrum. 
This spectrum is from the pixel that was observed with the two central detectors SLWC3 and SSWD4 (yellow and orange crosses) 
of the first jiggle observation of the Obs. ID = 1342219550. 
The spectra in red and blue represent detections and non-detections 
based on our threshold of $S/N_{\rm s}$ = 5 (statistical signal-to-noise ratio; Sect. \ref{s:FTS_CO_CI}).}
\end{figure}

\subsection{High energy photons}
\label{s:30dor_high_energy} 

30 Doradus has also been known as a notable source of high energy photons.
For example, \cite{Townsley06a,Townsley06b} presented \textit{Chandra} X-ray observations of 30 Doradus,  
where a convoluted network of diffuse structures (associated with superbubbles and supernova remnants (SNRs)), 
as well as $\sim$100 point sources (associated with O-type stars and W-R stars), were revealed.   
Thanks to the high spatial and spectral resolutions of the \textit{Chandra} observations, 
the authors were able to investigate the properties of the X-ray-emitting hot plasma in detail, 
estimating the temperature of (3--9) $\times$ 10$^{6}$ K and 
the surface brightness of (3--130) $\times$ 10$^{31}$ erg s$^{-1}$ pc$^{-2}$. 
In addition, \textit{Fermi} $\gamma$-ray observations recently showed that 
30 Doradus is the brightest source in the LMC with an emissivity of  
$\sim$3 $\times$ 10$^{-26}$ photons s$^{-1}$ sr$^{-1}$ per hydrogen atom (\citealt{Abdo10LMC}).
All in all, the presence of high energy photons in 30 Doradus suggests that 
strong stellar winds and SNe have injected 
a large amount of mechanical energy into the surrounding ISM, 
driving shocks and accelerating particles. 

\section{Data} 
\label{s:data}

In this section, we present the data used in our study. 
Some of the main characteristics of the datasets, including rest wavelengths, angular resolutions, and sensitivities, 
are listed in Table \ref{t:table1}. 

\subsection{Herschel SPIRE spectroscopic data} 
\label{s:FTS_data}

30 Doradus was observed with the SPIRE FTS in the high spectral resolution, intermediate spatial sampling mode 
(Obs. IDs: 1342219550, 1342257932, and 1342262908). 
The FTS consists of two bolometer arrays, SPIRE Long Wavelength (SLW) and SPIRE Short Wavelength (SSW), 
which cover the wavelength ranges of 303--671 $\mu$m and 194--313 $\mu$m. 
Depending on wavelength, the FTS beam size ranges from 17$''$ to 42$''$ 
(corresponding to 4--10 pc at the distance of the LMC; \citealt{Makiwa13}; \citealt{RWu15}). 
The SLW and SSW comprise 19 and 37 hexagonally packed detectors, which cover approximately 3$'$ $\times$ 3$'$. 
In the intermediate spatial sampling mode, these bolometer arrays are moved in a four-point jiggle with one beam spacing, 
resulting in sub-Nyquist-sampled data. 
Note that spectral lines are not resolved in our observations due to the insufficient frequency resolution of $\Delta f$ = 1.2 GHz 
(corresponding to the velocity resolution of $\Delta \varv$ $\sim$ 230--800 km s$^{-1}$ across the SLW and SSW). 

To derive integrated intensity images and their uncertainties, 
we essentially followed \cite{Lee16} and \cite{RWu15} and summarize our procedure here. 
First of all, we processed the FTS data using the \textit{Herschel} Interactive Processing Environment (HIPE) version 11.0, 
along with the SPIRE calibration version 11.0 (\citealt{Fulton10}; \citealt{Swinyard14}). 
As an example, the processed spectra from two central SLW and SSW detectors are presented in Fig. \ref{f:FTS_lines}, 
with the locations of the spectral lines observed with the SPIRE FTS. 
We then performed line measurement of point source calibrated spectra for each transition,  
where a linear combination of parabola and sinc functions was adopted to model the continuum and the emission line. 
The continuum subtracted spectra were eventually projected onto a 5$'$ $\times$ 5$'$ common grid with a pixel size of 15$''$ 
to construct a spectral cube. 
Finally, the integrated intensity ($I_{\rm CO}$, $I_{\rm CI}$, or $I_{\rm NII}$) was derived by carrying out line measurement of the constructed cube, 
and its final 1$\sigma$ uncertainty ($\sigma_{\rm f}$) was estimated by summing two sources of error in quadrature,  
$\sigma_{\rm f}$ = $\sqrt{\sigma_{\rm s}^2 + \sigma_{\rm c}^2}$,  
where $\sigma_{\rm s}$ is the statistical error derived from line measurement 
and $\sigma_{\rm c}$ is the calibration error of 10\% (SPIRE Observers' Manual\footnote{\url{http://herschel.esac.esa.int/Docs/SPIRE/html/spire\_om.html}}). 

Throughout our study, the FTS data were frequently combined with other tracers of gas and dust. 
To compare the different datasets at a common angular resolution, 
we then smoothed the FTS images to 42$''$, which corresponds to the FWHM of the FTS CO(4--3) observations,   
by employing the kernels from \cite{RWu15}. 
These kernels were created based on the fitting of a two-dimensional Hermite-Gaussian function to the FTS beam profile, 
taking into account the complicated dependence on wavelength. 
In addition, the smoothed images were rebinned to have a final pixel size of 30$''$, 
which roughly corresponds to the jiggle spacing of the SLW observations. 
We present the resulting integrated intensity maps in Fig. \ref{f:CO7_6} and Appendix \ref{s:appendix1} 
and refer to \cite{Lee16} and \cite{RWu15} for full details on the data reduction and map-making procedures. 
Line detections in our FTS observations are discussed in Sect. \ref{s:FTS_CO_CI}. 

We note that high resolution spectra of CO(4--3), CO(7--6), and \CI 370 $\mu$m ($\sim$25$''$--40$''$) were previously obtained for 30 Doradus 
by \cite{Pineda12} with the NANTEN2 telescope. 
The authors performed the observations as a single pointing toward 
($\alpha$, $\delta$)$_{\rm J2000}$ = (05$^{\rm h}$38$^{\rm m}$48.6$^{\rm s}$, $-69^{\circ}$04$'$43.2$''$),  
and we found that the NANTEN2-to-FTS ratios of the integrated intensities for this position are $\lesssim$ 1.2, 
suggesting that our intensity measurements are consistent with \cite{Pineda12} within 1$\sigma$ uncertainties.  


\subsection{Herschel PACS spectrosopic data} 
\label{s:PACS_data}

Following \cite{Lee16}, we used PACS \CII 158 $\mu$m and \OI 145 $\mu$m data of 30 Doradus.
These data were first presented in \cite{Chevance16}, 
and we here provide a brief summary on the observations and data reduction. 
Note that \OI 63 $\mu$m was not used for our study, since the line is optically thick throughout the mapped region 
(e.g., \OI 145 $\mu$m-to-\OI 63 $\mu$m ratio $>$ 0.1; \citealt{Tielens85a}; \citealt{Chevance16}). 

30 Doradus was mapped with the PACS spectrometer in the unchopped scan mode 
(Obs. IDs: 1342222085 to 1342222097 and 1342231279 to 1342231285). 
As an integral field spectrometer, the PACS consists of 25 (spatial) $\times$ 16 (spectral) pixels 
and covers 51--220 $\mu$m with a field-of-view of 47$''$ $\times$ 47$''$ (\citealt{Poglitsch10}). 
The \CII 158 $\mu$m and \OI 145 $\mu$m fine-structure lines were observed in 31 and 11 raster positions, 
covering approximately 4$'$ $\times$ 5$'$ over the sky.  
The beam size of the spectrometer at 160 $\mu$m is 12$''$ (PACS Observers' Manual\footnote{\url{http://herschel.esac.esa.int/Docs/PACS/html/pacs\_om.html}}). 
  
The obtained observations were reduced using the HIPE version 12.0 (\citealt{Ott10}) from Level 0 to Level 1. 
The reduced cubes were then processed with PACSman (\citealt{Lebouteiller12}) to derive integrated intensity maps. 
In essence, each spectrum was modeled with a combination of polynomial (baseline) and Gaussian (emission line) functions, 
and the measured line fluxes were projected onto a common grid with a pixel size of 3$''$. 
The final 1$\sigma$ uncertainty in the integrated intensity was then estimated by adding the statistical error from line measurement/map projection 
and the calibration error of 22\% in quadrature. 
For details on the observations, as well as the data reduction and map-making procedures, 
we refer to \cite{Lebouteiller12}, \cite{Cormier15}, and \cite{Chevance16}.    

In our study, the original PACS images were smoothed and rebinned to match the FTS resolution (42$''$) and pixel size (30$''$). 
This smoothing and rebinning procedure resulted in a total of 13 common pixels to work with 
(e.g., Fig. \ref{f:UV_sources}; mainly limited by the small coverage of the \OI 145 $\mu$m data), 
over which \CII 158 $\mu$m and \OI 145 $\mu$m were clearly detected with $S/N_{\rm s}$ $\gg$ 5. 

\begin{figure*}
\centering
\includegraphics[scale=0.22]{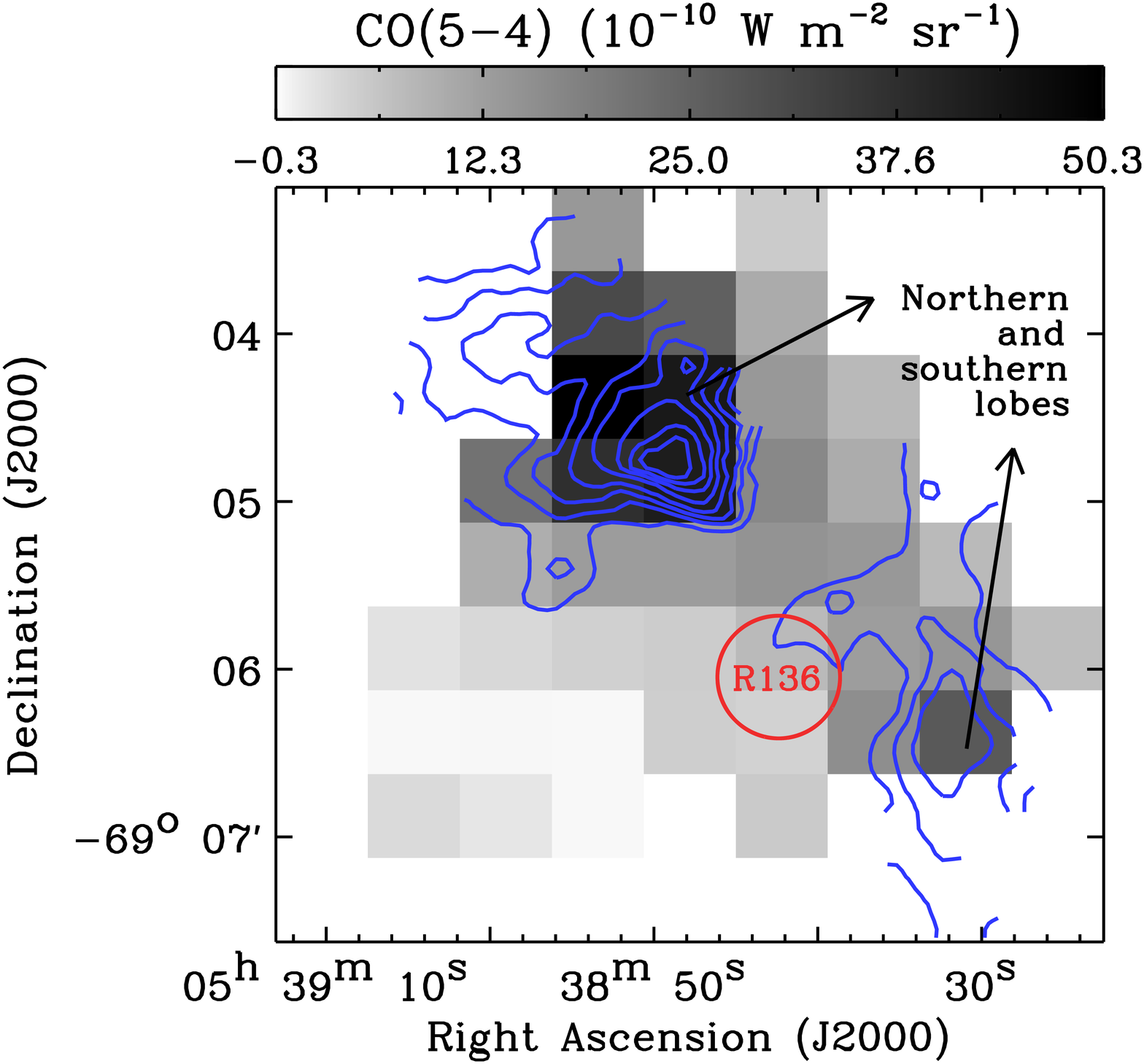}\hspace{1cm} 
\includegraphics[scale=0.22]{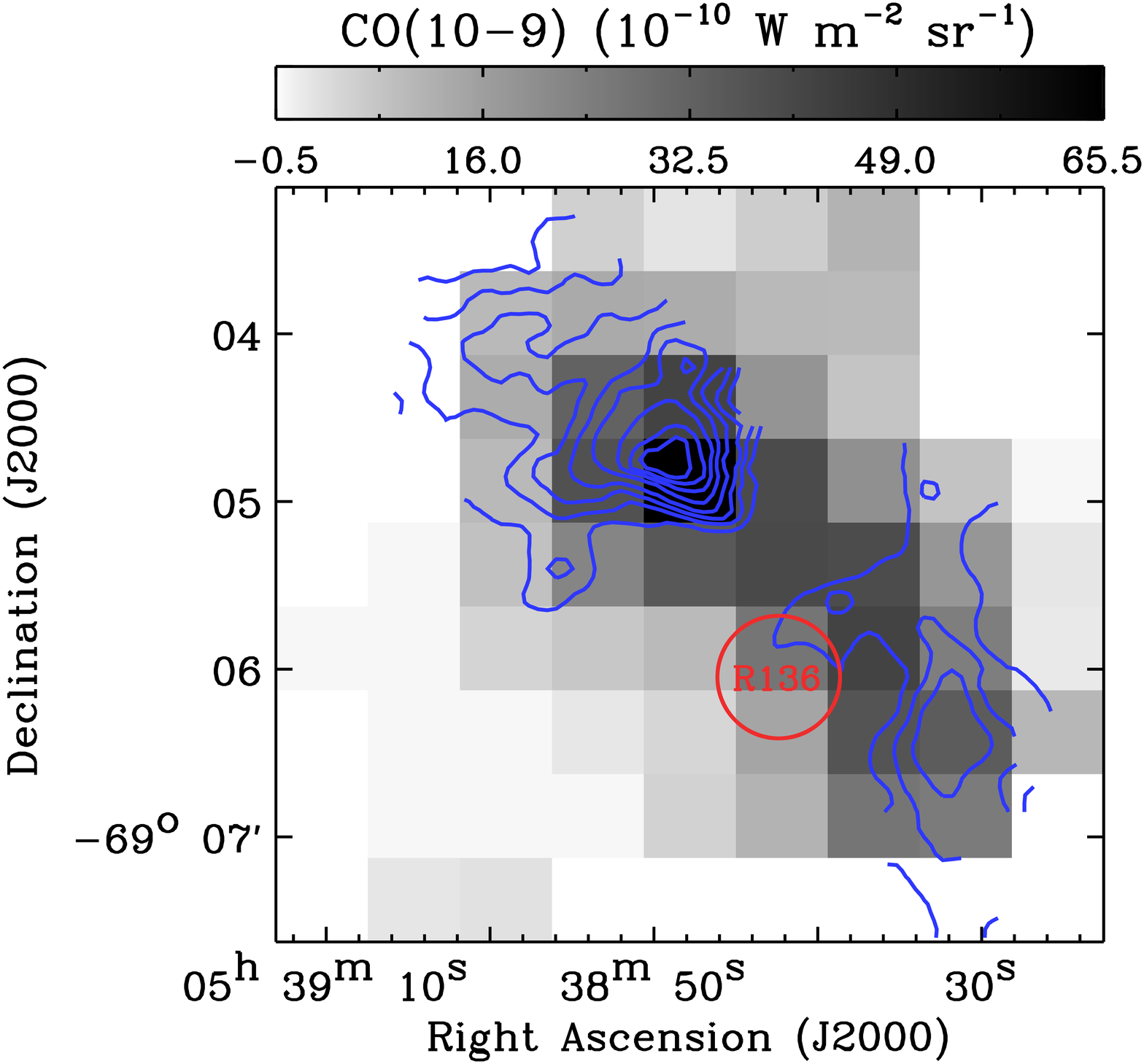}
\caption{\label{f:CO5_4_CII} Comparison between CO ($J$=5--4 and $J$=10--9 on the \textit{left} and \textit{right plots}) and \CII 158 $\mu$m (blue contours). 
The PACS \CII 158 $\mu$m data at the original resolution of 12$''$ are overlaid 
with levels ranging from 20\% to 90\% of the peak (2.2 $\times$ 10$^{-6}$ W m$^{-2}$ sr$^{-1}$) in 10\% steps.  
The location of the R136 cluster is indicated as the red circle.
Note that the grayscale bar goes below zero simply to show the pixels with low intensities.}
\end{figure*}

\subsection{Spitzer IRS H$_{2}$ data}
\label{s:IRS_data}  
 
In addition, we made use of \textit{Spitzer} InfraRed Spectrograph (IRS) observations of H$_{2}$ 0--0 S(3) in 30 Doradus.  
These observations were initially presented in \cite{Indebetouw09}, and we re-processed them as follows mainly to match the FTS resolution and pixel size.  
First, Basic Calibrated Data (BCD) products were downloaded 
from the \textit{Spitzer} Heritage Archive (SHA), and 
exposures were cleaned with IRSclean\footnote{\url{http://irsa.ipac.caltech.edu/data/SPITZER/docs/dataanalysistools/tools/irsclean/}} 
and combined using SMART\footnote{\url{http://irs.sirtf.com/IRS/SmartRelease}} (\citealt{Higdon04}; \citealt{Lebouteiller10}). 
The data were then imported into CUBISM 
(\citealt{Smith07}) for further cleaning and building a calibrated data cube 
with pixel sizes of 2$''$ and 5$''$ for the Short-Low (SL) and Long-Low (LL) modules. 

To produce a H$_{2}$ 0--0 S(3) map, we performed a Monte Carlo simulation  
where 100 perturbed cubes were created based on the calibrated data cube.   
These cubes were then convolved and resampled to have a resolution of 42$''$ and a pixel size of 30$''$, and 
spectral line fitting was performed using LMFIT (\citealt{Newville14}) for each cube. 
Finally, the line flux and associated uncertainty were calculated for each pixel  
using the median and median absolute deviation of the 100 measured flux values. 
While the resulting H$_{2}$ map is as large as the FTS CO maps, 
we found that the observations were not sensitive: only five pixels have detections with $S/N_{\rm s}$ $\sim$ 5.

\subsection{Ground-based CO data} 
\label{s:ground_based_CO} 

\begin{figure*}
\centering
\includegraphics[scale=0.53]{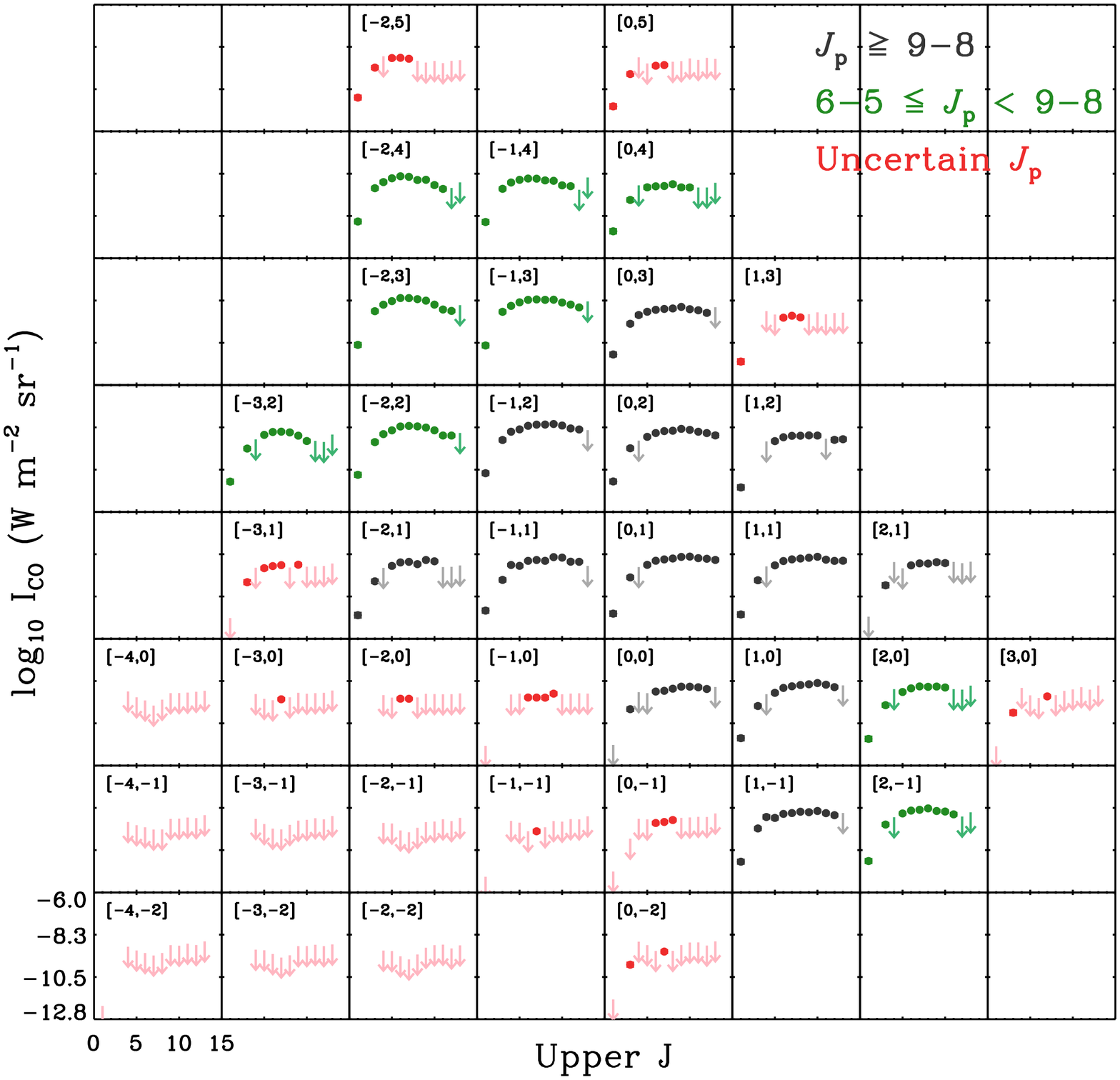} 
\caption{\label{f:obs_CO_SLEDs} CO SLEDs of 30 Doradus. 
Here each CO SLED was constructed by using all available CO transitions from $J_{\rm u}$ = 1 to 13 
for an individual data point in 30$''$ size. 
To indicate a location relative to the pixel that is closest to R136, 
a pair of numbers is shown in the \textit{top left} corner of each pixel, 
e.g., [0,0] corresponds to ($\alpha$, $\delta$)$_{\rm J2000}$ = (05$^{\rm h}$38$^{\rm m}$42$^{\rm s}$, $-69^{\circ}$05$'$53$''$). 
In addition, the circles and error bars (too small to be visible in many cases) show the measured intensities and 1$\sigma$ uncertainties for detections, 
while the downward arrows represent the upper limits based on $S/N_{\rm s}$ = 5 for non-detections. 
Finally, the CO SLEDs are presented in different colors depending on the transition they peak ($J_{\rm p}$): 
black ($J_{\rm p}$=9--8 or 10--9), green (6--5 $\leq$ $J_{\rm p}$ $<$ 9--8), and red (uncertain $J_{\rm p}$ due to many non-detections). 
The non-detections are then shown in lighter shades (gray, light green, and pink) to distinguish them from the detections.} 
\end{figure*}

We complemented our FTS CO observations with ground-based CO(1--0) and (3--2) data. 
The CO(1--0) data were taken from the MAGellanic Mopra Assessment (MAGMA) survey (\citealt{Wong11}), 
where the 22-m Mopra telescope was used to map CO(1--0) in the LMC on 45$''$ scales. 
Meanwhile, the CO(3--2) data were obtained by \cite{Minamidani08} on 22$''$ scales using the 10-m Atacama Submillimeter Telescope Experiment (ASTE) telescope. 
For both datasets, the final uncertainties in the integrated intensities were estimated in a similar manner as we did for our FTS CO observations: 
adding the statistical error derived from the root-mean-square (rms) noise per channel and the calibration error 
(25\% and 20\% for CO(1--0) and CO(3--2) respectively; \citealt{Lee16}) in quadrature. 
We smoothed and rebinned the CO(1--0)\footnote{In this paper, we used the CO(1--0) data at the original resolution of 45$''$, 
which is quite close to the FTS resolution of 42$''$, 
with a rebinned pixel size of 30$''$.}
and CO(3--2) maps to match the FTS data, 
leading to 31 and 26 pixels to work with respectively. 
Among these pixels, a majority (22 and 25 pixels for CO(1--0) and CO(3--2) respectively)  
had clear detections with $S/N_{\rm s}$ $>$ 5 (e.g., Fig. \ref{f:obs_CO_SLEDs}). 

\subsection{Derived dust and IR continuum properties} 
\label{s:dust_properties} 

Finally, we used the dust and IR continuum properties of 30 Doradus  
that were first estimated by \cite{Chevance16} at 12$''$ resolution 
based on the dust spectral energy distribution (SED) model of \cite{Galliano18}. 
The \cite{Galliano18} SED model employs the hierarchical Bayesian approach 
and considers realistic optical properties, stochastic heating, and the mixing of physical conditions in observed regions.
For our analyses, we essentially followed \cite{Chevance16} and 
constrained the far-IR luminosity (60--200 $\mu$m; $L_{\rm FIR}$) and $V$-band dust extinction ($A_{V}$) over the FTS coverage on 42$''$ scales. 
In our spatially resolved modeling of dust SEDs covering mid-IR to sub-mm, 
the amorphous carbon (AC) composition was considered along with the following free parameters: 
the total dust mass ($M_{\rm dust}$), PAH (polycyclic aromatic hydrocarbon)-to-dust mass ratio ($f_{\rm PAH}$), 
index for the power-law distribution of starlight intensities ($\alpha_{U}$), 
lower cut-off for the power-law distribution of starlight intensities ($U_{\rm min}$), 
range of starlight intensities ($\Delta U$), and mass of old stars ($M_{\rm star}$). 
For details on our dust SED modeling, we refer to \cite{Galliano18}.

\section{Results}
\label{s:results}

In this section, we mainly discuss the observed properties of the FTS lines, 
with a particular emphasis on CO and \CI emission. 
The spectra and integrated intensity images of the FTS lines are presented in Fig. \ref{f:CO7_6} and Appendix \ref{s:appendix1}.  

\subsection{Spatial distribution of CO and \CI emission} 
\label{s:FTS_CO_CI} 

Following \cite{Lee16}, we consider spectra with $S/N_{\rm s}$ 
(statistical signal-to-noise ratio; integrated intensity divided by $\sigma_{\rm s}$) $>$ 5 as detections 
and group CO transitions into three categories: 
low-$J$ for $J_{\rm u} \leq 5$, intermediate-$J$ for $6 \leq J_{\rm u} \leq 9$, and high-$J$ for $J_{\rm u} \geq 10$. 
In our FTS observations, all CO transitions from $J$=4--3 to $J$=13--12, as well as \CI 370 $\mu$m, were clearly detected. 
The sensitivity at $\sim$500 GHz, on the other hand, was not good enough for \CI 609 $\mu$m to be detected. 

In general, we found that CO ($J$=1--0 to 13--12; $J$=2--1 not included) and \CI 370 $\mu$m emission lines 
are distributed along the northern and southern lobes around R136, 
with primary and secondary peaks at ($\alpha$, $\delta$)$_{\rm J2000}$ $\sim$ (05$^{\rm h}$38$^{\rm m}$51$^{\rm s}$, $-69^{\circ}$04$'$38$''$)
and (05$^{\rm h}$38$^{\rm m}$38$^{\rm s}$, $-69^{\circ}$06$'$08$''$) (e.g., Fig \ref{f:CO5_4_CII}). 
This overall morphology is similar to that of PDR tracers, 
such as \CII 158 $\mu$m, \OI 145 $\mu$m, and PAH emission (\citealt{Chevance16}). 
A close examination, however, revealed that detailed distributions are slightly different between the transitions. 
For example, the region between the northern and southern lobes, 
($\alpha$, $\delta$)$_{\rm J2000}$ $\sim$ (05$^{\rm h}$38$^{\rm m}$45$^{\rm s}$, $-69^{\circ}$05$'$30$''$), 
becomes bright in intermediate- and high-$J$ CO emission,   
resulting in the declining correlation between CO lines and fine-structure lines. 
To be specific, we found that the Spearman rank correlation coefficient remains high ($\rho \sim$ 0.8--0.9) 
for \CII 158 $\mu$m and CO from $J$=1--0 to 8--7 ($J$=2--1 not included), 
while being low for $J$=9--8 and 10--9 ($\rho$ = 0.4 and 0.1).
\CI 370 $\mu$m was found to be strongly correlated with \CII 158 $\mu$m ($\rho$ = 0.9). 
For these estimates, we only considered detections and the transitions with sufficient number of detections. 
The decreasing correlation between \CII 158 $\mu$m and CO mainly results from the mid-region becoming bright in intermediate- and high-$J$ CO lines, 
indicating the spatial variations in CO SLEDs (Sect. \ref{s:observed_CO_SLEDs}). 
To illustrate this result, we show CO $J$=5--4 and 10--9 along with \CII 158 $\mu$m in Fig. \ref{f:CO5_4_CII}. 


\begin{figure*}
\centering 
\includegraphics[scale=0.2]{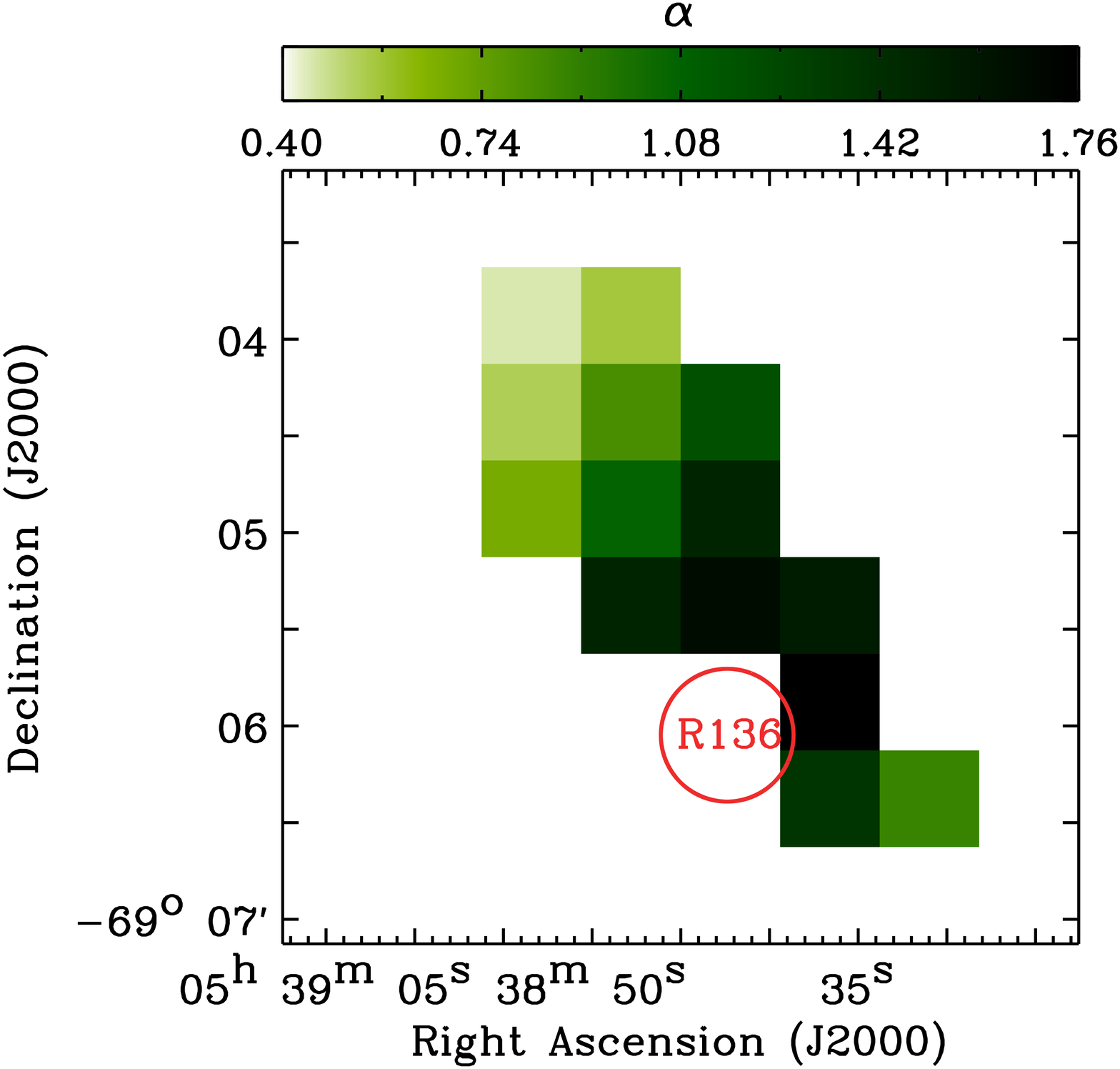}\hspace{1cm}
\includegraphics[scale=0.4]{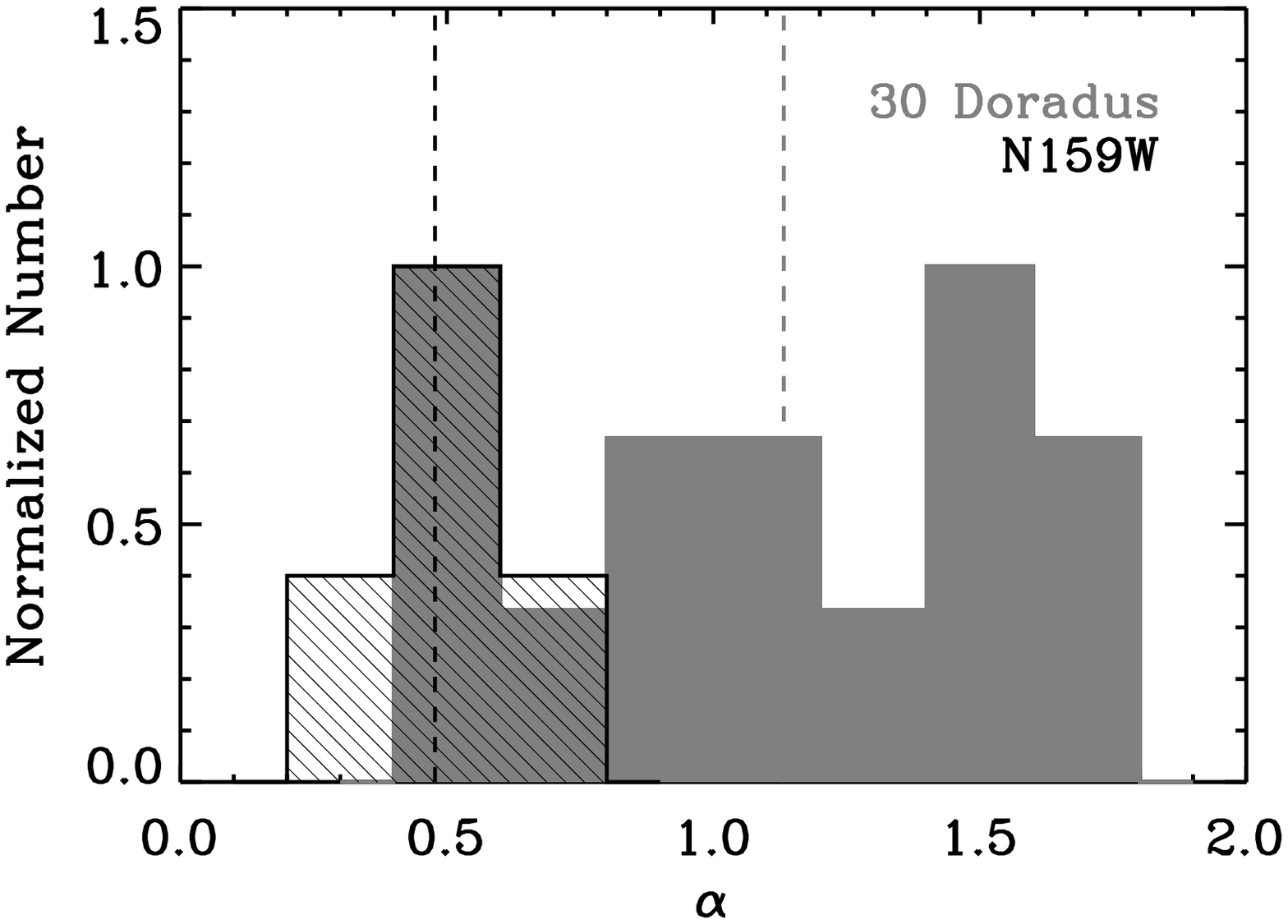} 
\caption{\label{f:alpha} \textit{Left}: High-to-intermediate-$J$ CO ratio ($\alpha$) for 30 Doradus.  
For the derivation of $\alpha$, 25 pixels were masked since they have non-detections for the required transitions. 
The location of R136 is also indicated as the red circle.
\textit{Right}: Comparison of $\alpha$ between 30 Doradus (gray, solid) and N159W (black, hatched). 
The $\alpha$ values of N159W were calculated by applying Eq. (\ref{eq:alpha}) to the data from \cite{Lee16}, 
and the median value of each histogram is shown as the dashed line.
For both 30 Doradus and N159W, the $\alpha$ values are on 42$''$ scales.}
\end{figure*}

\subsection{Observed CO SLEDs} 
\label{s:observed_CO_SLEDs}

The observed CO SLEDs of 30 Doradus are presented in Fig. \ref{f:obs_CO_SLEDs}. 
To construct these CO SLEDs, we first combined the FTS integrated intensity images 
with the ground-based CO(1--0) and (3--2) data at the common resolution of 42$''$. 
We then used different colors to indicate the CO SLEDs with different peak transitions 
(black, green, and red for $J_{\rm p} \geq$ 9--8, $J_{\rm p}$ $<$ 9--8, and uncertain $J_{\rm p}$;   
$J_{\rm p}$ = transition where a CO SLED peaks)
and marked the location of each pixel relative to R136. 

Fig. \ref{f:obs_CO_SLEDs} clearly shows that the shape of the CO SLEDs changes over the mapped region of 4$'$ $\times$ 4$'$ ($\sim$60 pc $\times$ 60 pc). 
For example, the majority (12 out of the total 21 pixels with certain $J_{\rm p}$) peak at $J$=9--8 or 10--9,  
while some have 6--5 $\leq$ $J_{\rm p}$ $<$ 9--8. 
In addition, the slope of the CO SLEDs varies substantially.  
To quantify the variation in the slopes, we then defined the high-to-intermediate-$J$ CO ratio ($\alpha$) as follows, 
\begin{equation}
\label{eq:alpha} 
\alpha = \frac{I_{\rm CO}(J\!=\!9\!-\!8) + I_{\rm CO}(J\!=\!10\!-\!9) + I_{\rm CO}(J\!=\!11\!-\!10)}{I_{\rm CO}(J\!=\!5\!-\!4) + I_{\rm CO}(J\!=\!6\!-\!5) + I_{\rm CO}(J\!=\!7\!-\!6)}, 
\end{equation} 
\noindent and estimated $\alpha$ on a pixel-by-pixel basis (Fig. \ref{f:alpha} \textit{Left}).
25 pixels were additionally masked in the process since they have non-detections for the required transitions. 
Note that we did not adopt the ``high-$J$ slope'', $\Delta I_{\rm CO,norm}\!=\![I_{\rm CO}(J_{\rm p}\!+\!3)\!-I_{\rm CO}(J_{\rm p})]\!/\!I_{\rm CO}(J_{\rm p})$,   
the parameter \cite{Lee16} used to characterize the observed CO SLEDs of N159W. 
This is because 
the high-$J$ slope, which measures a slope \textit{only beyond} $J_{\rm p}$, 
was found not to capture the more general shape around the peak of the CO SLEDs, 
e.g., our [$-1$,4] and [0,0] pixels would have comparable high-$J$ slopes of $-$0.24 despite their distinctly different CO SLEDs 
(the peak for [0,0] is much broader). 
In addition, we note that our $\alpha$ parameter is only slightly different from what \cite{Rosenberg15} adopted to classify 29 (U)LIRGs: 
we used CO $J$=9--8, 10--9, and 11--10 for the high-$J$ CO contribution instead of CO $J$=11--10, 12--11, and 13--12 
to better reflect the properties of the CO SLEDs observed in 30 Doradus, 
as well as to maximize the number of available transitions for the derivation of $\alpha$. 

We found that the derived $\alpha$ peaks around R136 with a value of 1.5--1.8 
and decreases radially down to $\sim$0.4, 
implying that the relative contribution of high-$J$ CO lines increases toward R136. 
Compared to N159W, another massive star-forming region in the LMC, 30 Doradus shows systematically higher $\alpha$ (Fig. \ref{f:alpha} \textit{right}). 
Specifically, the $\alpha$ values of N159W mostly trace the lower range of the 30 Doradus histogram with a factor of two lower median value ($\sim$0.5 vs. $\sim$1.1). 
This result indicates that the two regions have markedly different CO SLEDs, 
and we will revisit the shape of CO SLEDs as a probe of heating sources in Sect. \ref{s:CO_SLED_shape}. 

The varying $\alpha$, as well as the different $J_{\rm p}$ for the individual pixels, 
suggest that the properties of the CO-emitting gas change across 30 Doradus on 42$''$ or 10 pc scales. 
For example, the peak transition and slope of the CO SLEDs depend on the gas density and temperature, 
while the CO column density affects the overall line intensities. 
In the next sections, the physical conditions and excitation sources of the CO-emitting gas will be examined in a self-consistent manner 
based on state-of-the-art models of PDRs and shocks. 
In addition, the impact of high energy photons and particles, e.g., X-rays and cosmic-rays, on CO emission will also be assessed.  

\begin{figure*}
\centering
\includegraphics[scale=0.22]{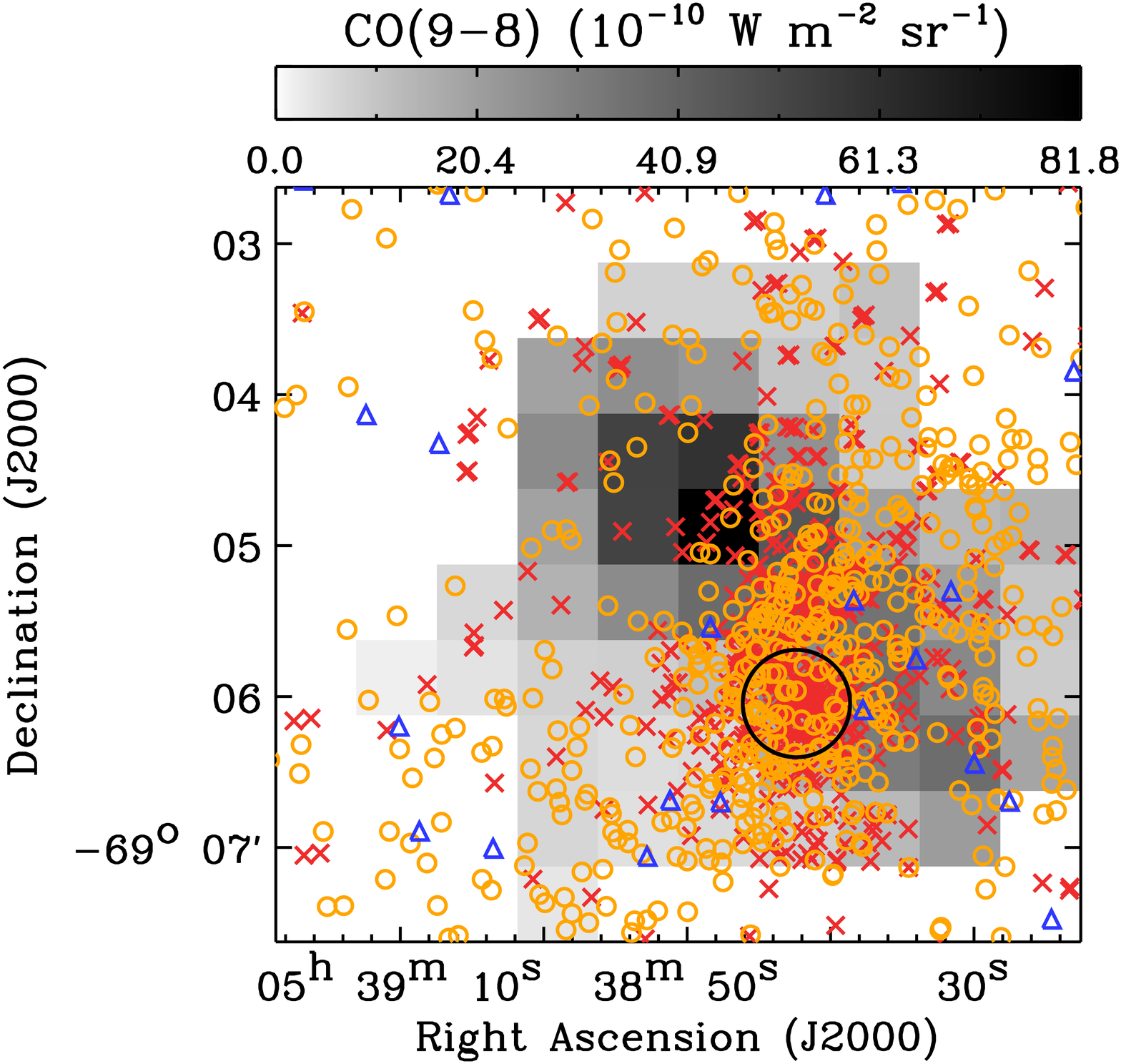}\hspace{1cm}
\includegraphics[scale=0.22]{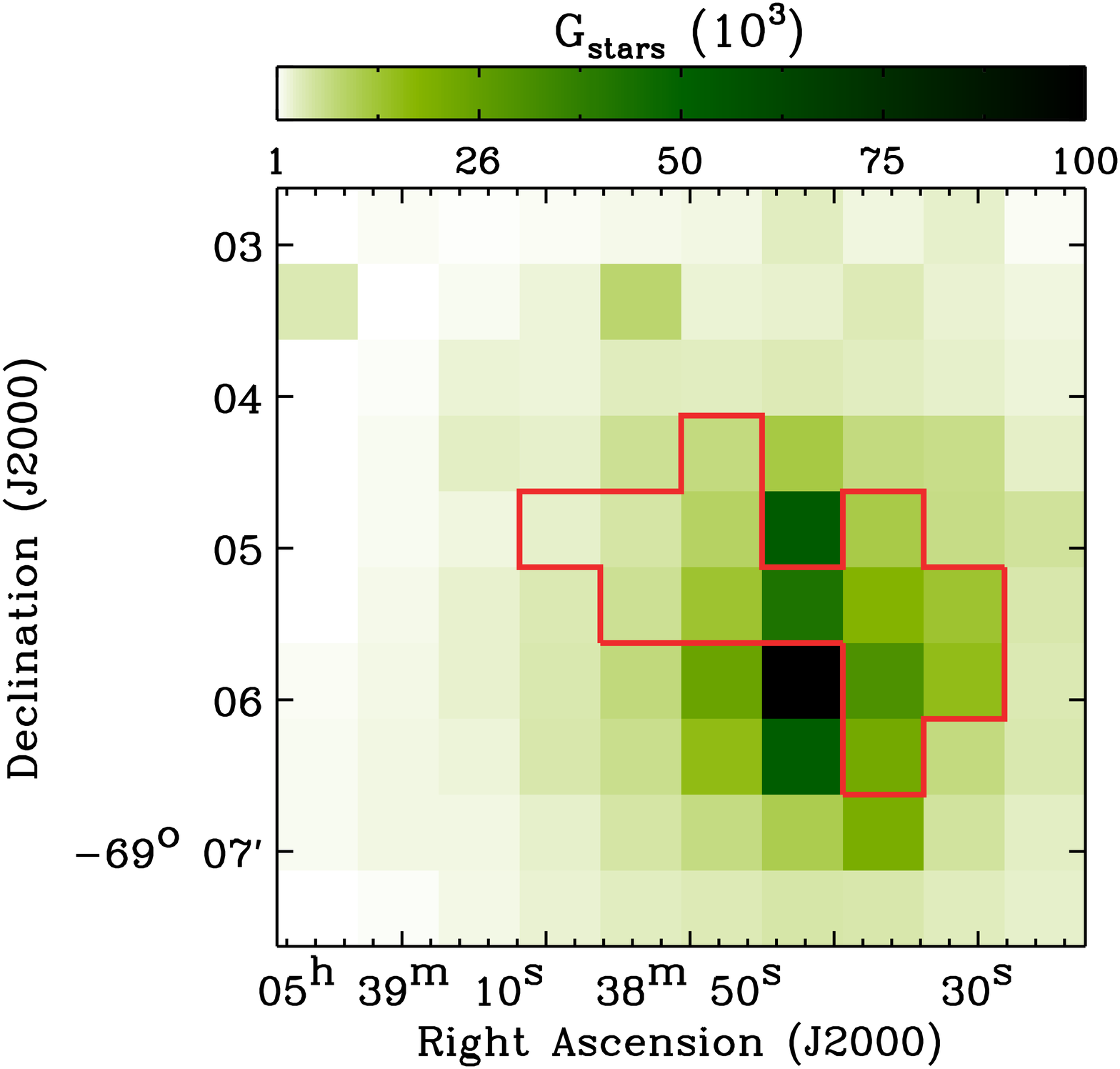} 
\caption{\label{f:UV_sources} \textit{Left}: UV sources overlaid on the CO(9--8) integrated intensity image. 
$\sim$1.3 $\times$ 10$^{4}$ stars we used for the derivation of $G_{\rm stars}$ (\textit{right}; Appendix \ref{s:appendix2} for details)
are presented here with different symbols depending on the stellar effective temperature $T_{\rm eff}$: 
$T_{\rm eff}$ $\geq$ 3 $\times$ 10$^{4}$ K (1406 stars as the red crosses; mostly O-type or W-R), 
10$^{4}$ K $\leq$ $T_{\rm eff}$ $<$ 3 $\times$ 10$^{4}$ K (9807 stars as the orange circles; mostly B-type) 
and $T_{\rm eff}$ $<$ 10$^{4}$ K (2116 stars as the blue triangles). 
The location of R136 is also indicated as the black circle.
\textit{Right}: Stellar UV radiation field $G_{\rm stars}$ on the plane of R136 (in units of 10$^{3}$ Mathis fields).
The pixels used for our PDR modeling are outlined in red.}
\end{figure*}

\section{Excitation sources for CO} 
\label{s:excitation_sources}

\subsection{Radiative source: UV photons} 
\label{s:UV_photons}

Far-UV ($E$ = 6--13.6 eV) photons from young stars have a substantial influence on the thermal and chemical structures of the surrounding ISM. 
As for gas heating, the following two mechanisms are then considered important: 
(1) photo-electric effect on large PAH molecules and small dust grains 
(far-UV photons absorbed by PAH molecules and grains create free electrons, which carry off excess kinetic energy of several eVs; 
e.g., \citealt{Bakes94}; \citealt{Weingartner01}; \citealt{Weingartner06}) and 
(2) far-UV pumping of H$_{2}$ molecules 
(far-UV pumped H$_{2}$ molecules mostly fluoresce back to a vibrational state in the electronic ground state, 
and these vibrationally excited H$_{2}$ molecules can heat the gas through collisional de-excitation; 
e.g., \citealt{Sternberg89}; \citealt{Burton90}). 

As the most extreme star-forming region in the Local Group of galaxies, 
30 Doradus hosts numerous massive stars producing an ample amount of UV photons (Sect. \ref{s:30dor_stellar_content}). 
In Fig. \ref{f:UV_sources}, such UV sources 
are overlaid on the integrated intensity image of CO(9--8), the transition where most of the observed CO SLEDs peak (Sect. \ref{s:observed_CO_SLEDs}). 
The strong concentration of the hot luminous stars in the central cluster R136 is particularly striking. 
In addition, we present the UV radiation field on the plane of R136  
(calculated by utilizing published catalogs of massive stars; Appendix \ref{s:appendix2} for details) in Fig. \ref{f:UV_sources}.
This UV radiation field $G_{\rm stars}$ on the plane of R136 ranges from $\sim$8 $\times$ 10$^{2}$ to $\sim$4 $\times$ 10$^{5}$ Mathis fields 
(its peak coincides well with R136) 
and can be considered as the maximum incident radiation field we would expect, since no absorption was taken into account. 
In the following sections, we evaluate the influence of the intense UV radiation field in 30 Doradus on CO emission by performing PDR modeling. 

\subsubsection{Meudon PDR model} 
\label{s:Meudon_PDR}

For our purpose, we used the Meudon PDR model (\citealt{LePetit06}). 
This one-dimensional stationary model essentially computes 
the thermal and chemical structures of a plane-parallel slab of gas and dust illuminated by a radiation field 
by solving radiative transfer, as well as thermal and chemical balance.
A chemical network of 157 species and 2916 reactions was adopted, and 
in particular H$_{2}$ formation was modeled based on the prescription by \cite{LeBourlot12}, 
which considers the Langmuir-Hinshelwood and Eley-Rideal mechanisms. 
While more sophisticated treatment of H$_{2}$ formation taking into account dust temperature fluctuations is important, as demonstrated by \cite{Bron14,Bron16}, 
we did not use this detailed model due to computing time reasons. 
Consideration of stochastic fluctuations in the dust temperature could increase H$_{2}$ formation in UV-illuminated regions,  
resulting in brighter emission of H$_{2}$ and other molecules that form once H$_{2}$ is present, e.g., CO. 
As for the thermal structure of the slab, the gas temperature was calculated in the stationary state considering the balance between heating and cooling.
The main heating processes were the photo-electric effect on grains and H$_{2}$ collisional de-excitation, 
and cooling rates were then derived by solving the non-LTE populations of main species such as C$^{+}$, C, O, CO, etc. 

\begin{table}[t]
\small
\begin{center}
\caption{\label{t:table2} Input parameters tailored for 30 Doradus$^{a}$.}
\begin{tabular}{l c} \toprule
\centering Parameter & Value \\ \midrule
Metallicity ($Z$) & 0.5 $Z_{\odot}$ \\
Dust-to-gas mass ratio ($M_{\rm dust}/M_{\rm gas}$) & 5 $\times$ 10$^{-3}$ \\
PAH fraction ($f_{\rm PAH}$) & 1\% \\ \toprule
Element & Gas phase abundance \\ 
 & (log$_{10}(n(X)/n({\rm H}))$) \\ \midrule 
He & $-1.05$ \\
C & $-4.30$ \\ 
N & $-4.91$ \\
O & $-3.75$ \\
Ne & $-4.36$ \\
Si & $-5.51$ \\
S & $-5.32$ \\
\bottomrule
\end{tabular}
\end{center} 
{$^{(a)}$ See \cite{Chevance16} for details on these parameters.}
\end{table}

In the Meudon PDR model, the following three parameters play an important role in controlling the structure of a PDR:   
(1) dust extinction $A_{V}$, 
(2) thermal pressure $P$, and 
(3) radiation field $G_{\rm UV}$. 
Specifically, the radiation field has the shape of the interstellar radiation field in the solar neighborhood 
as measured by \cite{Mathis83}, and its intensity scales with the factor $G_{\rm UV}$ 
($G_{\rm UV}$ = 1 corresponds to the integrated energy density of 6.0 $\times$ 10$^{-14}$ erg cm$^{-3}$ for $E$ = 6--13.6 eV). 
For our modeling of 30 Doradus, a plane-parallel slab of gas and dust with a constant $P$ and two-side illumination was then considered, and  
a large parameter space of 
$A_{V}$ = 1, 1.5, 2, 5, 7, 10, 25, 30, 35, and 40 mag, 
$P/k_{\rm B}$ = 10$^{4}$--10$^{9}$ K cm$^{-3}$, and 
$G_{\rm UV}$ = 1--10$^{5}$ was examined. 
For two-sided illumination, the varying $G_{\rm UV}$ = 1--10$^{5}$ was incident on the front side, 
while the fixed $G_{\rm UV}$ = 1 was used for the back side. 
In addition, following \cite{Chevance16}, we adopted the gas phase abundances, PAH fraction ($f_{\rm PAH}$), 
and dust-to-gas mass ratio ($M_{\rm dust}/M_{\rm gas}$) tailored for 30 Doradus as input parameters (Table \ref{t:table2}).
Finally, the cosmic-ray ionization rate $\zeta_{\rm CR}$ = 10$^{-16}$ s$^{-1}$ per H$_{2}$ molecule was used 
based on the observations of diffuse Galactic lines of sight, e.g., \cite{Indriolo12a} and \cite{Indriolo15}. 


\begin{figure*}
\centering
\includegraphics[scale=0.72]{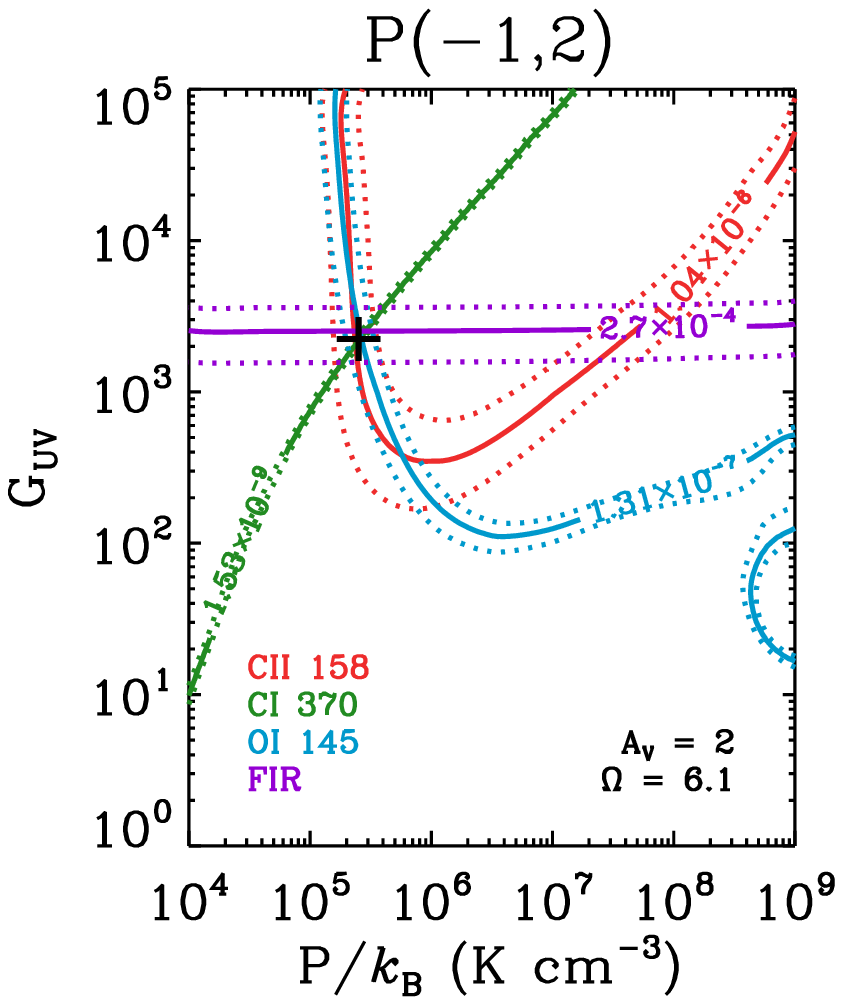} \hspace{1cm}
\includegraphics[scale=0.72]{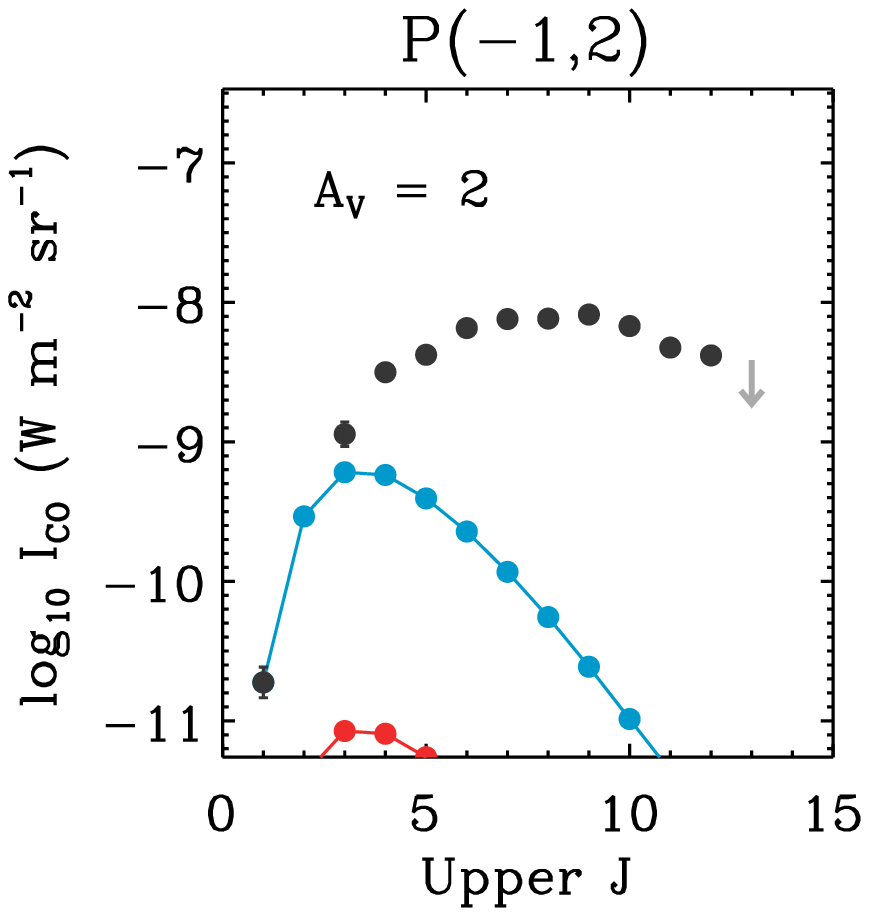}
\caption{\label{f:PDR_results1_1} Meudon PDR modeling of the three fine-structure lines and the FIR luminosity. 
The results presented here are for the pixel $[-1,2]$, which corresponds to   
($\alpha$, $\delta$)$_{\rm J2000}$ = (05$^{\rm h}$38$^{\rm m}$48$^{\rm s}$, $-69^{\circ}$04$'$53$''$),  
and the location of this pixel is indicated as the yellow star on Fig. \ref{f:PDR_results1_2}. 
On the \textit{left plot}, the observed values and their 1$\sigma$ uncertainties are shown as the solid and dotted lines 
(\CII 158 $\mu$m, \CI 370 $\mu$m, \OI 145 $\mu$m, and FIR luminosity in red, green, blue, and purple). 
The best-fit model with the minimum $\chi^{2}$ value is presented as the black cross, 
and the constrained $A_{V}$ and $\Omega$ are summarized in the \textit{bottom right} corner. 
The CO SLED predicted by the best-fit model (red; only partially shown since it is faint) 
is then compared with the observed CO SLED (dark and light gray for detections and non-detections) on the \textit{right plot}. 
In addition, for an easier comparison, the predicted CO SLED is scaled up by a factor of 72 to match the observed CO(1--0) integrated intensity and shown in blue.
The faint CO emission in the best-fit PDR model will be further examined in Sect. \ref{s:PDR_results2}}. 
\end{figure*}

\subsubsection{Strategy for PDR modeling} 
\label{s:PDR_strategy} 

The strategy for our PDR modeling was two-fold. 
First, we constrained $A_{V}$, $P$, and $G_{\rm UV}$ using \CII 158 $\mu$m, \CI 370 $\mu$m, \OI 145 $\mu$m, and FIR luminosity 
and assessed if the constrained conditions reproduce our CO observations. 
This is essentially what \cite{Lee16} did for N159W, except that integrated intensities, rather than line ratios, were employed for our model fitting. 
As we will show in Sect. \ref{s:PDR_results1}, however, 
the PDR component responsible for the fine-structure lines and FIR luminosity turned out to produce weak CO emission, 
and we hence further examined the conditions for CO emission by modeling CO transitions along with other observational constraints (Sect. \ref{s:PDR_results2}). 
This second step was motivated by recent studies of Galactic PDRs, i.e., \cite{Joblin18} for the Orion Bar and NGC 7023 NW and \cite{RWu18} for the Carina Nebula, 
where CO SLEDs up to $J_{\rm u}$ = 23 (for the Orion Bar) were successfully reproduced by the Meudon PDR model. 
These studies found that high-$J$ CO emission originates from the highly pressurized ($P/k_{\rm B} \sim 10^{8}$ K cm$^{-3}$) surface of PDRs, 
where hot chemistry characterized by fast ion--neutral reactions take place (e.g., \citealt{Goicoechea16, Goicoechea17}). 
Photoevaporation by strong UV radiation fields from young stars is considered to play a critical role in 
maintaining such high pressure at the edge of PDRs (e.g., \citealt{Bron18}). 
In the light of these new results on the physical, chemical, and dynamical processes in PDRs, 
we followed the approach by \cite{Joblin18} and \cite{RWu18} and searched for the conditions for CO by fitting CO lines up to $J_{\rm u}$ = 13.
For this, we employed the most up-to-date publicly available Meudon PDR model (version 1.5.2)\footnote{\label{ftn:model_url}\url{https://ism.obspm.fr/}}, 
as used by \cite{Joblin18} and \cite{RWu18}. 

\begin{figure*}
\centering
\includegraphics[scale=0.165]{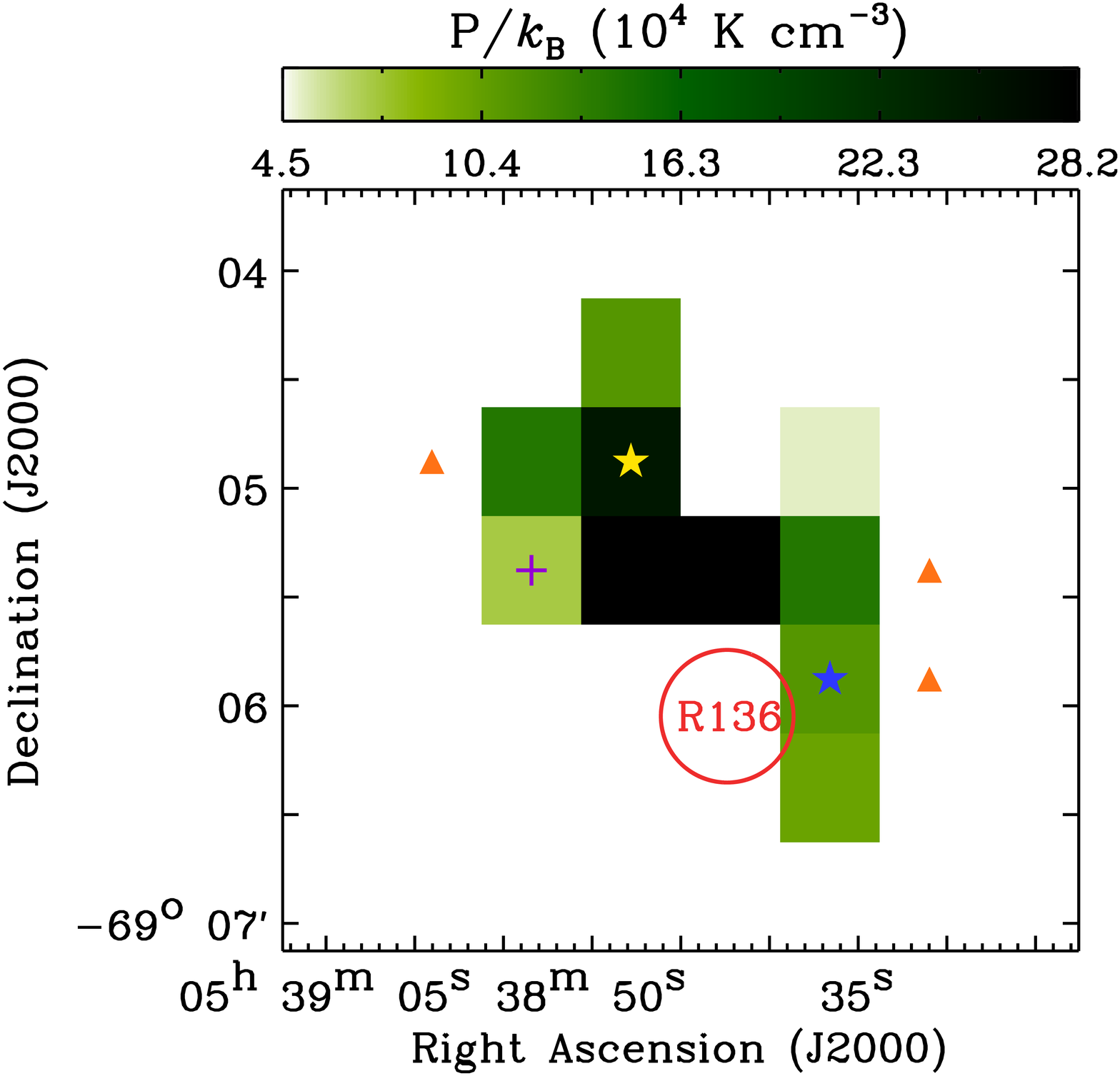}
\includegraphics[scale=0.165]{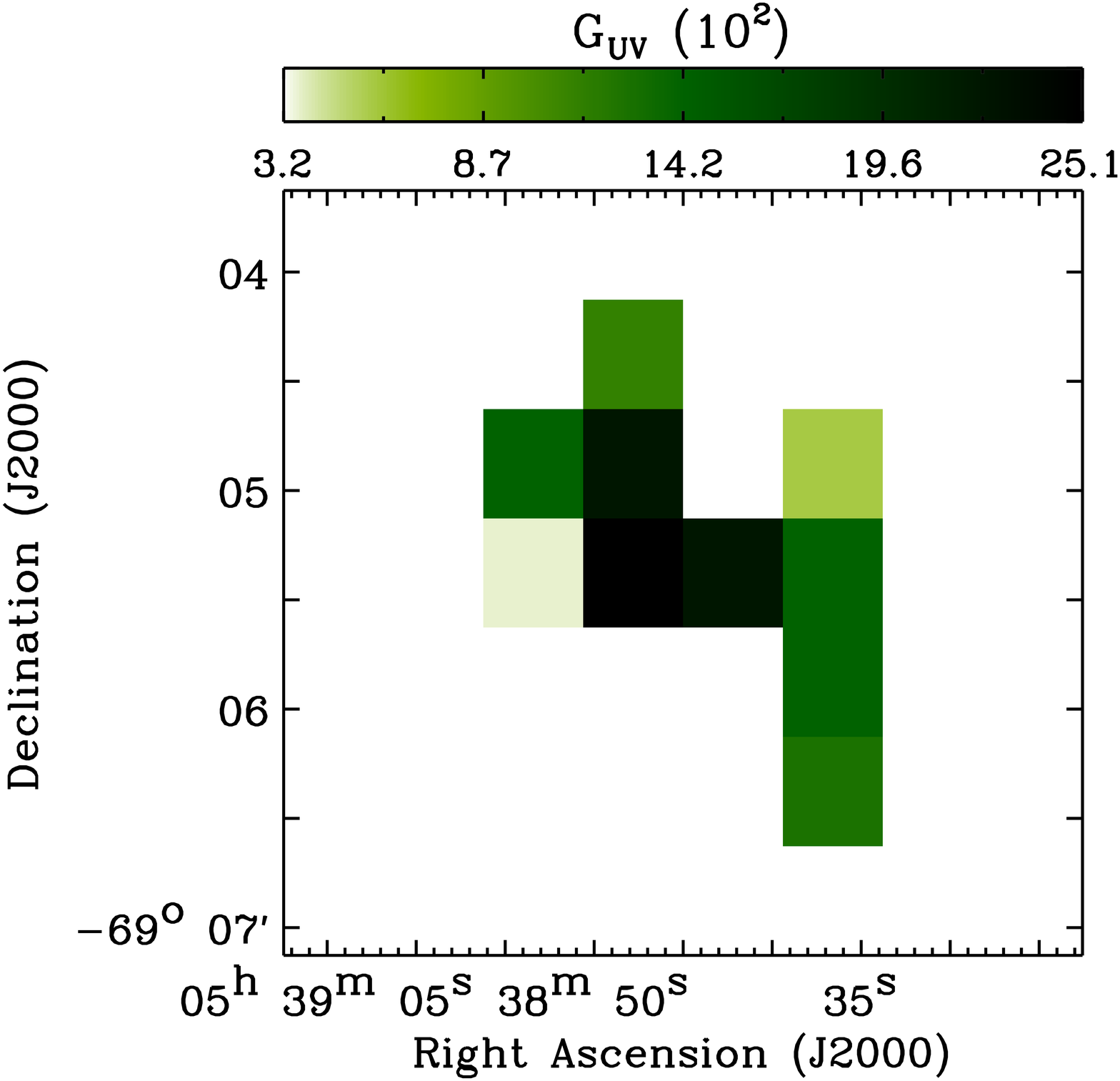}
\includegraphics[scale=0.165]{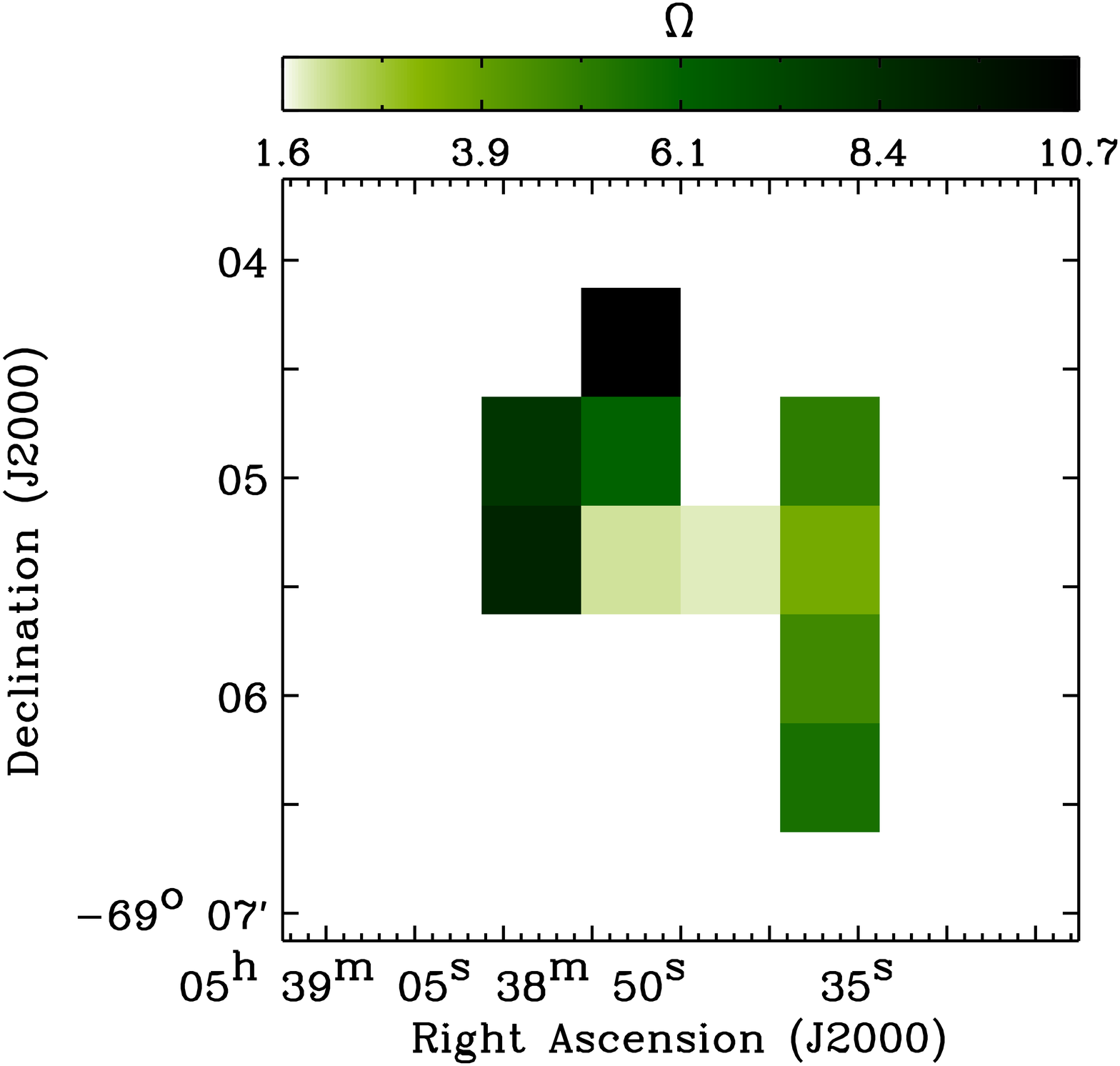}
\caption{\label{f:PDR_results1_2} Best-fit PDR solutions 
($P/k_{\rm B}$, $G_{\rm UV}$, and $\Omega$ on the \textit{left}, \textit{middle}, and \textit{right}). 
$A_{V}$ = 2 mag was constrained for all but one pixel, and the location of this pixel with the smaller $A_{V}$ = 1.5 mag 
is marked with the purple cross on the \textit{left panel}.
The three masked pixels with unreasonably high $\Omega$ $\sim$ 30--50 are denoted as the orange triangles (Sect. \ref{s:PDR_results1} for details),   
while the pixels for Fig. \ref{f:PDR_results1_1} and Fig. \ref{f:PDR_results2_1} are indicated as the yellow and blue stars respectively.  
Finally, the location of R136 is shown as the red circle.}
\end{figure*}

\subsubsection{Modeling: fine-structure lines and FIR emission}
\label{s:PDR_results1} 

We started PDR modeling by first examining the conditions for \CII 158 $\mu$m, \CI 370 $\mu$m, \OI 145 $\mu$m, and FIR emission. 
To do so, we used the PACS and SPIRE spectroscopic data on 42$''$ scales (Sect. \ref{s:data}), 
as well as the FIR luminosity map corrected for the contribution from the ionized medium 
($L_{\rm FIR}^{\rm PDR}$; Appendix \ref{s:appendix3} for details on the correction), 
and derived $\chi^{2}$ for 13 pixels
where all three fine-structure lines were detected (red outlined pixels in Fig. \ref{f:UV_sources}): 

\begin{equation}
\label{eq:chi2} 
\chi^{2} = \sum_{i}\left[\frac{I_{i,\rm{obs}} - (\Omega I_{i,\rm{mod}})}{\sigma_{i,\rm{obs,f}}}\right]^{2},
\end{equation}

\noindent where $I_{i,\rm{obs}}$ = observed integrated intensity, 
$\Omega I_{i,\rm{mod}}$ = model prediction scaled by the beam filling factor $\Omega$, 
and $\sigma_{i,\rm{obs,f}}$ = final 1$\sigma$ uncertainty in the observed integrated intensity. 
A large range of $\Omega$ = 10$^{-2}$--10$^{2}$ was considered in our $\chi^{2}$ calculation, 
and best-fit solutions were then identified as having minimum $\chi^{2}$ values. 
To demonstrate how our modeling was done, 
a plot of $G_{\rm UV}$ vs. $P/k_{\rm B}$ is presented in Fig. \ref{f:PDR_results1_1} for one pixel.
Note that $\Omega$ $>$ 1 implies the presence of multiple components along a line of sight, 
e.g., Sect. 5.1 of \cite{Chevance16} for more discussions.   

For 10 out of the total 13 pixels, we found that best-fit PDR models with 
$P/k_{\rm B}$ = $5 \times 10^{4}$$\sim$$3 \times 10^{5}$ K cm$^{-3}$, 
$G_{\rm UV}$ = 400$\sim$2500, 
$\Omega$ = 2$\sim$11, 
and $A_{V}$ = 1.5 or 2 mag reproduce well the observed fine-structure lines and FIR luminosity. 
These PDR solutions are presented in Fig. \ref{f:PDR_results1_2}. 
For the other three pixels, we then found that best-fit models have significantly higher $\Omega$ $\sim$ 30--50, 
as well as $P$ and $G_{\rm UV}$ that are not smooth across adjacent pixels. 
Our close examination, however, revealed that the observed fine-structure lines and FIR luminosity can still be reproduced 
within a factor of two by PDR models with 
$P/k_{\rm B}$ $\sim$ 10$^{5}$ K cm$^{-3}$, 
$G_{\rm UV}$ $\sim$ 10$^{3}$, 
$\Omega$ $\lesssim$ 10, 
and $A_{V}$ = 1.5 or 2 mag. 

The images of $P$ and $G_{\rm UV}$ in Fig. \ref{f:PDR_results1_2} show that 
both properties peak at the north of R136 and decline outward from there.
On the contrary, $\Omega$ has the minimum value of $\sim$2 at the regions where $P$ and $G_{\rm UV}$ peak 
and increases toward the outer edge of our coverage. 
While these spatial distributions of the PDR parameters are essentially consistent with what \cite{Chevance16} found, 
the absolute values are quite different, e.g., 
the maximum $P$ and $G_{\rm UV}$ values in our analysis are a factor of $\sim$10 lower than those in \cite{Chevance16}. 
There are a number of factors that could contribute to the discrepancy, 
and our detailed comparison suggests that the difference in the angular resolution (42$''$ vs. 12$''$) is most likely the primary factor (Appendix \ref{s:appendix4}). 
The same resolution effect was also noted by \cite{Chevance16},  
stressing the importance of high spatial resolution in the studies of stellar radiative feedback.
Finally, we note that the existence of several clouds whose individual $A_{V}$ is roughly 2 mag is indeed in agreement with
what we estimated from dust SED modeling ($A_{V}$ $\sim$ 8--20 mag; Sect. \ref{s:dust_properties}), 
implying that the PDR component for the fine-structure lines and FIR luminosity constitutes a significant fraction ($\gtrsim$ 50\%) 
of dust extinction along the observed lines of sight. 

Interestingly, we found that CO emission is quite faint in the constrained PDR conditions. 
To be specific, the PDR models underestimate the observed CO integrated intensities by at least a factor of 10, 
and the discrepancy becomes greater with increasing $J$, e.g., from a factor of $\sim$10--70 for CO(1--0) 
to a factor of $\sim$(2--5) $\times$ 10$^{5}$ for CO(13--12). 
The worsening discrepancy with increasing $J$ suggests that the shape of the observed CO SLEDs is not reproduced by the PDR models, 
and we indeed found that the predicted CO SLEDs peak at $J$=3--2, 
which is much lower than the observed $J_{\rm p}$ $\geq$ 6--5. 
This large discrepancy between our CO observations and the model predictions (in terms of both the amplitude and shape of the CO SLEDs) 
is clearly demonstrated in Fig. \ref{f:PDR_results1_1}.
Finally, we note that the H$_{2}$ 0--0 S(3) line is predicted to be as bright as $\sim$2 $\times$ (10$^{-9}$--10$^{-8}$) W m$^{-2}$ sr$^{-1}$, 
which is consistent with the measured upper limits based on 5$\sigma_{\rm s}$ 
(unfortunately, H$_{2}$ 0--0 S(3) is not detected over the 13 pixels where PDR modeling was performed).  

\begin{figure*}
\centering
\includegraphics[scale=0.47]{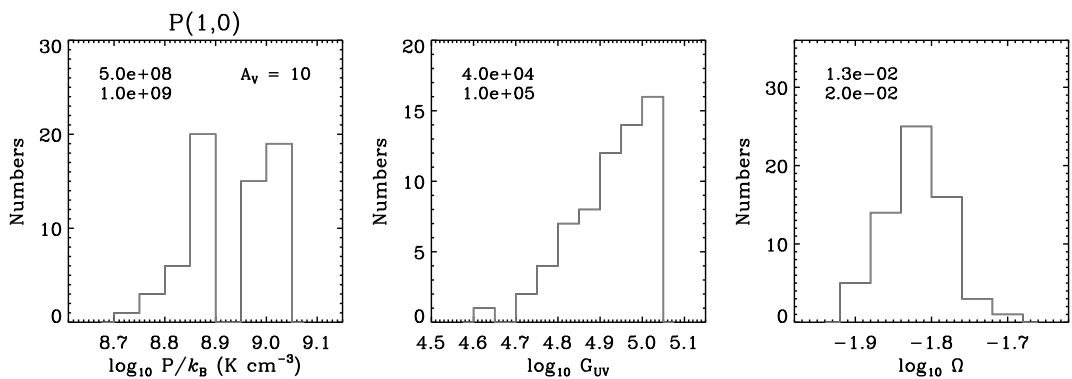}
\caption{\label{f:PDR_results2_1} PDR models with $A_{V}$ = 10 mag that were selected by the criteria in Sect. \ref{s:PDR_results2}
($P/k_{\rm B}$, $G_{\rm UV}$, and $\Omega$ on the \textit{left}, \textit{middle}, and \textit{right panels}; number of selected models = 64). 
This particular example is for the pixel $[1,0]$ 
(corresponding to ($\alpha$, $\delta$)$_{\rm J2000}$ = (05$^{\rm h}$38$^{\rm m}$37$^{\rm s}$, $-69^{\circ}$05$'$53$''$)), 
and the location of the pixel is indicated as the blue star in Fig. \ref{f:PDR_results1_2}. 
In each plot, the minimum and maximum values of the PDR parameter are shown in the \textit{top left} corner.} 
\end{figure*}

\begin{figure*}
\centering
\includegraphics[scale=0.52]{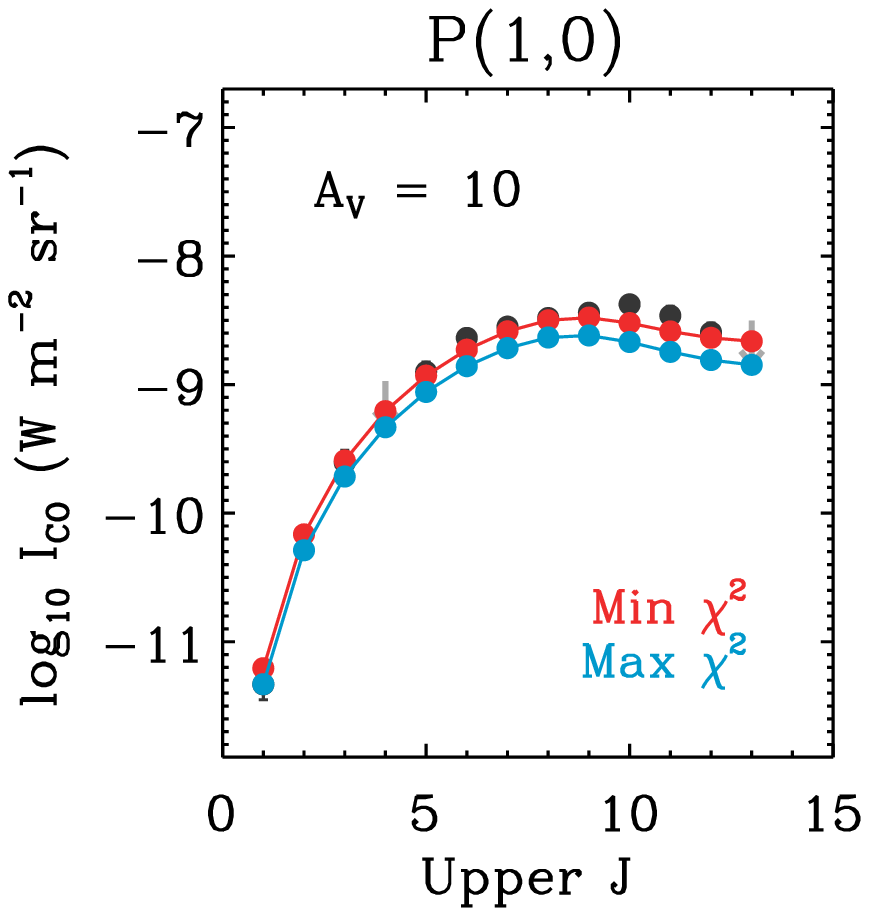} 
\includegraphics[scale=0.355]{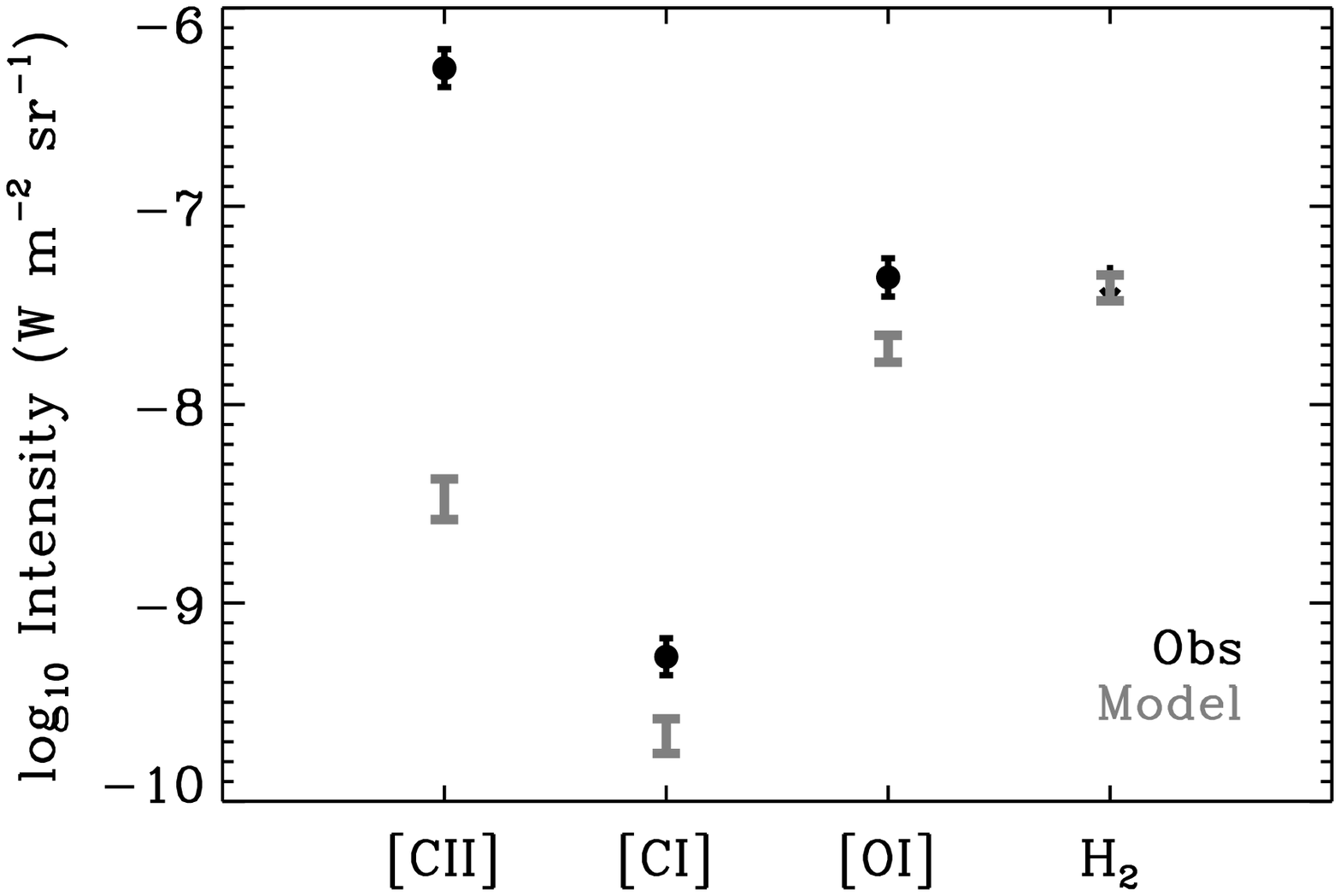} 
\includegraphics[scale=0.355]{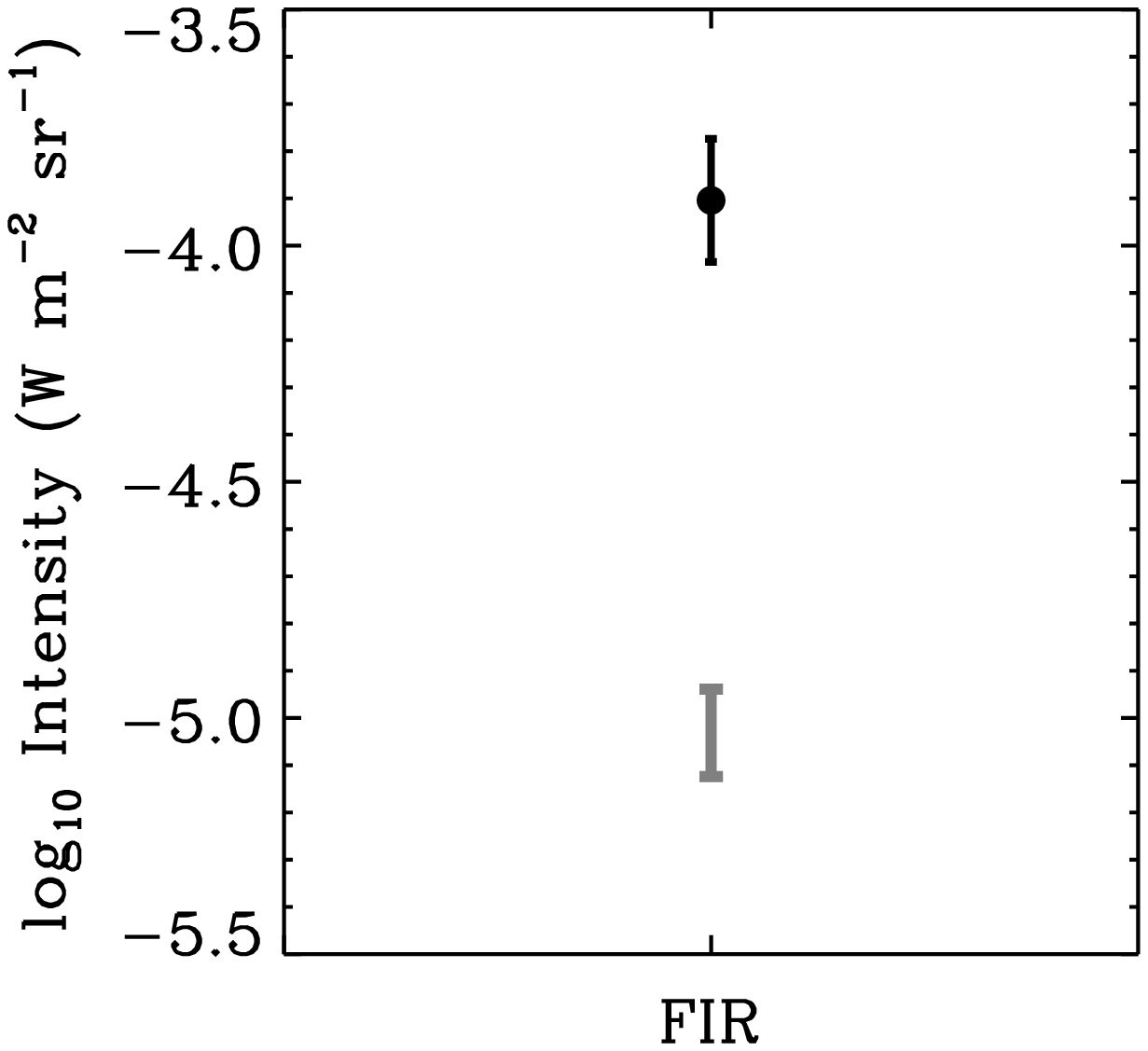} 
\caption{\label{f:PDR_results2_2} Comparison between the observations and the predictions from the PDR models in Fig. \ref{f:PDR_results2_1}. 
In the \textit{left plot}, the observed CO SLED (dark and light gray for detections and non-detections) is compared with 
the predictions from two models (those resulting in minimum and maximum $\chi^{2}$ values with respect to the observed CO lines are presented in red and blue). 
In the \textit{middle} and \textit{right plots}, the observed \CII 158 $\mu$m, \CI 370 $\mu$m, \OI 145 $\mu$m, H$_{2}$ 0--0 S(3), 
and FIR luminosity are shown (black) along with 
the ranges of the model predictions (from minimum to maximum values; gray).}
\end{figure*}

\subsubsection{Modeling: CO lines} 
\label{s:PDR_results2} 

Our modeling in Sect. \ref{s:PDR_results1} strongly suggests that CO emission in 30 Doradus arises from the conditions 
that are drastically different from those for the fine-structure lines and FIR luminosity, i.e.,   
$P/k_{\rm B}$ = a few (10$^{4}$--10$^{5}$) K cm$^{-3}$, $G_{\rm UV}$ = a few (10$^{2}$--10$^{3}$), and $A_{V}$ $\sim$ 2 mag. 
More precisely, the CO-emitting regions would most likely have higher densities and/or higher temperatures 
(to have $J_{\rm p}$ $\geq$ 6--5), 
as well as higher dust extinction (to form more CO molecules, leading to brighter emission), than the \CII 158 $\mu$m-emitting regions.  
This conclusion is essentially the same as what \cite{Lee16} found for N159W.   
We then went one step further by modeling the observed CO transitions, 
examining the PDR conditions from which the CO-emitting gas would arise. 

Initially, we began by computing $\chi^{2}$ using CO transitions up to $J$=13--12 and finding best-fit PDR models with minimum $\chi^{2}$ values. 
This exercise, however, revealed that the models become highly degenerate 
once high $A_{V}$ ($\gtrsim$ 5 mag), $P/k_{\rm B}$ ($\gtrsim$ 10$^{8}$ K cm$^{-3}$), and $G_{\rm UV}$ ($\gtrsim$ 10$^{3}$) are achieved. 
In addition, many best-fit models were incompatible with the observed fine-structure lines and FIR luminosity.
Specifically, the best-fit models always underestimate \CII 158 $\mu$m and FIR luminosity (model-to-observation ratio $\lesssim$ 0.1), 
while mostly reproducing \OI 145 $\mu$m and \CI 370 $\mu$m within a factor of four or less. 
As for H$_{2}$ 0--0 S(3), the best-fit models predict too bright emission in many cases.
This result hints that at least two components, the low-$P$ and high-$P$ PDRs, 
would be required to explain all the transitions we observed in 30 Doradus. 
To work around the degeneracy issue and exclude models with unreasonable predictions for the fine-structure lines and FIR luminosity, 
we then decided to evaluate a collection of PDR models that reproduce the observed CO reasonably well, 
rather than focusing on best-fit models, and employ other observations as secondary constraints. 
To this end, we selected the PDR models that satisfy the following criteria:
(1) The detected CO lines are reproduced within a factor of two. 
In the case of CO(1--0), the prediction is only required not to exceed twice the observed value, 
considering that CO(1--0) could trace different physical conditions than intermediate- and high-$J$ CO lines 
(e.g., \citealt{Joblin18}; \citealt{RWu18}).  
(2) The model predictions agree with the measured upper limits when the CO lines are not detected. 
(3) For \CII 158 $\mu$m, \CI 370 $\mu$m, \OI 145 $\mu$m, and FIR luminosity, 
the model predictions plus the contributions from the low-$P$ PDR component in Sect. \ref{s:PDR_results1} 
are within a factor of two from the observed values. 
(4) Finally, the model prediction plus the contribution from the low-$P$ PDR component
should be consistent with the H$_{2}$ 0--0 S(3) upper limit. 
These four criteria were applied to the 10 pixels 
where we constrained the best-fit PDR models for the fine-structure lines and FIR luminosity (Fig. \ref{f:PDR_results1_2}), 
along with a large range of $\Omega$ = 10$^{-4}$--1 in consideration.\footnote{For CO emission, 
we examined beam filling factors that are smaller than those in Sect. \ref{s:PDR_results1}, 
primarily based on the ALMA CO(2--1) observations by \cite{Indebetouw13} 
(where CO clumps in 30 Doradus were found much smaller than our 30$''$ pixels).}
Since bright CO($J$ $\gtrsim$ 4--3) emission mostly arises from a relatively narrow range of physical conditions 
($A_{V}$ $\gtrsim$ 5 mag, $P/k_{\rm B}$ $\gtrsim$ 10$^{8}$ K cm$^{-3}$, and $G_{\rm UV}$ $\gtrsim$ 10$^{3}$) in the Meudon PDR model,  
slight changes in modeling, e.g., 
removing the (3) and (4) criteria or modeling CO lines with $J$ $\gtrsim$ 4--3 only, 
do not make a large impact on the constrained parameters.  
Finally, we note that our modeling 
with two components of gas is simplistic, given that multiple components would likely be mixed on $\sim$10 pc scales.  
Nevertheless, our analyses would still provide average physical conditions of the components within the beam. 

\begin{figure*}
\centering
\includegraphics[scale=0.35]{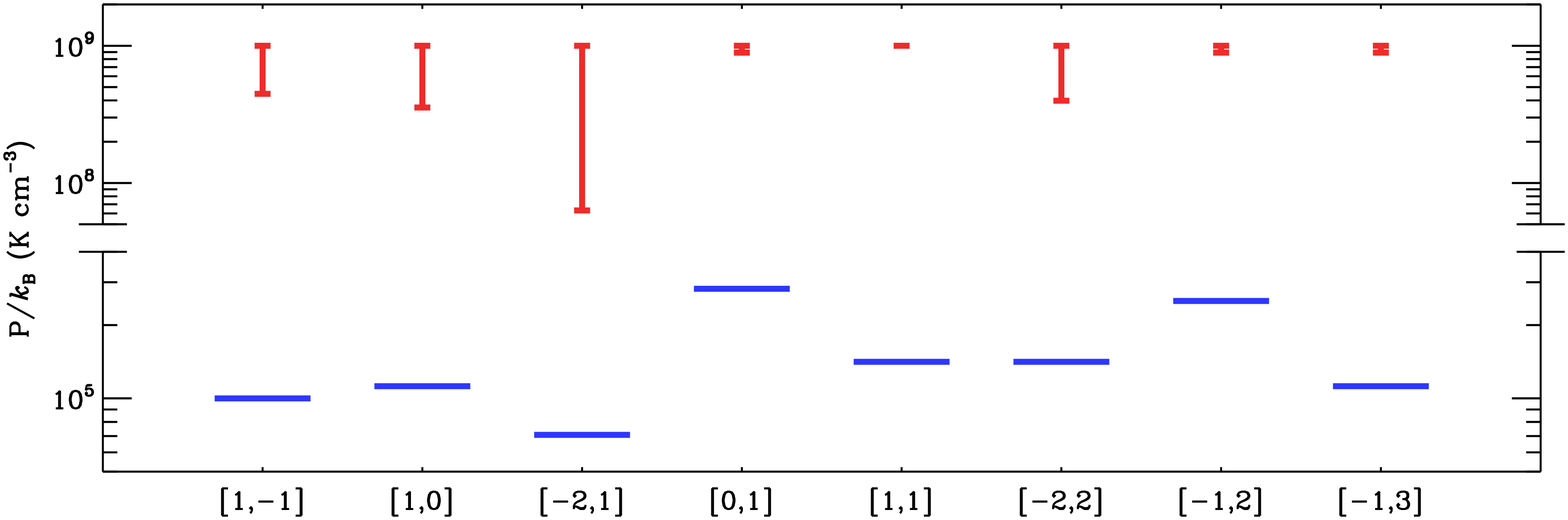}
\includegraphics[scale=0.35]{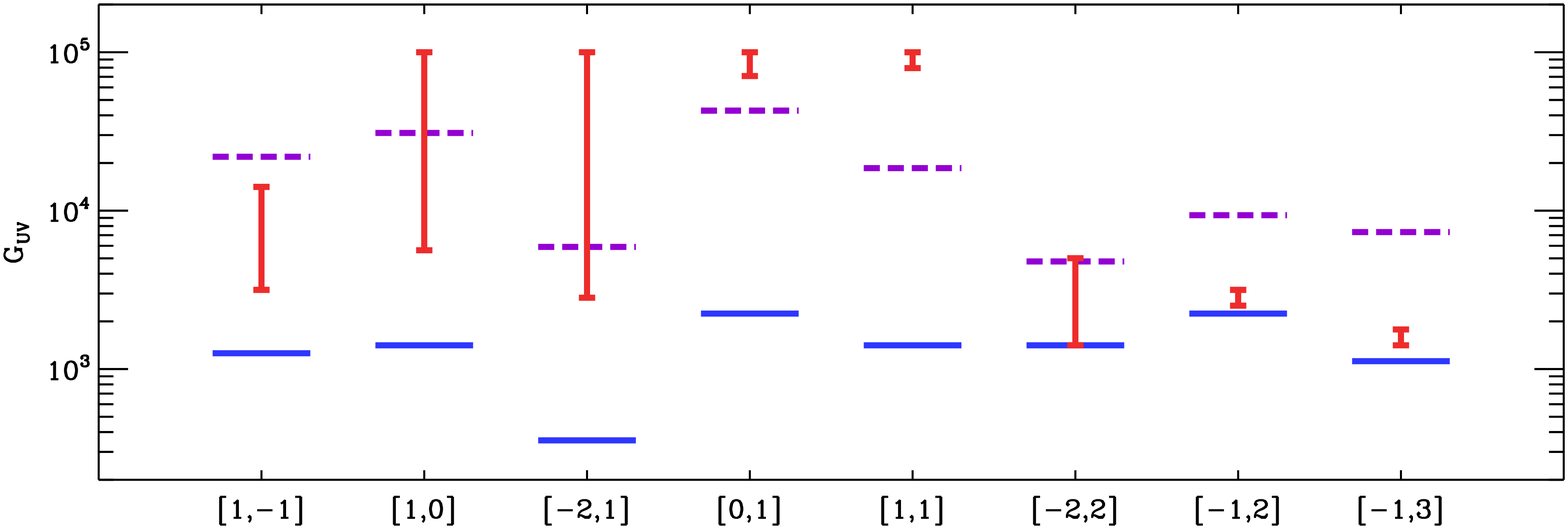}
\includegraphics[scale=0.35]{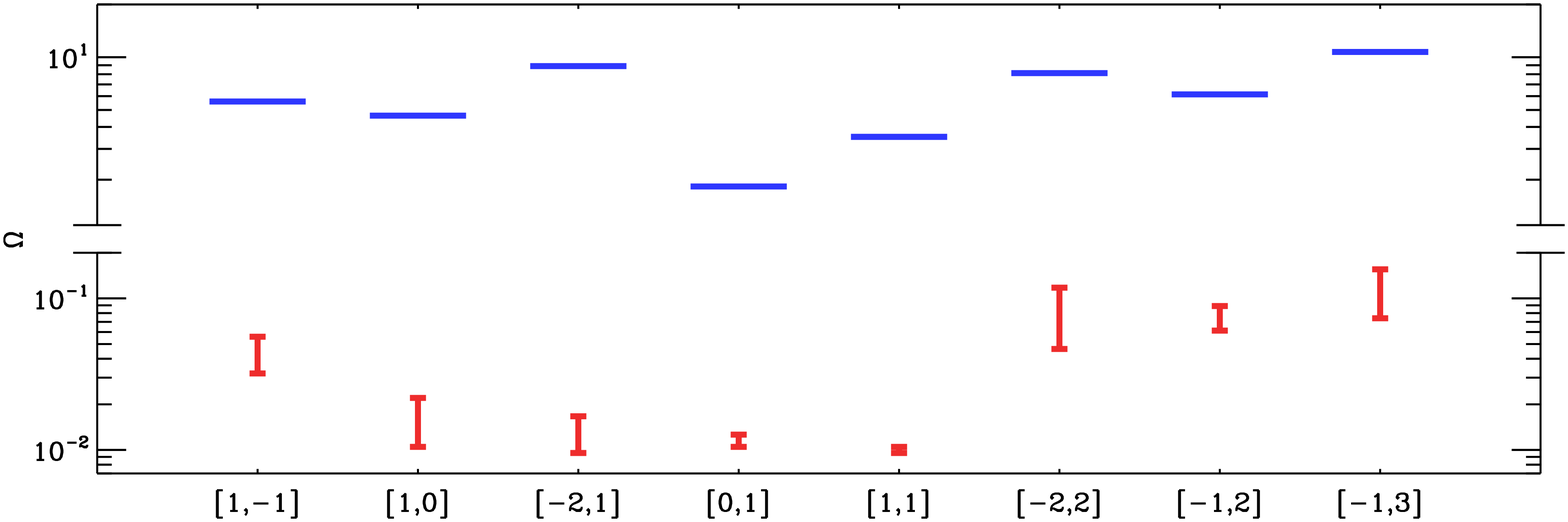}
\caption{\label{f:PDR_results2_3} Comparison between the low- and high-$P$ PDR models  
($P/k_{\rm B}$, $G_{\rm UV}$, and $\Omega$ on the \textit{top}, \textit{middle}, and \textit{bottom panels}). 
In each plot, the low-$P$ models we constrained using the fine-structure lines and FIR luminosity (Sect. \ref{s:PDR_results1}) are indicated as the blue solid lines, 
while the high-$P$ models for the CO lines are presented as the red bars (ranging from the minimum to maximum values). 
In the $G_{\rm UV}$ plot, $G_{\rm stars}$ in Fig. \ref{f:UV_sources} is also shown as the purple dashed lines.  
In total, eight pixels where we found reasonably good high-$P$ models are shown in each plot, 
and the location of each pixel can be inferred from, e.g., Fig. \ref{f:CO7_6} and Fig. \ref{f:obs_CO_SLEDs}.
Finally, note that the $P$ and $\Omega$ plots have broken $y$-axes to show a wide range of the data.}   
\end{figure*}

Overall, we were able to find reasonably good PDR solutions that meet the above selection criteria for eight out of the total 10 pixels
($[-1,1]$ and $[1,2]$ do not have solutions). 
The constrained parameters were then as follows: 
$A_{V}$ = 5--40 mag, $P/k_{\rm B}$ = $\sim$10$^{8}$--10$^{9}$ K cm$^{-3}$, $G_{\rm UV}$ = $\sim$10$^{3}$--10$^{5}$, and $\Omega$ = $\sim$0.01--0.1. 
Note that $A_{V}$ was not well constrained, since increasing $A_{V}$ beyond 5 mag only increases the size of the cold CO core ($<$ 50 K) in a PDR, 
not the warm layer ($\gtrsim$ 50--100 K) where most of intermediate- and high-$J$ CO emission originates (e.g., Sect. \ref{s:UV_shocks}). 
In addition, $\Omega$ $\sim$ 0.01--0.1 implies that the CO-emitting clumps would be $\sim$0.7--2 pc in size, 
which is consistent with the ALMA CO(2--1) observations 
where CO emission was found to primarily arise from $\sim$0.3--1 pc size structures  (\citealt{Indebetouw13}). 
In Fig. \ref{f:PDR_results2_1} and Fig. \ref{f:PDR_results2_2}, we then present the selected PDR models for one pixel, 
as well as the predicted line and continuum intensities, as an example. 
 
Interestingly, we found that the constrained PDR models significantly underestimate \CII 158 $\mu$m and FIR luminosity (e.g., Fig. \ref{f:PDR_results2_2}): 
the discrepancy with our data ranges from $\sim$100 to $\sim$10$^{3}$ for \CII 158 $\mu$m
and from $\sim$10 to $\sim$100 for FIR luminosity. 
On the other hand, \OI 145 $\mu$m and \CI 370 $\mu$m were marginally reproduced (within a factor of four or less) in most cases:
four out of the eight pixels for \OI 145 $\mu$m and seven out of the eight pixels for \CI 370 $\mu$m. 
The measured H$_{2}$ 0--0 S(3) upper limits were also consistent with the model predictions. 
All in all, these results indicate that at least two PDR components are needed to explain all the observational constraints we have for 30 Doradus: 
(1) the low-$P$ (10$^{4}$--10$^{5}$ K cm$^{-3}$) component that provides most of the dust extinction along the observed lines of sight 
and emits intensely in \CII 158 $\mu$m and FIR continuum  
and (2) the high-$P$ (10$^{8}$--10$^{9}$ K cm$^{-3}$) component that is mainly responsible for CO emission. 
For \OI 145 $\mu$m, \CI 370 $\mu$m, and H$_{2}$ 0--0 S(3), both components contribute.
We indeed confirmed that the sum of the two components fully reproduces the observations in our study (including CO $J$=1--0). 

To understand how different the two components are in terms of their physical properties,
we then made a comparison between the constrained PDR parameters on a pixel-by-pixel basis (Fig. \ref{f:PDR_results2_3}). 
Our comparison revealed first of all that the high-$P$ component indeed has significantly higher $P$ than the low-$P$ component 
(a factor of $\sim$10$^{3}$--10$^{4}$). 
Combined with the fact that the high-$P$ models have much smaller $\Omega$ than the low-$P$ models (a factor of $\sim$10$^{2}$--10$^{3}$),
this result implies that the CO-emitting regions in 30 Doradus are more compact, as well as warmer and/or denser, 
than the [C~\textsc{ii}]-emitting regions.  
The relative distribution of the two regions can then be inferred from the $G_{\rm UV}$ values. 
For most of the pixels in our consideration, we found that the UV radiation incident onto the surface of the CO-emitting regions is more intense 
than that for the [C~\textsc{ii}]-emitting regions (by up to a factor of $\sim$300). 
These pixels also have $G_{\rm stars}$ that is comparable to or slightly higher than $G_{\rm UV}$ for the high-$P$ component. 
Considering that the UV radiation field would be most intense on the plane of R136 ($G_{\rm stars}$)
and decrease as the distance $r$ from R136 increases ($\propto$ 1/$r^{2}$ if no absorption is taken into account),  
our results imply that the CO-emitting regions would likely be either in between R136 and the [C~\textsc{ii}]-emitting regions 
or much closer to R136. 
For the pixel $[1,1]$ though,  $G_{\rm UV}$ for the high-$P$ component is higher than $G_{\rm stars}$ by up to a factor of $\sim$5,
a somewhat large discrepancy even considering uncertainties in $G_{\rm UV}$ and $G_{\rm stars}$ 
(so the PDR solution could be unreasonable). 

In summary, we conclude that the observed CO transitions in 30 Doradus (up to $J$=13--12) could be powered by UV photons 
and likely originate from highly compressed ($P/k_{\rm B}$ $\sim$ 10$^{8}$--10$^{9}$ K cm$^{-3}$),  
highly illuminated ($G_{\rm UV}$ $\sim$ 10$^{3}$--10$^{5}$) clumps with a scale of $\sim$0.7--2 pc. 
These clumps are also partially responsible for the observed \CI 370 $\mu$m and \OI 145 $\mu$m,  
but emit quite faintly in \CII 158 $\mu$m and FIR continuum emission, 
hinting the presence of another component with drastically different physical properties. 
Our PDR modeling then suggests that this additional component 
indeed has lower $P$ (a few (10$^{4}$--10$^{5}$) K cm$^{-3}$) and $G_{\rm UV}$ (a few (10$^{2}$--10$^{3}$))  
and likely fills a large fraction of our 30$''$-size pixels. 
Interestingly, the constrained PDR parameters imply that the two distinct components are likely not co-spatial 
(the high-$P$ PDR component closer to R136), which is a somewhat unusual geometry. 
More detailed properties of the two components, e.g., gas density and temperature, will be discussed in Sect. \ref{s:UV_shocks}, 
along with another viable heating source for CO, shocks.  

\begin{figure*}
\centering
\includegraphics[scale=0.59]{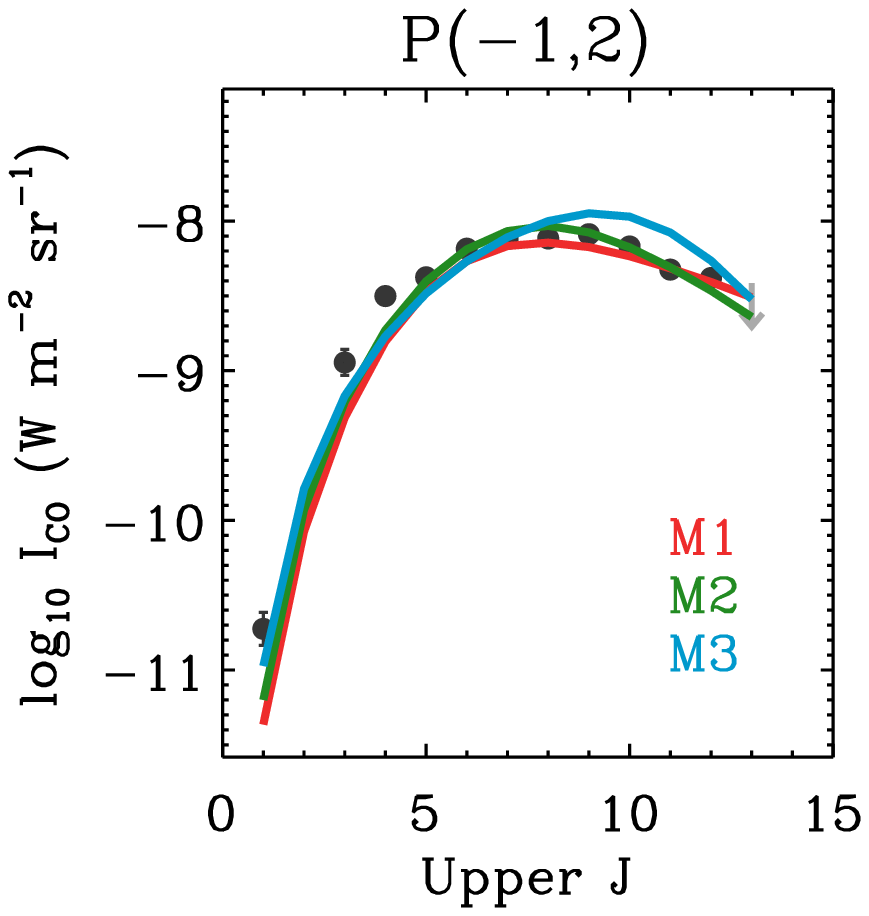} \hspace{1cm}
\includegraphics[scale=0.41]{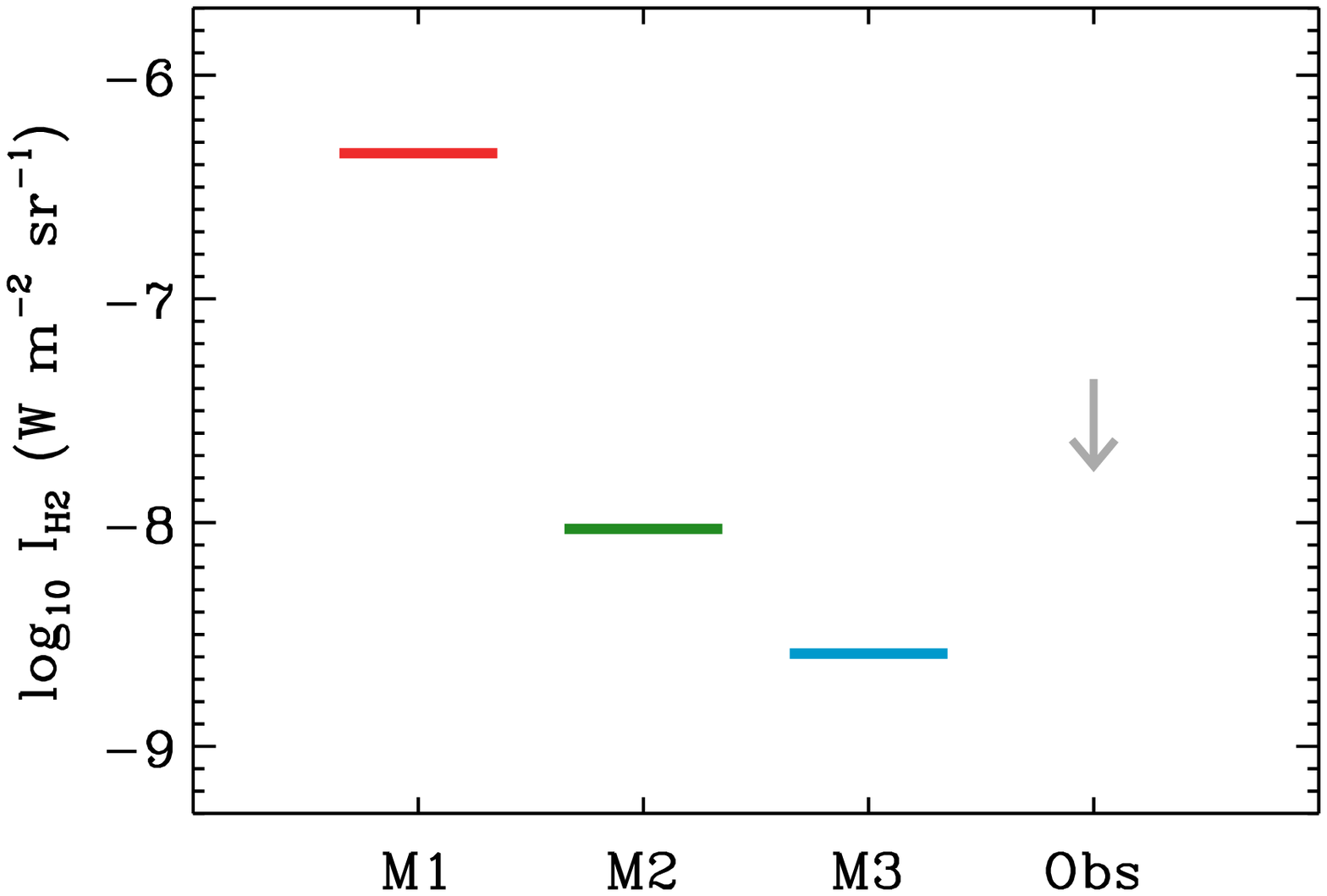} 
\caption{\label{f:shock_deg1} Degeneracy in $n_{\rm pre}$ and $\varv_{\rm s}$. 
To illustrate this issue, the observed CO and H$_{2}$ 0--0 S(3) transitions of the pixel $[-1,2]$ are shown in the \textit{left} and \textit{right plots} 
along with three different shock models ($``$M1$"$ in red: $n_{\rm pre}$ = 10$^{4}$ cm$^{-3}$ and $\varv_{\rm s}$ = 28 km s$^{-1}$; 
$``$M2$"$ in green: $n_{\rm pre}$ = 5 $\times$ 10$^{4}$ cm$^{-3}$ and $\varv_{\rm s}$ = 7.5 km s$^{-1}$; 
$``$M3$"$ in blue: $n_{\rm pre}$ = 10$^{6}$ cm$^{-3}$ and $\varv_{\rm s}$ = 4 km s$^{-1}$). 
For both plots, dark and light gray colors are used to represent detections and non-detections, 
and $G'_{\rm UV}$ = 0 and $\Omega$ $\sim$ 0.1 are adopted for the three shock models.}
\end{figure*}

\subsection{Radiative source: X-rays and cosmic-rays} 
\label{s:x_rays_cosmic_rays}

As described in Sect. \ref{s:30dor_high_energy}, 
abundant X-rays and cosmic-rays exist in 30 Doradus.  
These high energy photons and particles can play an important role in gas heating 
(mainly through photoionization of atoms and molecules), 
and yet we evaluated that their impact on the observed CO lines is negligible. 
Our evaluation was based on \cite{Lee16} and can be summarized as follows. 

In \cite{Lee16}, we examined the influence of X-rays by considering the most luminous X-ray source in the LMC, LMC X-1 (a black hole binary). 
The maximum incident X-ray flux of 10$^{-2}$ erg s$^{-1}$ cm$^{-2}$ (maximum since absorption between LMC X-1 and N159W was not taken into account) 
was incorporated into PDR modeling, 
and we found that X-rays make only a factor of three or so change in the total CO integrated intensity. 
Considering that the X-ray flux in 30 Doradus is much lower 
(up to 10$^{-4}$ erg s$^{-1}$ cm$^{-2}$ only) than the maximum case for N159W, 
we then concluded that X-rays most likely provide only a minor contribution to CO heating in 30 Doradus.

As for cosmic-ray heating, we again followed the simple calculation by \cite{Lee16}. 
In this calculation, H$_{2}$ cooling (primary cooling process for the warm and dense medium; e.g., \citealt{LeBourlot99}) 
was equated with cosmic-ray heating to estimate the cosmic-ray ionization rate of $\zeta_{\rm CR}$ $\gtrsim$ 3 $\times$ 10$^{-13}$ s$^{-1}$ 
that is required to fully explain the warm CO in N159W. 
While this cosmic-ray ionization rate is higher than 
the typical value for the diffuse ISM in the solar neighborhood  
by more than a factor of 1000 (e.g., \citealt{Indriolo15}), 
the measured $\gamma$-ray emissivity of N159W ($\sim$10$^{26}$ photons s$^{-1}$ sr$^{-1}$ per hydrogen atom) 
is comparable to the local ISM value (e.g., \citealt{Abdo09LocalISM}), 
suggesting that the cosmic-ray density in N159W is not atypical. 
Similarly, considering that the CO-emitting gas in 30 Doradus is warm and dense as in N159W (Sect. \ref{s:UV_shocks} for details), 
yet the $\gamma$-ray emissivity is only $\sim$3 $\times$ 10$^{26}$ photons s$^{-1}$ sr$^{-1}$ per hydrogen atom (\citealt{Abdo10LMC}), 
it is again likely that cosmic-rays in 30 Doradus are not abundant enough for CO heating. 

\subsection{Mechanical source: shocks} 
\label{s:shocks} 

Shocks are ubiquitous in the ISM, being continuously driven by various energetic processes, 
e.g., stellar activities such as outflows (YSOs and red giant stars), winds (OB and W-R stars), and explosions (novae and supernovae), 
as well as non-stellar activities such as colliding clouds and spiral density waves (e.g., \citealt{Hollenbach89}). 
These shocks can be an important source of heating, 
since they effectively transform the bulk of the injected mechanical energy into thermal energy.  
In the particular case of the dense and magnetized medium with a low fractional ionization  
(essentially corresponding to star-forming regions such as 30 Doradus),
C-type shocks can develop, whose main characteristics include: 
(1) molecules are accelerated without being thermally dissociated; and  
(2) the shocked medium radiates primarily in rotation-vibration transitions of molecules, 
as well as fine-structure lines of atoms and ions (\citealt{Draine93}). 
The emission from C-type shocks largely appears at IR wavelengths 
and provides a powerful means to probe 
the physical properties of the shocks and the ambient medium. 

\begin{figure*}
\centering
\includegraphics[scale=0.53]{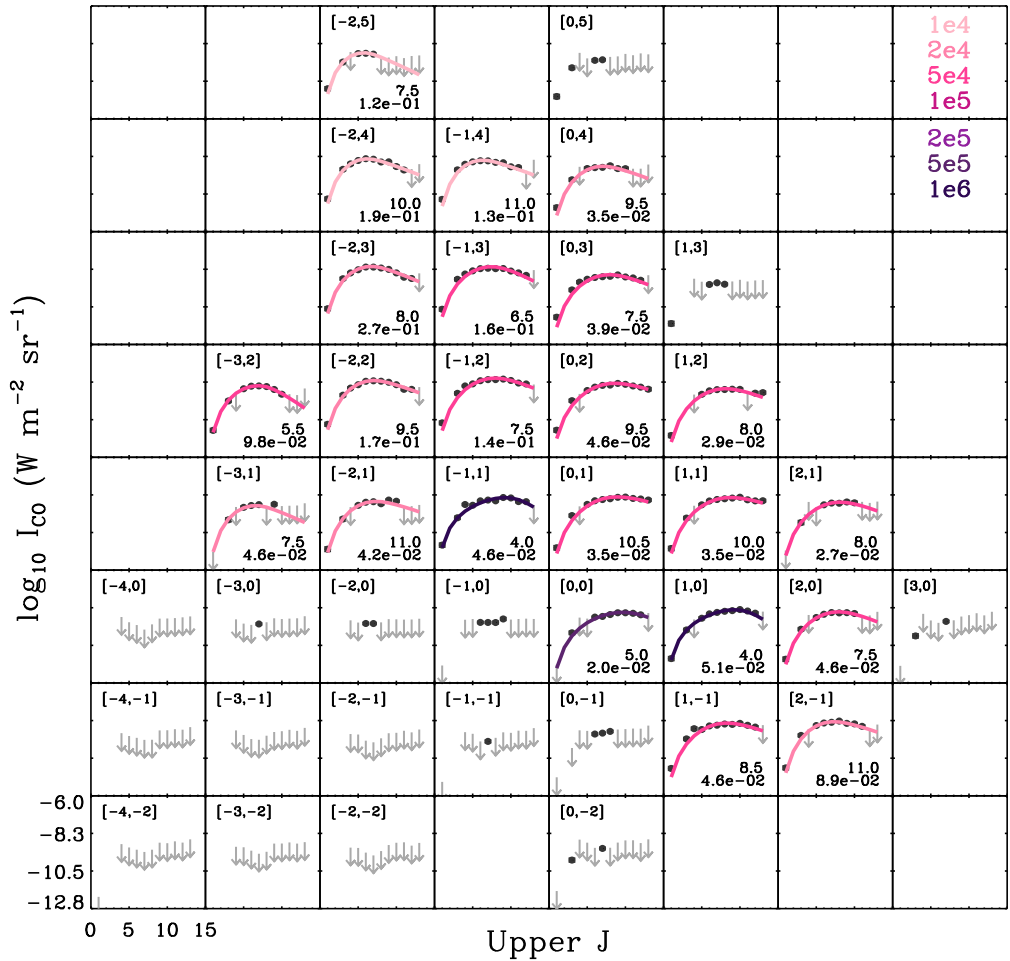} 
\caption{\label{f:CO_SLEDs_shocks} 
Constrained shock models overlaid with the observed CO SLEDs. 
As in Fig. \ref{f:obs_CO_SLEDs}, the circles and downward arrows represent the measured intensities and 5$\sigma_{\rm s}$-based upper limits. 
A location with respect to the pixel that is closest to R136 is also indicated as a pair of numbers in the \textit{top left} corner of each pixel 
(e.g., [0,0] corresponds to ($\alpha$, $\delta$)$_{\rm J2000}$ = (05$^{\rm h}$38$^{\rm m}$42$^{\rm s}$, $-69^{\circ}$05$'$53$''$), 
and each pixel is in 30$''$ size). 
The shock models with different $n_{\rm pre}$ are presented in different colors (darker shades for higher $n_{\rm pre}$), 
and the exact $n_{\rm pre}$ values (in cm$^{-3}$) are summarized in the \textit{top right} corner of this figure. 
Finally, the constrained shock velocities (in km s$^{-1}$) and beam filling factors (unitless)
are shown in the \textit{bottom right} corner of each pixel.}
\end{figure*}

\subsubsection{Paris-Durham shock model} 
\label{s:PD_model} 

Motivated by the results from \cite{Lee16} for N159W, 
we evaluated whether low-velocity C-type shocks could be another important source of heating for CO in 30 Doradus 
by comparing the observed line emission with predictions from the Paris-Durham shock model\textsuperscript{\ref{ftn:model_url}} (\citealt{Flower15}). 
This one-dimensional stationary model simulates the propagation of a shock wave (J- or C-type) through a layer of gas and dust 
and calculates the physical, chemical, and dynamical properties of the shocked layer.
For our analysis, we used the modified version by \cite{Lesaffre13} to model UV-irradiated shocks 
and created a grid of models with the following parameters: 
(1) pre-shock density $n_{\rm pre}$ = (1, 2, and 5) $\times$ 10$^{4}$, (1, 2, and 5) $\times$ 10$^{5}$, and 10$^{6}$ cm$^{-3}$, 
(2) shock velocity $\varv_{\rm s}$ = 4--11 km s$^{-1}$ with 0.5 km s$^{-1}$ steps 
and 12--30 km s$^{-1}$ with 2 km s$^{-1}$ steps 
(a finer grid was produced for $\varv_{\rm s}$ = 4--11 km s$^{-1}$ to properly sample 
a factor of $\sim$100 variation in H$_{2}$ 0--0 S(3) over this velocity range), 
(3) UV radiation field $G'_{\rm UV}$ (defined as a scaling factor relative to the \citealt{Draine78} radiation field) = 0 and 1, 
(4) dimensionless magnetic field parameter $b$ = 1 (defined as ($B$/$\mu$G)/($n_{\rm pre}$/cm$^{-3}$)$^{1/2}$,    
where $B$ is the strength of the magnetic field transverse to the direction of shock propagation), 
and (5) same gas and dust properties as used in our PDR modeling (Table \ref{t:table2} for details).  
In our grid of models, the magnetosonic speed varies from $\sim$20 km s$^{-1}$ ($G'_{\rm UV}$ = 1 case) 
to $\sim$80 km s$^{-1}$ ($G'_{\rm UV}$ = 0 case), 
and the post-shock pressure (roughly determined by the ram-pressure of the pre-shock medium) 
has a range of $\sim$10$^{5}$--10$^{9}$ K cm$^{-3}$. 
All our models fall into the C-type shock category. 
Finally, the calculated abundances of atoms and molecules were post-processed via the LVG method by \cite{Gusdorf12,Gusdorf15}
to compute level populations, line emissivities, and integrated intensities.

\subsubsection{Strategy for shock modeling} 
\label{s:shock_strategy} 

To examine the properties of shocks that could possibly heat CO in 30 Doradus, 
we went one step further than \cite{Lee16} 
by fitting the observed CO transitions with the shock models. 
In an attempt to break the degeneracy between the model parameters, 
we then considered other constraints as well, such as H$_{2}$ 0--0 S(3) and \CI 370 $\mu$m, in our shock modeling. 

\begin{figure*}
\centering
\includegraphics[scale=0.48]{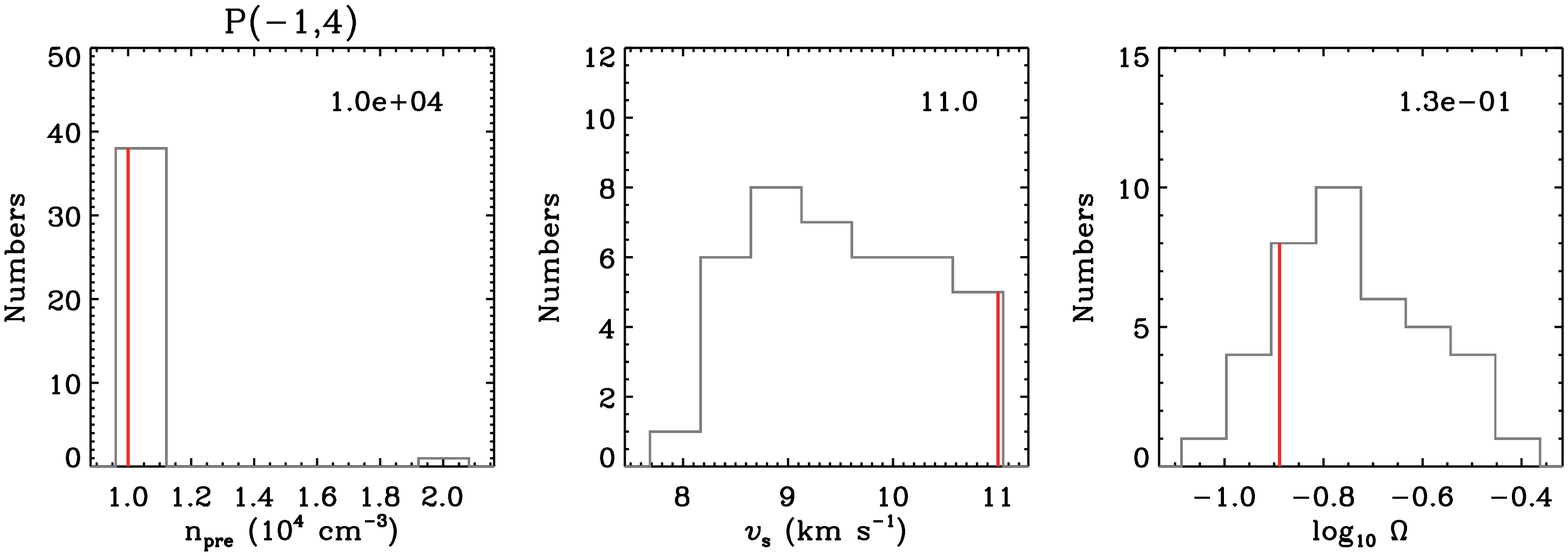}
\includegraphics[scale=0.48]{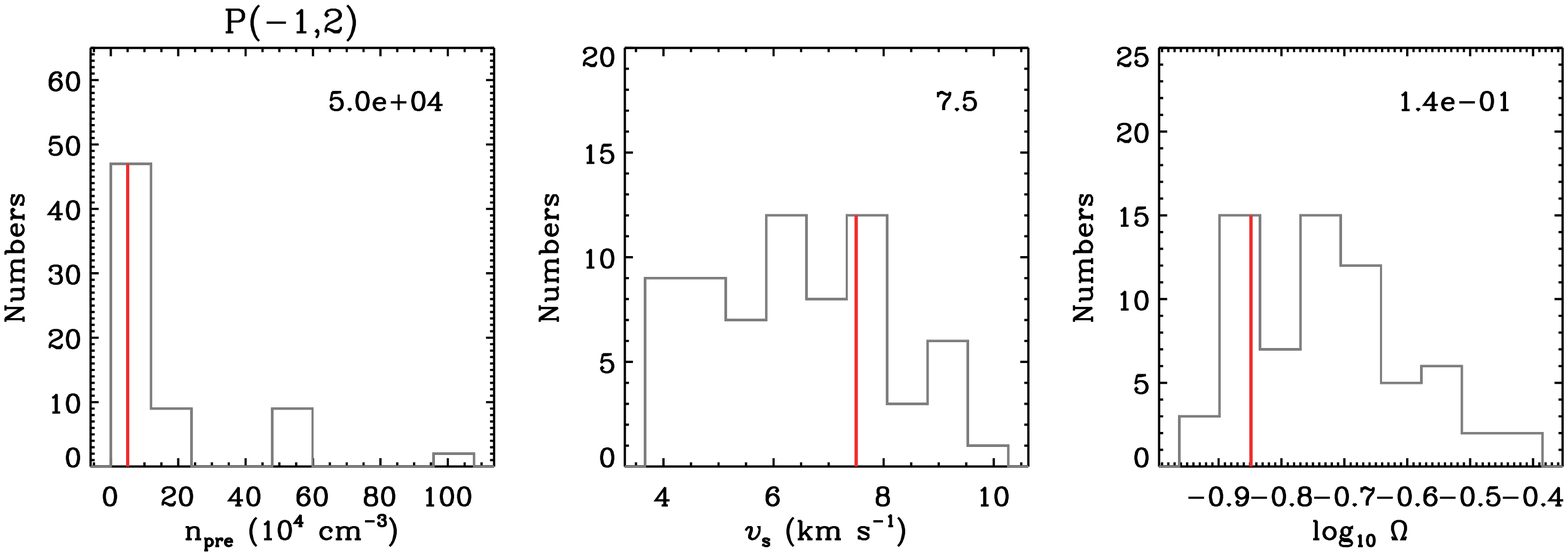}
\caption{\label{f:shock_deg2} Another illustration of the degeneracy in our shock modeling. 
Here the selected ``good'' shock models that reproduce our CO and H$_{2}$ 0--0 S(3) observations within a factor of two 
are presented for two pixels, $[-1,4]$ and $[-1,2]$ (\textit{top} and \textit{bottom}; number of the selected models = 39 and 67 respectively), 
as the gray histograms. 
For each histogram, the final shock parameter we constrained in Sect. \ref{s:shock_results1} is shown as the red solid line 
and summarized on the \textit{top right} corner. 
A comparison between the \textit{top} and \textit{bottom} histograms clearly shows that 
$[-1,4]$ has a quite narrow distribution of $n_{\rm pre}$.}
\end{figure*}

\subsubsection{Modeling: CO lines}
\label{s:shock_results1}

We began shock modeling by first deriving $\chi^{2}$ using the observed CO ($J$=3--2 to 13--12) and H$_{2}$ 0--0 S(3) transitions 
for 23 pixels that contain more than five CO detections
(so that the number of constraints $\geq$ 1 + the number of model parameters, $n_{\rm pre}$, $\varv_{\rm s}$, $G'_{\rm UV}$, and $\Omega$). 
Our $\chi^{2}$ calculation was essentially based on Eq. (\ref{eq:chi2}), but with an additional consideration for non-detections. 
Specifically, we set $[I_{i,\rm{obs}} - (\Omega I_{i,\rm{mod}})]$/$\sigma_{i,\rm{obs,f}}$ = 0
for the transitions whose 5$\sigma_{\rm s}$-based upper limits are consistent with model predictions. 
When the model predictions are higher than the upper limits, we then simply excluded such bad models from our analysis 
(so the non-detections were used only to provide hard limits on the models). 
In our $\chi^{2}$ analysis, CO(1--0) was not included to consider a possible presence of some cold pre-shock gas 
that could emit brightly in CO(1--0) (e.g., \citealt{Lee16}). 
Finally, the same $\Omega$ = 10$^{-4}$--1 as used in our PDR modeling (Sect. \ref{s:PDR_results2}) was examined.  

The inclusion of H$_{2}$ 0--0 S(3) in our $\chi^{2}$ analysis was intended to mitigate the degeneracy between $n_{\rm pre}$ and $\varv_{\rm s}$. 
In particular, we found that H$_{2}$ 0--0 S(3), even with upper limits, can effectively differentiate
high-density ($>$ 10$^{4}$ cm$^{-3}$), low-velocity ($\lesssim$ 10 km s$^{-1}$) shocks 
from low-density ($\sim$10$^{4}$ cm$^{-3}$), high-velocity ($\gtrsim$ 20 km s$^{-1}$) shocks. 
To demonstrate this, we show the observed CO and H$_{2}$ 0--0 S(3) for the pixel $[-1,2]$ in Fig. \ref{f:shock_deg1}, 
along with three different shock models
($n_{\rm pre}$ = 10$^{4}$, 5 $\times$ 10$^{4}$, and 10$^{6}$ cm$^{-3}$;  
$\varv_{\rm s}$ = 28, 7.5, and 4 km s$^{-1}$; $G'_{\rm UV}$ = 0; $\Omega$ $\sim$ 0.1). 
These shock models all reproduce the observed CO SLED within a factor of two,  
while showing a factor of $\sim$200 difference in H$_{2}$ 0--0 S(3). 
Specifically, the highest-velocity shock produces the brightest H$_{2}$ 0--0 S(3) of $\sim$4 $\times$ 10$^{-7}$ W m$^{-2}$ sr$^{-1}$ 
(primarily due to the high temperature of $\sim$10$^{3}$ K that is achieved by strong compression),
and the measured upper limit clearly rules out this model. 
On the other hand, the other two models have relatively low temperatures of $\sim$10$^{2}$ K 
and show an insignificant difference in H$_{2}$ 0--0 S(3) emission (a factor of four). 
Our H$_{2}$ observations, unfortunately, are not sensitive enough to discriminate this level of minor difference 
(e.g., only two out of the total 23 pixels have detections with $S/N_{\rm s}$ $\sim$ 5), 
resulting in the degeneracy in 5 $\times$ 10$^{4}$ cm$^{-3}$ $\lesssim$ $n_{\rm pre}$ $\lesssim$ 10$^{6}$ cm$^{-3}$ 
and $\varv_{\rm s}$ $\lesssim$ 10 km s$^{-1}$ in our shock analysis. 

\begin{figure*}
\centering
\includegraphics[scale=0.341,valign=t]{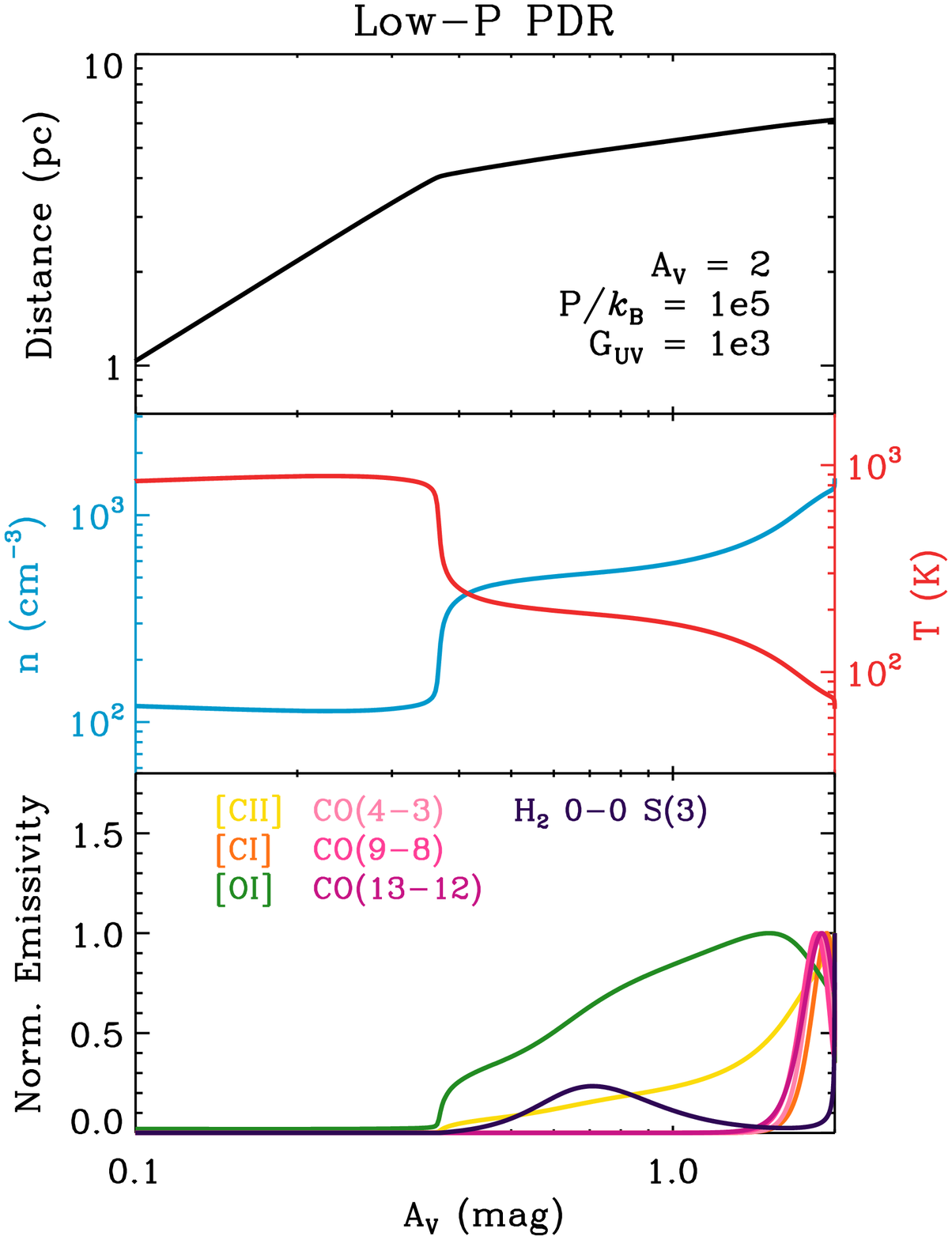}
\includegraphics[scale=0.341,valign=t]{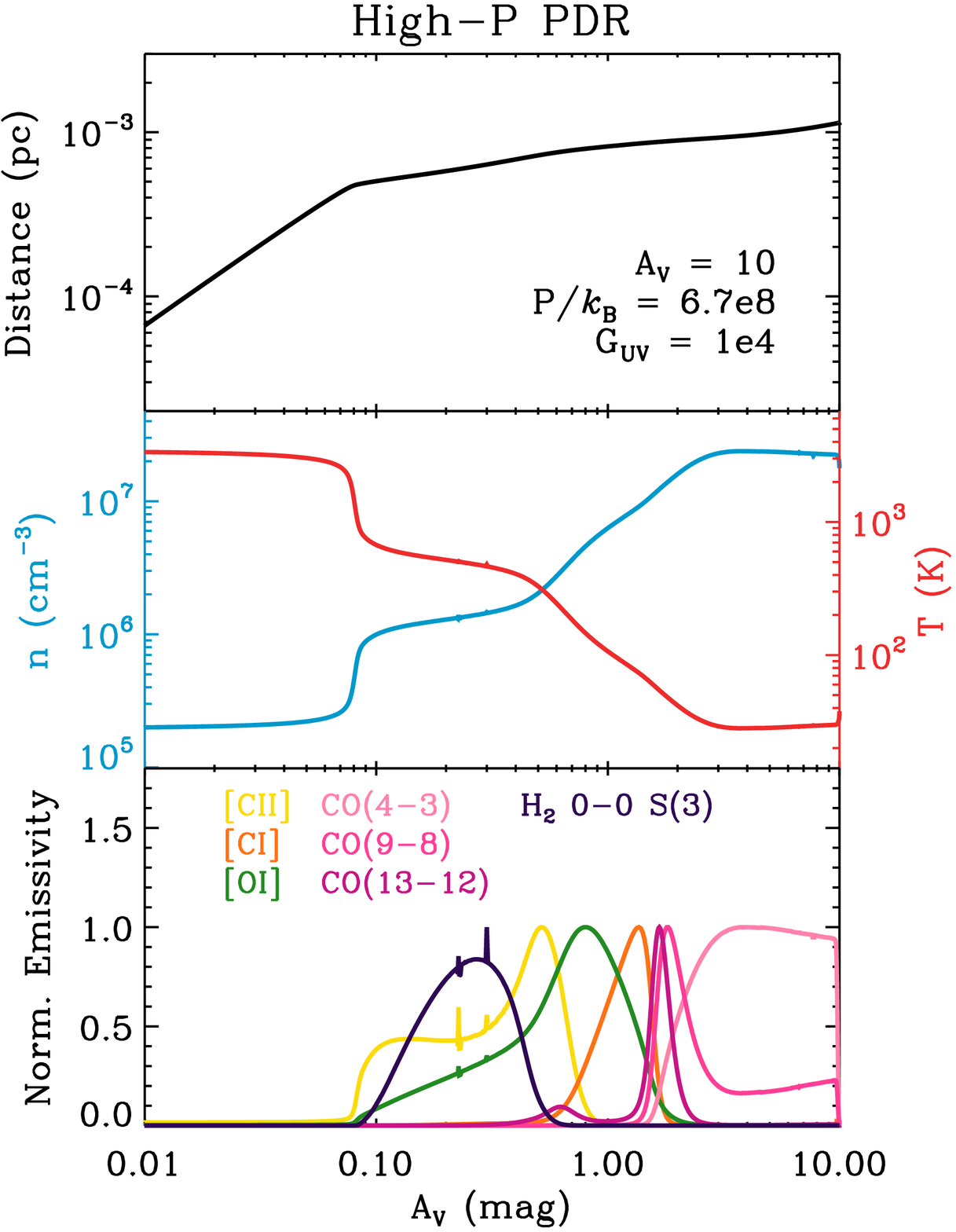}
\includegraphics[scale=0.341,valign=t]{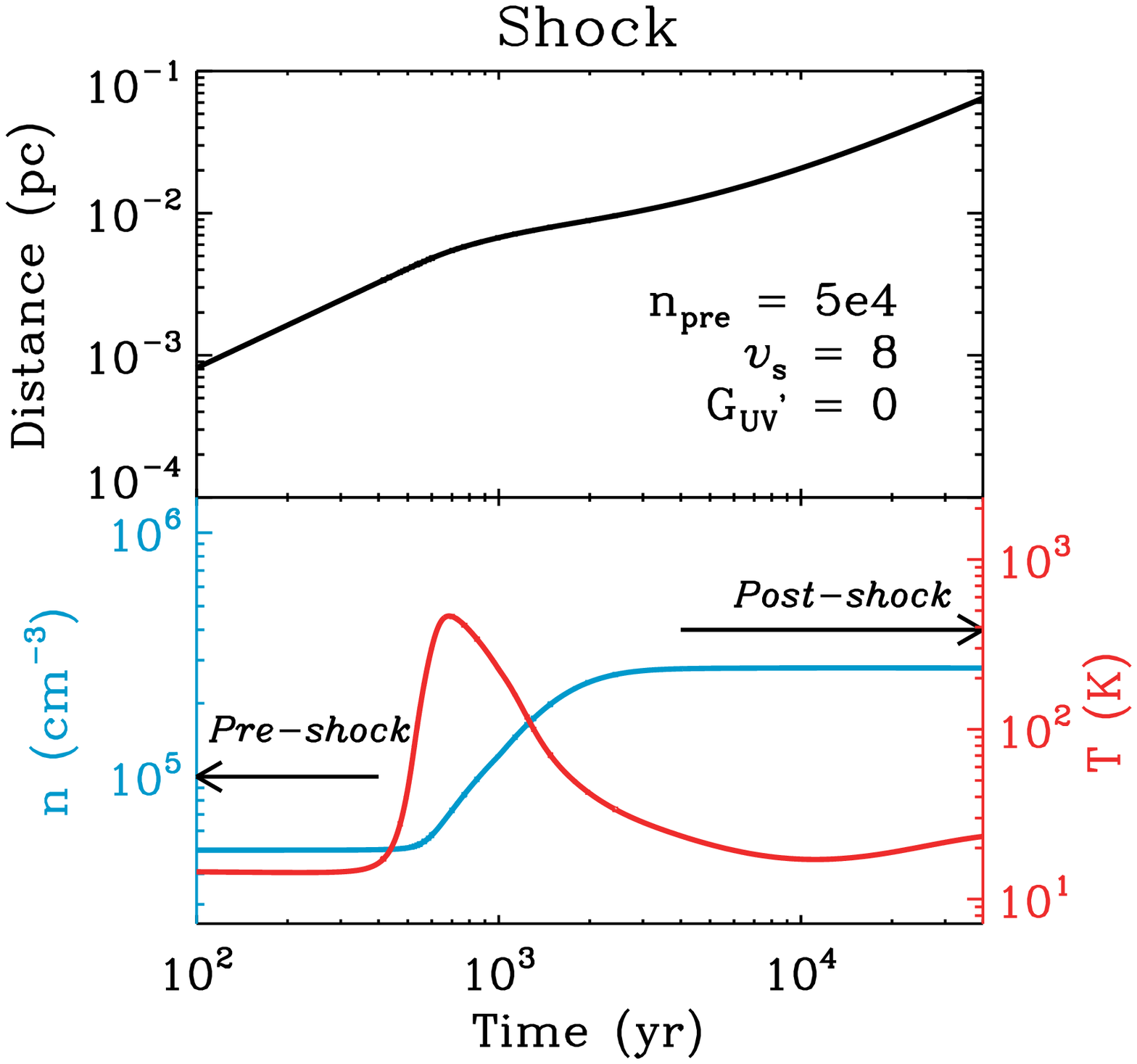} 
\caption{\label{f:UV_shock_comparison} Physical properties of the PDR and shock components in 30 Doradus.  
The following three models are presented as examples:  
(1) low-$P$ PDR with $A_{V}$ = 2 mag, $P/k_{\rm B}$ = 10$^{5}$ K cm$^{-3}$, and $G_{\rm UV}$ = 10$^{3}$ (\textit{left}), 
(2) high-$P$ PDR with $A_{V}$ = 10 mag, $P/k_{\rm B}$ = 6.7 $\times$ 10$^{8}$ K cm$^{-3}$, and $G_{\rm UV}$ = 10$^{4}$ (\textit{middle}), and 
(3) shock with $n_{\rm pre}$ = 5 $\times$ 10$^{4}$ cm$^{-3}$, $\varv_{\rm s}$ = 8 km s$^{-1}$, and $G'_{\rm UV}$ = 0 (\textit{right}). 
For the two PDR models, several physical quantities (\textit{top}: distance; \textit{middle}: density and temperature; 
\textit{bottom}: normalized emissivities of \CII 158 $\mu$m, \CI 370 $\mu$m, and \OI 145 $\mu$m, 
as well as selected CO and H$_{2}$ transitions) are shown as a function of dust extinction. 
For the shock model, the distance (measured along the direction of propagation of the shock wave; \textit{top}), 
as well as the density and temperature of the neutral fluid (\textit{bottom}), 
are plotted as a function of flow time.}
\end{figure*}

In addition to $n_{\rm pre}$ and $\varv_{\rm s}$, $G'_{\rm UV}$ and $\Omega$ are degenerate as well in the shock models 
($G'_{\rm UV}$ = 1 would dissociate more CO molecules than the $G'_{\rm UV}$ = 0 case, 
requiring a larger beam filling factor to achieve the same level of CO emission), 
and we tried to mitigate this degeneracy by considering the observed \CI 370 $\mu$m emission. 
For example, eight out of the total 23 pixels in our shock analysis have best-fit models with $G'_{\rm UV}$ = 1,   
and these $G'_{\rm UV}$ = 1 models overpredict \CI 370 $\mu$m by a factor of $\sim$10--20\footnote{
\CII 158 $\mu$m and \OI 145 $\mu$m in these models are still much fainter than those in our observations.}. 
In addition, the constrained $\Omega$ for these models is close to one, 
which is not compatible with what the high resolution ALMA CO(2--1) observations suggest ($\Omega$ $\lesssim$ 0.1). 
Considering that this is indeed a general trend 
(shock models with $G'_{\rm UV}$ = 1 that reproduce our CO and H$_{2}$ 0--0 S(3) observations 
tend to overpredict \CI 370 $\mu$m with unreasonably larger beam filling factors of $\sim$1), 
we then determined final shock properties by selecting $G'_{\rm UV}$ = 0 models with minimum $\chi^{2}$ values 
and present them in Fig. \ref{f:CO_SLEDs_shocks}.

Overall, we found that single shock models with $n_{\rm pre}$ $\sim$ 10$^{4}$--10$^{6}$ cm$^{-3}$, 
$\varv_{\rm s}$ $\sim$ 5--10 km s$^{-1}$, $\Omega$ $\sim$ 0.01--0.1, and no UV radiation field 
reproduce our CO and H$_{2}$ 0--0 S(3) observations reasonably well. 
The constrained $\varv_{\rm s}$ and $\Omega$ values seem to be consistent with previous studies as well  
($\varv_{\rm s}$: the 30$''$-scale CO(7--6) spectrum obtained by \cite{Pineda12} shows a linewdith of $\sim$10 km s$^{-1}$; 
$\Omega$: the ALMA CO(2--1) observations by \cite{Indebetouw13} suggest $\Omega$ $\lesssim$ 0.1),   
implying that our final shock models are reasonable. 
Considering the degeneracy that persists in our modeling, 
we will not discuss the shock parameters on a pixel-by-pixel basis
(our solutions in Fig. \ref{f:CO_SLEDs_shocks} should be considered as approximative models)
and instead focus on large-scale trends. 
For example, we examined the parameter space of ``good'' shock models that reproduce our CO and H$_{2}$ 0--0 S(3) observations within a factor of two 
and found that top pixels, e.g., $[-2,5]$, $[-2,4]$, and $[-1,4]$, indeed likely have a lower density of $\sim$10$^{4}$ cm$^{-2}$ compared to other pixels,  
based on a narrow distribution of $n_{\rm pre}$ (Fig. \ref{f:shock_deg2}). 

While reproducing the observed CO 
(including $J$=1--0; the shocked medium produces $\sim$30--80\% of the observed emission) 
and H$_{2}$ 0--0 S(3) transitions reasonably well, 
the shock models predict quite faint fine-structure lines. 
Specifically, we found that the model-to-observed line ratios are only $\lesssim$ 4 $\times$ 10$^{-6}$, $\lesssim$ 0.01, $\lesssim$ 0.03 
for \CII 158 $\mu$m, \CI 370 $\mu$m, and \OI 145 $\mu$m.
This result implies that at least two ISM components are required to fully explain our observations of 30 Doradus,
and the likely possibility in the shock scenario for CO would then be: 
(1) the low-$P$ PDR component (10$^{4}$--10$^{5}$ K cm$^{-3}$; Sect. \ref{s:PDR_results1}) 
that primarily contributes to the observed \CII 158 $\mu$m, \CI 370 $\mu$m, \OI 145 $\mu$m, and FIR emission 
and (2) the high-$P$ shock component (10$^{7}$--10$^{8}$ K cm$^{-3}$; Sect. \ref{s:UV_shocks} for details)
that radiates intensely mainly in CO. 
In the case of H$_{2}$ 0--0 S(3), both components contribute,  
and the combined contributions are still consistent with the measured upper limits. 
The shock-to-PDR ratio significantly varies from $\sim$0.1 to $\sim$6 for H$_{2}$ 0--0 S(3),  
which could be partly due to the degeneracy we still have in $n_{\rm pre}$ and $\varv_{\rm s}$. 

In short, we conclude that low-velocity C-type shocks with $n_{\rm pre}$ $\sim$ 10$^{4}$--10$^{6}$ cm$^{-3}$ 
and $\varv_{\rm s}$ $\sim$ 5--10 km s$^{-1}$ could be another important source of excitation for CO in 30 Doradus. 
The shock-compressed ($P/k_{\rm B}$ $\sim$ 10$^{7}$--10$^{8}$ K cm$^{-3}$) CO-emitting clumps are likely $\sim$0.7--2 pc in size 
and embedded within some low-$P$ ($P/k_{\rm B}$ $\sim$ 10$^{4}$--10$^{5}$ K cm$^{-3}$), 
UV-irradiated ($G_{\rm UV}$ $\sim$ 10$^{2}$--10$^{3}$) ISM component 
that produces bright \CII 158 $\mu$m, \CI 370 $\mu$m, \OI 145 $\mu$m, and FIR continuum emission. 
This low-$P$ PDR component fills a large fraction of our 30$''$-size pixels and provides up to $A_{V}$ $\sim$ 4--20 mag, 
shielding the shocked CO clumps from the dissociating UV radiation field.  
In the next sections, we will then present more detailed physical properties (e.g., density, temperature, etc.) 
of these shock and low-$P$ PDR components 
and compare them to those of the high-$P$ PDR component (Sect. \ref{s:PDR_results2}), 
with an aim of probing the origin of CO emission in 30 Doradus.  

\section{Discussions} 
\label{s:discussions} 

\subsection{Physical conditions of the neutral gas}
\label{s:UV_shocks} 

\subsubsection{Low thermal pressure component}
\label{s:UV_shocks_low_P}

We start our discussion by first presenting several physical quantities ($n$, $T$, and line emissivities) of 
a representative low-$P$ PDR model ($A_{V}$ = 2 mag, $P/k_{\rm B}$ = 10$^{5}$ K cm$^{-3}$, and $G_{\rm UV}$ = 10$^{3}$) 
as a function of $A_{V}$ in Fig. \ref{f:UV_shock_comparison}. 
As described throughout Sect. \ref{s:excitation_sources}, this low-$P$ PDR component is primarily bright in 
\CII 158 $\mu$m, \CI 370 $\mu$m, \OI 145 $\mu$m, and FIR continuum emission and 
essential to fully reproduce our multi-wavelength data of 30 Doradus. 

A close examination of the radial profiles in Fig. \ref{f:UV_shock_comparison} suggests that 
CO emission mostly originates from a diffuse and relatively warm medium with $n$ $\sim$ a few 100 cm$^{-3}$ and $T$ $\lesssim$ 100 K. 
The CO abundance in this diffuse and extended (line-of-sight depth of $\sim$6 pc) PDR component is low ($N$(CO) $\sim$ a few 10$^{13}$ cm$^{-2}$), 
which likely results from the following two aspects: 
(1) The slab of gas with relatively low dust extinction is illuminated by a strong UV radiation field. 
(2) The density is low. 
This low CO abundance is likely the primary reason for why the low-$P$ PDR component is so faint in CO emission. 

\subsubsection{High thermal pressure component}
\label{s:UV_shocks_high_P} 

Our analysis above indicates that high densities and/or temperatures would be needed for the observed bright CO emission, 
and we found that it is indeed the case for the high-$P$ PDR and shock components. 
For an illustration, we then again select representative high-$P$ PDR 
($A_{V}$ = 10 mag, $P/k_{\rm B}$ = 6.7 $\times$ 10$^{8}$ K cm$^{-3}$, and $G_{\rm UV}$ = 10$^{4}$)
and shock ($n_{\rm pre}$ = 5 $\times$ 10$^{4}$ cm$^{-3}$, $\varv_{\rm s}$ = 8 km s$^{-1}$, and $G'_{\rm UV}$ = 0) models 
and present their profiles in Fig. \ref{f:UV_shock_comparison}. 
Note that the shock profiles are different from those for the PDR models, 
in a way that they are shown as a function of the flow time through the shock structure (from the pre- to post-shock medium), rather than dust extinction.

The profiles in Fig. \ref{f:UV_shock_comparison} clearly show that 
the high-$P$ PDR has quite different conditions for CO emission compared to the low-$P$ PDR. 
For example, we found that low-$J$ CO lines are emitted from a highly dense and cold medium with $n$ $\gtrsim$ 10$^{7}$ cm$^{-3}$ and $T$ $\sim$ 30 K, 
while intermediate- and high-$J$ CO lines mainly arise from a relatively warm layer with $n$ $\sim$ a few 10$^{6}$ cm$^{-3}$ and $T$ $\sim$ 100 K. 
The bright CO emission from this dense and highly compressed PDR component (line-of-sight depth of $\sim$10$^{-3}$ pc)
is likely due to abundant CO molecules ($N$(CO) $\sim$ a few 10$^{18}$ cm$^{-2}$), 
which result from the sufficient dust extinction ($A_{V}$ $\gtrsim$ 5 mag) to protect CO from photodissociation, 
as well as from the high density. 

The physical properties of the high-$P$ PDR component seem slightly different from those of the low-velocity C-type shocks as well. 
Specifically, for the constrained shock models in Fig. \ref{f:CO_SLEDs_shocks}, 
we found that the shock-compressed CO-emitting layer (line-of-sight depth of $\sim$10$^{-2}$ pc) is less dense ($n$ $\sim$ a few (10$^{4}$--10$^{6}$) cm$^{-3}$) 
and less CO abundant ($N$(CO) $\sim$ a few (10$^{16}$--10$^{17}$) cm$^{-2}$), while having a higher temperature ($T$ $\sim$ 100--500 K), 
than the high-$P$ PDR counterpart. 

\subsection{Source of the high thermal pressure} 
\label{s:high_P_src} 

Our analyses suggest that the observed CO emission in 30 Doradus most likely originates from strongly compressed regions, 
whose high pressure ($P/k_{\rm B}$ $\sim$ 10$^{7}$--10$^{9}$ K cm$^{-3}$) could be driven by either UV photons or shocks.
Here we examine the likelihood of each case based on the known characteristics of 30 Doradus. 

\subsubsection{UV photons}  
\label{s:high_P_src_UV}

\begin{figure}
\centering
\includegraphics[scale=0.5]{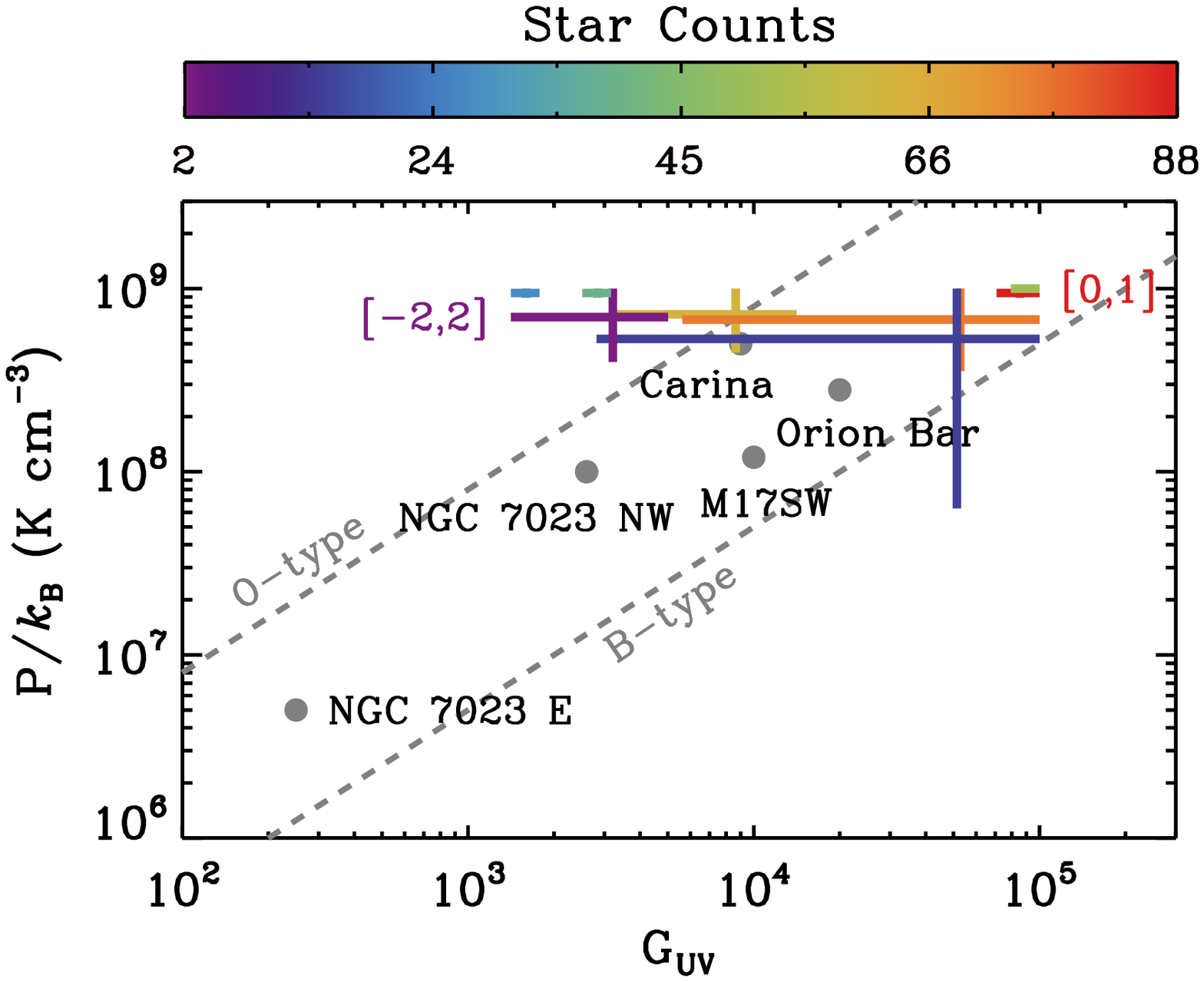}
\caption{\label{f:P_Guv_cor} $P/k_{\rm B}$ as a function of $G_{\rm UV}$ for various Galactic and extragalactic sources. 
The high-$P$ PDR conditions for CO emission in 30 Doradus are presented as the bars (same as in Fig \ref{f:PDR_results2_3}) 
in different colors depending on star counts, 
while other sources are shown as the gray circles
(Orion Bar and NGC 7023 NW from \citealt{Joblin18}; 
Carina Nebula from \citealt{RWu18};
NGC 7023 E from \citealt{Kohler14}; 
M17SW from \citealt{JP10}). 
In addition, the two pixels we discuss in the main text, [$-$2,2] and [0,1], are indicated, 
along with the predictions from \cite{Bron18} for B- and O-type stars
(gray dashed lines; $P/k_{\rm B}$-to-$G_{\rm UV}$ ratio = 5 $\times$ 10$^{3}$ and 8 $\times$ 10$^{4}$ respectively).}
\end{figure}

If UV photons are the dominant source of the high thermal pressure in the CO-emitting regions, 
we would expect somehow a correlation between stellar properties (e.g., spectral type, stellar density, etc.) and the constrained PDR conditions.  
Such a correlation was indeed predicted recently by the photoevaporating PDR model of \cite{Bron18}, 
where one-dimensional hydrodynamics, UV radiative transfer, and time-dependent thermo-chemical evolution are calculated simultaneously 
for a molecular cloud exposed to an adjacent massive star. 
In this model, the UV-illuminated surface of the cloud can freely evaporate into the surrounding gas, 
and this photoevaporation at the ionization and dissociation fronts produces high pressure (up to $\sim$10$^{9}$ K cm$^{-3}$). 
One of the predicted aspects of the photoevaporating PDR was a linear relation between $P/k_{\rm B}$ and $G_{\rm UV}$, 
whose slope depends on the spectral type of the star, 
e.g., the $P/k_{\rm B}$-to-$G_{\rm UV}$ ratios of $\sim$5 $\times$ 10$^{3}$ and $\sim$8 $\times$ 10$^{4}$ for B- and O-type stars (higher ratios for hotter stars).
This prediction seems to reproduce the observations of several Galactic PDRs (e.g., \citealt{Joblin18}; \citealt{RWu18}) 
and is shown in Fig. \ref{f:P_Guv_cor}. 

To evaluate whether UV photons are indeed responsible for the high thermal pressure in the CO-emitting regions, 
we examined the constrained high-$P$ PDR solutions in combination with the observed stellar properties. 
As an illustration, the minimum and maximum values of $P/k_{\rm B}$ and $G_{\rm UV}$ are indicated in Fig. \ref{f:P_Guv_cor} 
as bars in different colors depending on star counts. 
Here the star counts were estimated by counting the number of stars that fall into each FTS pixel in 30$''$ size 
($\sim$1.3 $\times$ 10$^{4}$ stars we used for the derivation of $G_{\rm stars}$ were considered; Fig. \ref{f:UV_sources} and Appendix \ref{s:appendix2})
and were found to vary by a factor of $\sim$40 from 2 to 88 for the eight pixels in our consideration.  
The measured $P/k_{\rm B}$ and $G_{\rm UV}$ values of 30 Doradus appear to be in reasonably good agreement with the predictions from \cite{Bron18}, 
but a close examination revealed that some of the observed trends are actually against expectations for UV-driven high pressure. 
For example, the pixels [$-$2,2] and [0,1] have the minimum and maximum star count respectively, 
yet their thermal pressures are comparable ($\sim$(0.5--1) $\times$ 10$^{9}$ K cm$^{-3}$). 
Considering that the high-$P$ PDR components of both pixels are equally likely quite close to the plane of R136 
(inferred from similar $G_{\rm UV}$ and $G_{\rm stars}$ values; Fig. \ref{f:PDR_results2_3}), 
it is indeed difficult to reconcile the comparable thermal pressures with a factor of $\sim$40 different star counts. 
In addition, [0,1] has a factor of $\sim$20 lower $P/k_{\rm B}$-to-$G_{\rm UV}$ ratio than [$-$2,2], 
even though it has more hotter stars, i.e., 13 stars with $T_{\rm eff}$ $\geq$ 4 $\times$ 10$^{4}$ K exist for [0,1], while none for [$-$2,2]. 
This result is in contrast with what \cite{Bron18} predicts. 

In summary, we conclude that while the Meudon PDR model reproduces the observed CO lines, 
the constrained PDR conditions are not in line with what we would expect based on the stellar content of 30 Doradus. 
This conclusion, however, is tentative and needs further examinations, 
since our current analyses have several limitations.  
For example, we analyzed the CO and fine-structure line observations of 30 Doradus using two PDR components. 
In reality, there would be a complicated mixture of multiple components on $\sim$10 pc scales, 
and spatially- and spectrally-resolved observations of CO and other neutral gas tracers (e.g., HCO$^{+}$ and HCN as dense gas tracers)
would be a key to fully assess the impact of UV photons on CO in 30 Doradus.
In addition, we compared the PDR properties of 30 Doradus with \cite{Bron18}, 
whose predictions are mainly for individual photoevaporating PDRs. 
To thoroughly examine whether UV photons are the dominant source of the high pressure for CO in 30 Doradus, 
a collective role of UV photons on larger scales must be considered, 
which would require simulations of multiple star clusters.  

\begin{figure*}
\centering
\includegraphics[scale=0.2]{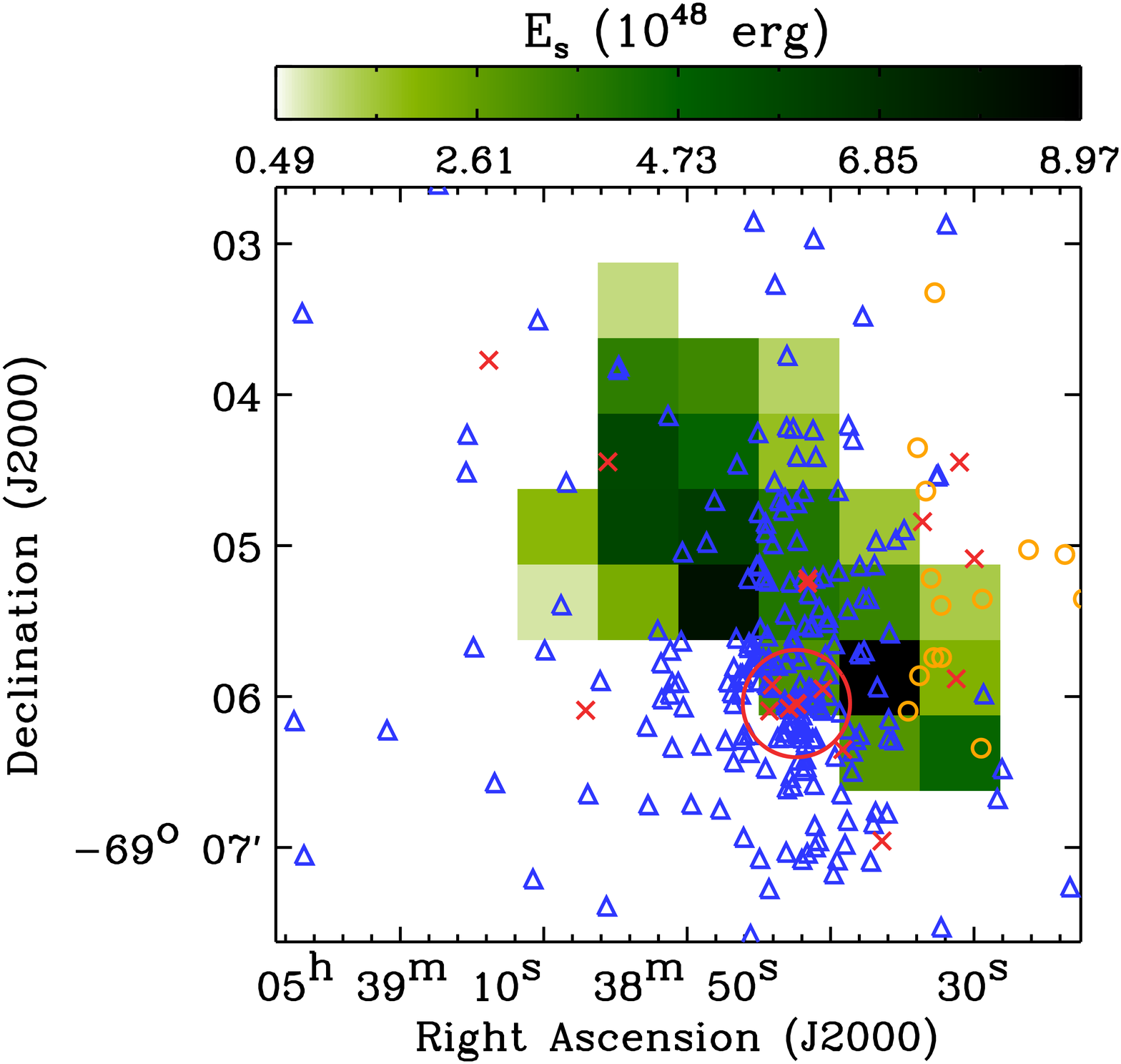}\hspace{1cm}
\includegraphics[scale=0.4]{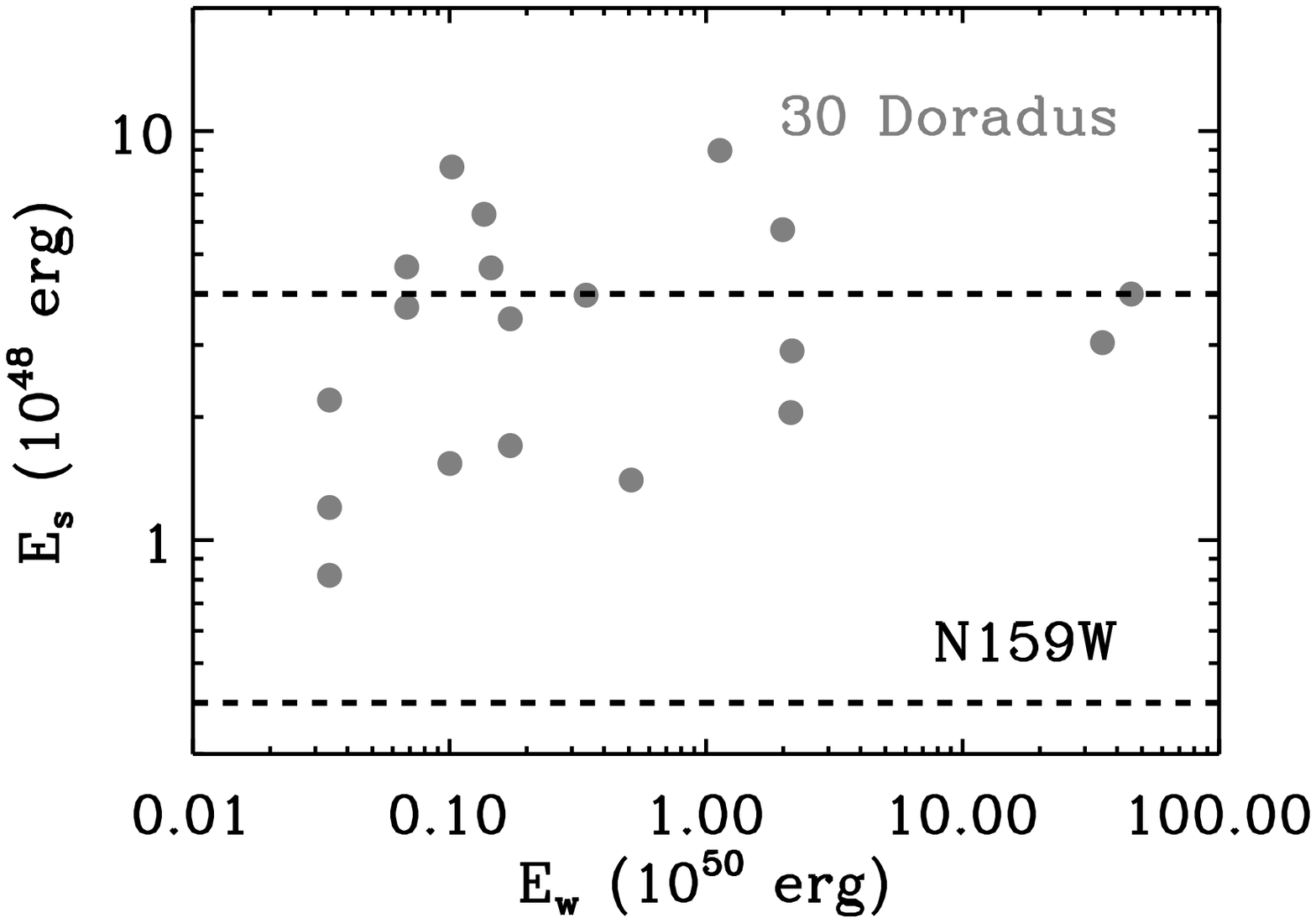}
\caption{\label{f:Es_Ew_plots} \textit{Left}: $E_{\rm s}$ of the final shock models (shown in Fig. \ref{f:CO_SLEDs_shocks}) is presented along with
$\sim$500 stars for which $E_{\rm w}$ estimates are available ($E_{\rm w} \geq 10^{50}$ erg as the red crosses, 
10$^{49}$ erg $\leq$ $E_{\rm w}$ $<$ 10$^{50}$ erg as the orange circles, 
and $E_{\rm s}$ $<$ 10$^{49}$ erg as the blue triangles; \citealt{Doran13}). 
The location of R136 is also indicated as the red circle. 
\textit{Right}: $E_{\rm s}$ as a function of $E_{\rm w}$. 
The data points for 30 Doradus are shown in gray, 
while the range of $E_{\rm s}$ estimated by \cite{Lee16} for N159W is overlaid as the black dashed line.}
\end{figure*}

\subsubsection{Low-velocity shocks}  
\label{s:high_P_src_shocks}

In the case where low-velocity shocks are the origin of the high thermal pressure in the CO-emitting regions, 
stellar winds from hot luminous stars could provide the required mechanical energy to drive the shocks. 
To examine such a possibility, we calculated the total energy dissipated by shocks ($E_{\rm s}$) 
using our constrained models in Sect. \ref{s:shock_results1} (\citealt{Lee16} for details on the calculation)  
and compared it with the stellar wind energy ($E_{\rm w}$) of $\sim$500 W-R and OB-type stars from \cite{Doran13} (Fig. \ref{f:Es_Ew_plots} \textit{Left}). 
For our comparison, the shock timescale of 0.1 Myr (typical time needed for the shocked medium to return to equilibrium)
and the wind timescale of 2 Myr (average OB stellar lifetime) were assumed, 
essentially following \cite{Doran13} and \cite{Lee16}. 
When considering $\sim$150 stars that fall into the 23 pixels where our shock solutions exist,  
we found that stellar winds from these stars can inject the total mechanical energy of $\sim$10$^{52}$ erg, 
which would be sufficient to drive the low-velocity shocks dissipating $\sim$10$^{50}$ erg in the region. 
These total energies of $\sim$10$^{52}$ erg and $\sim$10$^{50}$ erg were estimated by simply summing up $E_{\rm w}$ and $E_{\rm s}$ 
over the 23 pixels in our consideration. 
Interestingly, Fig. \ref{f:Es_Ew_plots} \textit{Left} shows that the shock and wind energy have contrasting spatial distributions:
$E_{\rm s}$ varies smoothly across the region, while $E_{\rm w}$ is highly concentrated around R136.
To quantify this difference, we then calculated $E_{\rm w}$ on a pixel-by-pixel basis 
by summing $E_{\rm w}$ values of all W-R and OB-type stars that fall into each FTS pixel
and compared it to $E_{\rm s}$ (Fig. \ref{f:Es_Ew_plots} \textit{Right}). 
As we just discussed, Fig. \ref{f:Es_Ew_plots} \textit{Right} clearly demonstrates that 
$E_{\rm s}$ is relatively uniform with a factor of $\sim$10 variations, 
while $E_{\rm w}$ changes by a factor of $\sim$2000 ($E_{\rm w}$ $>$ 10$^{51}$ erg coincides with R136 and its adjacent pixels).    
This highly concentrated distribution of $E_{\rm w}$ was also noted by \cite{Doran13}, 
i.e., 75\% of the total wind luminosity is contained within 20 pc of R136, 
and suggests that stellar winds are likely not the main driver of the low-velocity shocks. 

In addition to stellar winds from hot luminous stars, 
SNe can inject a large amount of mechanical energy into the surrounding ISM 
($\sim$10$^{51}$ erg per SNe; e.g., \citealt{Hartmann99}). 
So far 59 SNRs have been identified in the LMC (\citealt{Maggi16}), 
and 30 Doradus harbors only one SNR, N157B at 
($\alpha$, $\delta$)$_{\rm J2000}$ = (05$^{\rm h}$37$^{\rm m}$47$^{\rm s}$, $-69^{\circ}$10$'$20$''$) (\citealt{Chen06}). 
Considering that 30 Doradus hosts $\sim$25\% of the massive stars in the LMC (\citealt{Kennicutt86})
and core-collapsed SNRs closely follow active star formation (25 such SNRs in the LMC; \citealt{Maggi16}), however, 
we would expect to find more SNRs in 30 Doradus and roughly estimate the expected number of 25 $\times$ 0.25 $\sim$ 6 SNRs
(which could have been missed previously due to their low surface brightness and/or the crowdedness in the 30 Doradus region). 
While our estimate is uncertain,  
it is indeed consistent with the roughly half-dozen high-velocity ionized bubbles in 30 Doradus 
(likely blown up by SNe; e.g., \citealt{Chu94})
and implies that SNe could provide sufficient energy to drive the low-velocity shocks.  
But again, as in the case of stellar winds, 
the relatively uniform distribution of $E_{\rm s}$ would be difficult to explain in the framework of SNe-driven shocks. 

Our results so far suggest that the low-velocity shocks in 30 Doradus likely originate from non-stellar sources.  
This conclusion is also consistent with the fact that 30 Doradus and N159W have comparable $E_{\rm s}$ on $\sim$10 pc scales 
(our FTS pixel size; Fig. \ref{f:Es_Ew_plots} \textit{Right}) despite a large difference in the number of massive young stars: 
$\sim$1100 in 30 Doradus vs. $\sim$150 in N159W (\citealt{Farina09}; \citealt{Doran13}).
The comparable $E_{\rm s}$ values between 30 Doradus and N159W would in turn suggest that large-scale processes 
($\gtrsim$ 600 pc; distance between 30 Doradus and N159W) are likely the major source of the low-velocity shocks, 
and the kpc-scale injection of significant energy into the Magellanic Clouds has been indeed suggested by previous power spectrum analyses 
(e.g., \citealt{Elmegreen01}; \citealt{Nestingen-Palm17}). 
One of the possible processes for energy injection on kpc-scales is the complicated interaction between the Milky Way and the Magellanic Clouds. 
While the dynamics of the entire Magellanic System 
(two Magellanic Clouds and gaseous structures surrounding them, i.e., the Stream, the Bridge, and the Leading Arm)
is still a subject of active research, 
it has been well known that the southeastern \HI overdensity region where 30 Doradus and N159W are located (Fig. \ref{f:intro}) 
is strongly perturbed (e.g., \citealt{Luks92}) and likely influenced by tidal and/or ram-pressure stripping 
between the Milky Way and the Magellanic Clouds (e.g., \citealt{D'onghia16}). 
Such an energetic interplay between galaxies can deposit a large amount of mechanical energy,  
which would then cascade down to small scales and low velocities, 
as witnessed in both local and high-redshift interacting systems 
(e.g., \citealt{Appleton17}; \citealt{Falgarone17}). 

Finally, we note that low-velocity shocks would be pervasive in the LMC if they indeed arise from kpc-scale processes. 
These shocks would have a negligible impact on the low-$P$ PDR component in 30 Doradus though, 
since the shocks would compress only a fraction of this diffuse and extended component, 
e.g., line-of-sight depth of $\sim$10$^{-2}$ pc and $\sim$6 pc  
for the high-$P$ shock and low-$P$ PDR component respectively (Sect. \ref{s:UV_shocks}). 

\subsection{CO SLEDs as a probe of the excitation mechanisms of warm molecular gas}  
\label{s:CO_SLED_shape} 

So far we analyzed the observed CO SLEDs of 30 Doradus 
with an aim of examining the excitation mechanisms of warm molecular gas in star-forming regions, 
and our results clearly show that CO SLEDs alone cannot differentiate heating sources.
For example, the observed CO SLEDs significantly change across 30 Doradus (Sect. \ref{s:observed_CO_SLEDs}), 
e.g., $J_{\rm p}$ ranging from 6--5 to 10--9 and $\alpha$ ranging from $\sim$0.4 to $\sim$1.8, 
and our PDR and shock modeling suggest that 
these varying CO SLEDs mostly reflect the changes in physical conditions (e.g., temperature and density), 
rather than underlying excitation mechanisms. 
The fact that N159W has systematically different CO SLEDs (e.g., $J_{\rm p}$ = 4--3 to 7--6 and $\alpha$ = 0.3--0.7; \citealt{Lee16}), 
yet likely shares the same excitation mechanism as 30 Doradus, also supports our conclusion. 
In addition to CO lines, complementary observational constraints, 
e.g., fine-structure lines, FIR luminosity, and properties of massive young stars, were then found to be highly essential 
to examine the excitation mechanisms in detail and evaluate their feasibility. 
All in all, our study demonstrates that one should take a comprehensive approach 
when interpreting multi-transition CO observations in the context of probing the excitation sources of warm molecular gas
(e.g., \citealt{Mashian15}; \citealt{Indriolo17}; \citealt{Kamenetzky17}). 


Another key result from our work is the crucial role of shocks in CO heating. 
As described in Sect. \ref{s:intro}, both Galactic and extragalactic studies have highlighted the importance of mechanical heating for CO,  
and our N159W and 30 Doradus analyses show that mechanical heating by low-velocity shocks ($\sim$10 km s$^{-1}$) 
is indeed a major contributor to the excitation of molecular gas on $\sim$10 pc scales. 
What remains relatively uncertain is the source of shocks. 
While we concluded that low-velocity shocks in N159W and 30 Doradus likely originate from large-scale processes 
such as the complex interaction between the Milky Way and the Magellanic Clouds, 
this hypothesis should be verified by observing independent shock tracers, e.g., SiO and H$_{2}$ transitions, throughout the LMC. 
Such observations would be possible with current and upcoming facilities, e.g., ALMA, \textit{SOFIA}, and \textit{JWST}, 
providing insights into the injection and dissipation of mechanical energy in the ISM. 
These observations will also further test our tentative rejection of UV photons as the main heating source for CO (Sect. \ref{s:high_P_src_UV}).


\section{Summary}
\label{s:summary} 

In this paper, we present \textit{Herschel} SPIRE FTS observations of 30 Doradus, 
the most extreme starburst region in the Local Universe harboring more than 1000 massive young stars. 
To examine the physical conditions and excitation mechanisms of molecular gas, 
we combined the FTS CO observations (CO $J$=4--3 to $J$=13--12) with other tracers of gas and dust 
and analyzed them on 42$''$ or $\sim$10 pc scales using the state-of-the-art Meudon PDR and Paris-Durham shock models. 
Our main results are as follows. 

\begin{enumerate}
\item In our FTS observations, important cooling lines in the ISM, 
such as CO rotational transitions (from $J$=4--3 to $J$=13--12), \CI 370 $\mu$m, and \NII 205 $\mu$m, were clearly detected. 

\item We constructed CO SLEDs on a pixel-by-pixel basis by combining the FTS observations with ground-based CO(1--0) and CO(3--2) data 
and found that the CO SLEDs vary considerably across 30 Doradus. 
These variations include the changes in the peak transition $J_{\rm p}$ (from $J$=6--5 to $J$=10--9), 
as well as in the slope characterized by the high-to-intermediate $J$ ratio $\alpha$ (from $\sim$0.4 to $\sim$1.8). 

\item To evaluate the impact of UV photons on CO,  
we performed Meudon PDR modeling and showed that CO emission in 30 Doradus could arise 
from $\sim$0.7--2 pc scale PDR clumps with $A_{V}$ $\gtrsim$ 5 mag, 
$P/k_{\rm B}$ $\sim$ 10$^{8}$--10$^{9}$ K cm$^{-3}$, and $G_{\rm UV}$ $\sim$ 10$^{3}$--10$^{5}$. 
Interestingly, these PDR clumps are quite faint in \CII 158 $\mu$m and FIR dust continuum emission, 
and we found that another PDR component with lower $A_{V}$ $\sim$ 2 mag, $P/k_{\rm B}$ $\sim$ a few (10$^{4}$--10$^{5}$) K cm$^{-3}$, 
and $G_{\rm UV}$ $\sim$ a few (10$^{2}$--10$^{3}$) (filling a large fraction of $\sim$10 pc-size FTS pixels) 
is required to explain the observed fine-structure lines 
(\CII 158 $\mu$m, \CI 370 $\mu$m, and \OI 145 $\mu$m) and FIR luminosity. 
The constrained properties of the high-$P$ PDR clumps, however, are not consistent with what we would expect based on the stellar content of 30 Doradus, 
and we thus tentatively concluded that UV photons are likely not the primary heating source for CO. 

\item Based on the observed X-ray and $\gamma$-ray properties of 30 Doradus,  
we concluded that X-rays and cosmic-rays likely play a minor role in CO heating. 

\item Our Paris-Durham shock modeling showed that the observed CO SLEDs of 30 Doradus can be reproduced by 
low-velocity C-type shocks with $n_{\rm pre}$ $\sim$ 10$^{4}$--10$^{6}$ cm$^{-3}$ and $\varv_{\rm s}$ $\sim$ 5--10 km s$^{-1}$. 
The shock-compressed ($P/k_{\rm B}$ $\sim$ 10$^{7}$--10$^{8}$ K cm$^{-3}$) CO-emitting clumps on $\sim$0.7--2 pc scales 
are likely well-shielded from dissociating UV photons and embedded within the low-$P$ PDR component 
that emits brightly in \CII 158 $\mu$m, \CI 370 $\mu$m, \OI 145 $\mu$m, and FIR continuum emission. 
Considering the properties of massive young stars in 30 Doradus, 
we excluded the stellar origin of low-velocity shocks
and concluded that the shocks are likely driven by large-scale processes 
such as the interaction between the Milky Way and the Magellanic Clouds.

\item Our analyses suggest that the significant variations in the observed CO SLEDs of 30 Doradus mostly reflect 
the changes in physical conditions (e.g., temperature and density), rather than underlying excitation mechanisms.
This implies that the shape of CO SLEDs alone cannot be used as a probe of heating sources. 
\end{enumerate}

While large-scale low-velocity shocks were suggested as the dominant heating source of CO in 30 Doradus, 
we note that our conclusion was based on $\sim$10 pc scale analyses. 
As \cite{Indriolo17} pointed out, CO SLEDs strongly depend on spatial scales,  
and how the spatial scale of CO observations affects the interpretation of heating sources is currently unclear. 
For a more comprehensive picture on the nature of warm molecular gas in individual star-forming regions, 
it would hence be critical to analyze high spatial and spectral resolution CO observations over a large area 
in combination with complementary constraints, e.g., PDR and shock tracers, as well as stellar properties. 

\begin{acknowledgements} 
We would like to thank the anonymous referee for constructive comments that improved this work.  
We also thank E. Bron, B. Godard, C. Matzner, E. Roueff, A. Tielens, and S. Viti for helpful discussions 
and acknowledge support from the SYMPATICO grant of the French Agence Nationale de la Recherche (ANR-11-BS56-0023), 
the PCMI program of the French Centre National de la Recherche Scientifique,  
the sub-project A6 of the Collaborative Research Council 956 funded by the Deutsche Forschungsgemeinschaft,
and the Emmy Noether grant KR4801/1-1 funded by the Deutsche Forschungsgemeinschaft. 
PACS has been developed by a consortium of institutes led by MPE (Germany) and including UVIE (Austria); 
KU Leuven, CSL, IMEC (Belgium); CEA, LAM (France); MPIA (Germany); INAF-IFSI/OAA/OAP/OAT, LENS, SISSA (Italy); IAC (Spain).
This development has been supported by the funding agencies BMVIT (Austria), ESA-PRODEX (Belgium), CEA/CNES (France), DLR (Germany), ASI/INAF (Italy), and CICYT/MCYT (Spain). 
SPIRE has been developed by a consortium of institutes led by Cardiff University (UK) and including Univ. Lethbridge (Canada); NAOC (China); 
CEA, LAM (France); IFSI, Univ. Padua (Italy); IAC (Spain); Stockholm Observatory (Sweden); Imperial College London, RAL, UCL-MSSL, UKATC, Univ. Sussex (UK); 
and Caltech, JPL, NHSC, Univ. Colorado (USA).
This development has been supported by national funding agencies: CSA (Canada); NAOC (China); CEA, CNES, CNRS (France); ASI (Italy); MCINN (Spain); SNSB (Sweden); STFC, UKSA (UK); and NASA (USA).
\end{acknowledgements} 

\bibliographystyle{aa}
\bibliography{/media/mlee/Bibtex/myref}

\begin{thebibliography}{102}
\expandafter\ifx\csname natexlab\endcsname\relax\def\natexlab#1{#1}\fi

\bibitem[{{Abdo} {et~al.}(2009){Abdo}, {Ackermann}, {Ajello}, {Atwood},
  {Axelsson}, {Baldini}, {Ballet}, {Barbiellini}, {Bastieri}, {Baughman},
  {Bechtol}, {Bellazzini}, {Berenji}, {Bloom}, {Bonamente}, {Borgland},
  {Bregeon}, {Brez}, {Brigida}, {Bruel}, {Burnett}, {Caliandro}, {Cameron},
  {Caraveo}, {Carlson}, {Casandjian}, {Cecchi}, {{\c C}elik}, {Chekhtman},
  {Cheung}, {Ciprini}, {Claus}, {Cohen-Tanugi}, {Conrad}, {Cutini}, {Dermer},
  {de Angelis}, {de Palma}, {Digel}, {Silva}, {Drell}, {Dubois}, {Dumora},
  {Farnier}, {Favuzzi}, {Fegan}, {Focke}, {Frailis}, {Fukazawa}, {Funk},
  {Fusco}, {Gargano}, {Gasparrini}, {Gehrels}, {Germani}, {Giebels},
  {Giglietto}, {Giordano}, {Glanzman}, {Godfrey}, {Grenier}, {Grondin},
  {Grove}, {Guillemot}, {Guiriec}, {Hanabata}, {Harding}, {Hayashida}, {Hays},
  {Hughes}, {J{\'o}hannesson}, {Johnson}, {Johnson}, {Johnson}, {Kamae},
  {Katagiri}, {Kawai}, {Kerr}, {Kn{\"o}dlseder}, {Kocian}, {Kuehn}, {Kuss},
  {Lande}, {Latronico}, {Lemoine-Goumard}, {Longo}, {Loparco}, {Lott},
  {Lovellette}, {Lubrano}, {Makeev}, {Mazziotta}, {McEnery}, {Meurer},
  {Michelson}, {Mitthumsiri}, {Mizuno}, {Moiseev}, {Monte}, {Monzani},
  {Morselli}, {Moskalenko}, {Murgia}, {Nolan}, {Norris}, {Nuss}, {Ohsugi},
  {Okumura}, {Omodei}, {Orlando}, {Ormes}, {Ozaki}, {Paneque}, {Panetta},
  {Parent}, {Pepe}, {Pesce-Rollins}, {Piron}, {Pohl}, {Porter}, {Rain{\`o}},
  {Rando}, {Razzano}, {Reimer}, {Reimer}, {Reposeur}, {Ritz}, {Rochester},
  {Rodriguez}, {Ryde}, {Sadrozinski}, {Sanchez}, {Sander}, {Saz Parkinson},
  {Schalk}, {Sellerholm}, {Sgr{\`o}}, {Smith}, {Smith}, {Spandre}, {Spinelli},
  {Starck}, {Stecker}, {Strickman}, {Strong}, {Suson}, {Tajima}, {Takahashi},
  {Takahashi}, {Tanaka}, {Thayer}, {Thayer}, {Thompson}, {Tibaldo}, {Torres},
  {Tosti}, {Tramacere}, {Uchiyama}, {Usher}, {Vasileiou}, {Vilchez}, {Vitale},
  {Waite}, {Wang}, {Winer}, {Wood}, {Ylinen}, \& {Ziegler}}]{Abdo09LocalISM}
{Abdo}, A.~A., {Ackermann}, M., {Ajello}, M., {et~al.} 2009, \apj, 703, 1249

\bibitem[{{Abdo} {et~al.}(2010){Abdo}, {Ackermann}, {Ajello}, {Atwood},
  {Baldini}, {Ballet}, {Barbiellini}, {Bastieri}, {Baughman}, {Bechtol},
  {Bellazzini}, {Berenji}, {Blandford}, {Bloom}, {Bonamente}, {Borgland},
  {Bregeon}, {Brez}, {Brigida}, {Bruel}, {Burnett}, {Buson}, {Caliandro},
  {Cameron}, {Caraveo}, {Casandjian}, {Cecchi}, {{\c C}elik}, {Chekhtman},
  {Cheung}, {Chiang}, {Ciprini}, {Claus}, {Cohen-Tanugi}, {Cominsky}, {Conrad},
  {Cutini}, {Dermer}, {de Angelis}, {de Palma}, {Digel}, {Silva}, {Drell},
  {Dubois}, {Dumora}, {Farnier}, {Favuzzi}, {Fegan}, {Focke}, {Fortin},
  {Frailis}, {Fukazawa}, {Fusco}, {Gargano}, {Gasparrini}, {Gehrels},
  {Germani}, {Giavitto}, {Giebels}, {Giglietto}, {Giordano}, {Glanzman},
  {Godfrey}, {Gotthelf}, {Grenier}, {Grondin}, {Grove}, {Guillemot}, {Guiriec},
  {Hanabata}, {Harding}, {Hayashida}, {Hays}, {Horan}, {Hughes}, {Jackson},
  {Jean}, {J{\'o}hannesson}, {Johnson}, {Johnson}, {Johnson}, {Johnson},
  {Kamae}, {Katagiri}, {Kataoka}, {Kawai}, {Kerr}, {Kn{\"o}dlseder}, {Kocian},
  {Kuss}, {Lande}, {Latronico}, {Lemoine-Goumard}, {Longo}, {Loparco}, {Lott},
  {Lovellette}, {Lubrano}, {Madejski}, {Makeev}, {Marshall}, {Martin},
  {Mazziotta}, {McConville}, {McEnery}, {Meurer}, {Michelson}, {Mitthumsiri},
  {Mizuno}, {Moiseev}, {Monte}, {Monzani}, {Morselli}, {Moskalenko}, {Murgia},
  {Nolan}, {Norris}, {Nuss}, {Ohsugi}, {Omodei}, {Orlando}, {Ormes}, {Paneque},
  {Parent}, {Pelassa}, {Pepe}, {Pesce-Rollins}, {Piron}, {Porter}, {Rain{\`o}},
  {Rando}, {Razzano}, {Reimer}, {Reimer}, {Reposeur}, {Ritz}, {Rodriguez},
  {Romani}, {Roth}, {Ryde}, {Sadrozinski}, {Sanchez}, {Sander}, {Saz
  Parkinson}, {Scargle}, {Sellerholm}, {Sgr{\`o}}, {Siskind}, {Smith}, {Smith},
  {Spandre}, {Spinelli}, {Starck}, {Strickman}, {Strong}, {Suson}, {Tajima},
  {Takahashi}, {Tanaka}, {Thayer}, {Thayer}, {Thompson}, {Tibaldo}, {Torres},
  {Tosti}, {Tramacere}, {Uchiyama}, {Usher}, {Vasileiou}, {Venter}, {Vilchez},
  {Vitale}, {Waite}, {Wang}, {Weltevrede}, {Winer}, {Wood}, {Ylinen}, \&
  {Ziegler}}]{Abdo10LMC}
{Abdo}, A.~A., {Ackermann}, M., {Ajello}, M., {et~al.} 2010, \aap, 512, A7

\bibitem[{{Appleton} {et~al.}(2017){Appleton}, {Guillard}, {Togi}, {Alatalo},
  {Boulanger}, {Cluver}, {Pineau des For{\^e}ts}, {Lisenfeld}, {Ogle}, \&
  {Xu}}]{Appleton17}
{Appleton}, P.~N., {Guillard}, P., {Togi}, A., {et~al.} 2017, \apj, 836, 76

\bibitem[{{Bakes} \& {Tielens}(1994)}]{Bakes94}
{Bakes}, E.~L.~O. \& {Tielens}, A.~G.~G.~M. 1994, \apj, 427, 822

\bibitem[{{Bron} {et~al.}(2018){Bron}, {Ag{\'u}ndez}, {Goicoechea}, \&
  {Cernicharo}}]{Bron18}
{Bron}, E., {Ag{\'u}ndez}, M., {Goicoechea}, J.~R., \& {Cernicharo}, J. 2018,
  ArXiv e-prints [\eprint[arXiv]{1801.01547}]

\bibitem[{{Bron} {et~al.}(2014){Bron}, {Le Bourlot}, \& {Le Petit}}]{Bron14}
{Bron}, E., {Le Bourlot}, J., \& {Le Petit}, F. 2014, \aap, 569, A100

\bibitem[{{Bron} {et~al.}(2016){Bron}, {Le Petit}, \& {Le Bourlot}}]{Bron16}
{Bron}, E., {Le Petit}, F., \& {Le Bourlot}, J. 2016, \aap, 588, A27

\bibitem[{{Burton} {et~al.}(1990){Burton}, {Hollenbach}, \&
  {Tielens}}]{Burton90}
{Burton}, M.~G., {Hollenbach}, D.~J., \& {Tielens}, A.~G.~G.~M. 1990, \apj,
  365, 620

\bibitem[{{Chen} {et~al.}(2006){Chen}, {Wang}, {Gotthelf}, {Jiang}, {Chu}, \&
  {Gruendl}}]{Chen06}
{Chen}, Y., {Wang}, Q.~D., {Gotthelf}, E.~V., {et~al.} 2006, \apj, 651, 237

\bibitem[{{Chevance} {et~al.}(2016){Chevance}, {Madden}, {Lebouteiller},
  {Godard}, {Cormier}, {Galliano}, {Hony}, {Indebetouw}, {Le Bourlot}, {Lee},
  {Le Petit}, {Pellegrini}, {Roueff}, \& {Wu}}]{Chevance16}
{Chevance}, M., {Madden}, S.~C., {Lebouteiller}, V., {et~al.} 2016, \aap, 590,
  A36

\bibitem[{{Chu} \& {Kennicutt}(1994)}]{Chu94}
{Chu}, Y.-H. \& {Kennicutt}, Jr., R.~C. 1994, \apj, 425, 720

\bibitem[{{Cormier} {et~al.}(2015){Cormier}, {Madden}, {Lebouteiller}, {Abel},
  {Hony}, {Galliano}, {R{\'e}my-Ruyer}, {Bigiel}, {Baes}, {Boselli},
  {Chevance}, {Cooray}, {De Looze}, {Doublier}, {Galametz}, {Hughes},
  {Karczewski}, {Lee}, {Lu}, \& {Spinoglio}}]{Cormier15}
{Cormier}, D., {Madden}, S.~C., {Lebouteiller}, V., {et~al.} 2015, \aap, 578,
  A53

\bibitem[{{Crowther} \& {Dessart}(1998)}]{Crowther98}
{Crowther}, P.~A. \& {Dessart}, L. 1998, \mnras, 296, 622

\bibitem[{{Crowther} {et~al.}(2010){Crowther}, {Schnurr}, {Hirschi}, {Yusof},
  {Parker}, {Goodwin}, \& {Kassim}}]{Crowther10}
{Crowther}, P.~A., {Schnurr}, O., {Hirschi}, R., {et~al.} 2010, \mnras, 408,
  731

\bibitem[{{de Graauw} {et~al.}(2010){de Graauw}, {Helmich}, {Phillips},
  {Stutzki}, {Caux}, {Whyborn}, {Dieleman}, {Roelfsema}, {Aarts}, {Assendorp},
  {Bachiller}, {Baechtold}, {Barcia}, {Beintema}, {Belitsky}, {Benz}, {Bieber},
  {Boogert}, {Borys}, {Bumble}, {Ca{\"i}s}, {Caris}, {Cerulli-Irelli},
  {Chattopadhyay}, {Cherednichenko}, {Ciechanowicz}, {Coeur-Joly}, {Comito},
  {Cros}, {de Jonge}, {de Lange}, {Delforges}, {Delorme}, {den Boggende},
  {Desbat}, {Diez-Gonz{\'a}lez}, {di Giorgio}, {Dubbeldam}, {Edwards},
  {Eggens}, {Erickson}, {Evers}, {Fich}, {Finn}, {Franke}, {Gaier}, {Gal},
  {Gao}, {Gallego}, {Gauffre}, {Gill}, {Glenz}, {Golstein}, {Goulooze},
  {Gunsing}, {G{\"u}sten}, {Hartogh}, {Hatch}, {Higgins}, {Honingh}, {Huisman},
  {Jackson}, {Jacobs}, {Jacobs}, {Jarchow}, {Javadi}, {Jellema}, {Justen},
  {Karpov}, {Kasemann}, {Kawamura}, {Keizer}, {Kester}, {Klapwijk}, {Klein},
  {Kollberg}, {Kooi}, {Kooiman}, {Kopf}, {Krause}, {Krieg}, {Kramer},
  {Kruizenga}, {Kuhn}, {Laauwen}, {Lai}, {Larsson}, {Leduc}, {Leinz}, {Lin},
  {Liseau}, {Liu}, {Loose}, {L{\'o}pez-Fernandez}, {Lord}, {Luinge}, {Marston},
  {Mart{\'{\i}}n-Pintado}, {Maestrini}, {Maiwald}, {McCoey}, {Mehdi}, {Megej},
  {Melchior}, {Meinsma}, {Merkel}, {Michalska}, {Monstein}, {Moratschke},
  {Morris}, {Muller}, {Murphy}, {Naber}, {Natale}, {Nowosielski}, {Nuzzolo},
  {Olberg}, {Olbrich}, {Orfei}, {Orleanski}, {Ossenkopf}, {Peacock}, {Pearson},
  {Peron}, {Phillip-May}, {Piazzo}, {Planesas}, {Rataj}, {Ravera}, {Risacher},
  {Salez}, {Samoska}, {Saraceno}, {Schieder}, {Schlecht}, {Schl{\"o}der},
  {Schm{\"u}lling}, {Schultz}, {Schuster}, {Siebertz}, {Smit}, {Szczerba},
  {Shipman}, {Steinmetz}, {Stern}, {Stokroos}, {Teipen}, {Teyssier}, {Tils},
  {Trappe}, {van Baaren}, {van Leeuwen}, {van de Stadt}, {Visser}, {Wildeman},
  {Wafelbakker}, {Ward}, {Wesselius}, {Wild}, {Wulff}, {Wunsch}, {Tielens},
  {Zaal}, {Zirath}, {Zmuidzinas}, \& {Zwart}}]{deGraauw10}
{de Graauw}, T., {Helmich}, F.~P., {Phillips}, T.~G., {et~al.} 2010, \aap, 518,
  L6

\bibitem[{{D'Onghia} \& {Fox}(2016)}]{D'onghia16}
{D'Onghia}, E. \& {Fox}, A.~J. 2016, \araa, 54, 363

\bibitem[{{Doran} {et~al.}(2013){Doran}, {Crowther}, {de Koter}, {Evans},
  {McEvoy}, {Walborn}, {Bastian}, {Bestenlehner}, {Gr{\"a}fener}, {Herrero},
  {K{\"o}hler}, {Ma{\'{\i}}z Apell{\'a}niz}, {Najarro}, {Puls}, {Sana},
  {Schneider}, {Taylor}, {van Loon}, \& {Vink}}]{Doran13}
{Doran}, E.~I., {Crowther}, P.~A., {de Koter}, A., {et~al.} 2013, \aap, 558,
  A134

\bibitem[{{Draine}(1978)}]{Draine78}
{Draine}, B.~T. 1978, \apjs, 36, 595

\bibitem[{{Draine} \& {McKee}(1993)}]{Draine93}
{Draine}, B.~T. \& {McKee}, C.~F. 1993, \araa, 31, 373

\bibitem[{{Elmegreen} {et~al.}(2001){Elmegreen}, {Kim}, \&
  {Staveley-Smith}}]{Elmegreen01}
{Elmegreen}, B.~G., {Kim}, S., \& {Staveley-Smith}, L. 2001, \apj, 548, 749

\bibitem[{{Evans} {et~al.}(2011){Evans}, {Taylor}, {H{\'e}nault-Brunet},
  {Sana}, {de Koter}, {Sim{\'o}n-D{\'{\i}}az}, {Carraro}, {Bagnoli}, {Bastian},
  {Bestenlehner}, {Bonanos}, {Bressert}, {Brott}, {Campbell}, {Cantiello},
  {Clark}, {Costa}, {Crowther}, {de Mink}, {Doran}, {Dufton}, {Dunstall},
  {Friedrich}, {Garcia}, {Gieles}, {Gr{\"a}fener}, {Herrero}, {Howarth},
  {Izzard}, {Langer}, {Lennon}, {Ma{\'{\i}}z Apell{\'a}niz}, {Markova},
  {Najarro}, {Puls}, {Ramirez}, {Sab{\'{\i}}n-Sanjuli{\'a}n}, {Smartt},
  {Stroud}, {van Loon}, {Vink}, \& {Walborn}}]{Evans11}
{Evans}, C.~J., {Taylor}, W.~D., {H{\'e}nault-Brunet}, V., {et~al.} 2011, \aap,
  530, A108

\bibitem[{{Falgarone} {et~al.}(2017){Falgarone}, {Zwaan}, {Godard}, {Bergin},
  {Ivison}, {Andreani}, {Bournaud}, {Bussmann}, {Elbaz}, {Omont}, {Oteo}, \&
  {Walter}}]{Falgarone17}
{Falgarone}, E., {Zwaan}, M.~A., {Godard}, B., {et~al.} 2017, \nat, 548, 430

\bibitem[{{Fari{\~n}a} {et~al.}(2009){Fari{\~n}a}, {Bosch}, {Morrell},
  {Barb{\'a}}, \& {Walborn}}]{Farina09}
{Fari{\~n}a}, C., {Bosch}, G.~L., {Morrell}, N.~I., {Barb{\'a}}, R.~H., \&
  {Walborn}, N.~R. 2009, \aj, 138, 510

\bibitem[{{Flower} \& {Pineau des For{\^e}ts}(2015)}]{Flower15}
{Flower}, D.~R. \& {Pineau des For{\^e}ts}, G. 2015, \aap, 578, A63

\bibitem[{{Fulton} {et~al.}(2010){Fulton}, {Baluteau}, {Bendo}, {Benielli},
  {Gastaud}, {Griffin}, {Guest}, {Imhof}, {Lim}, {Lu}, {Naylor}, {Panuzzo},
  {Polehampton}, {Schwartz}, {Surace}, {Swinyard}, \& {Xu}}]{Fulton10}
{Fulton}, T.~R., {Baluteau}, J.-P., {Bendo}, G., {et~al.} 2010, in Society of
  Photo-Optical Instrumentation Engineers (SPIE) Conference Series, Vol. 7731,
  Society of Photo-Optical Instrumentation Engineers (SPIE) Conference Series

\bibitem[{{Galliano}(2018)}]{Galliano18}
{Galliano}, F. 2018, \mnras, 476, 1445

\bibitem[{{Goicoechea} {et~al.}(2017){Goicoechea}, {Cuadrado}, {Pety}, {Bron},
  {Black}, {Cernicharo}, {Chapillon}, {Fuente}, \& {Gerin}}]{Goicoechea17}
{Goicoechea}, J.~R., {Cuadrado}, S., {Pety}, J., {et~al.} 2017, \aap, 601, L9

\bibitem[{{Goicoechea} {et~al.}(2016){Goicoechea}, {Pety}, {Cuadrado},
  {Cernicharo}, {Chapillon}, {Fuente}, {Gerin}, {Joblin}, {Marcelino}, \&
  {Pilleri}}]{Goicoechea16}
{Goicoechea}, J.~R., {Pety}, J., {Cuadrado}, S., {et~al.} 2016, \nat, 537, 207

\bibitem[{{Griffin} {et~al.}(2010){Griffin}, {Abergel}, {Abreu}, {Ade},
  {Andr{\'e}}, {Augueres}, {Babbedge}, {Bae}, {Baillie}, {Baluteau}, {Barlow},
  {Bendo}, {Benielli}, {Bock}, {Bonhomme}, {Brisbin}, {Brockley-Blatt},
  {Caldwell}, {Cara}, {Castro-Rodriguez}, {Cerulli}, {Chanial}, {Chen},
  {Clark}, {Clements}, {Clerc}, {Coker}, {Communal}, {Conversi}, {Cox},
  {Crumb}, {Cunningham}, {Daly}, {Davis}, {de Antoni}, {Delderfield}, {Devin},
  {di Giorgio}, {Didschuns}, {Dohlen}, {Donati}, {Dowell}, {Dowell}, {Duband},
  {Dumaye}, {Emery}, {Ferlet}, {Ferrand}, {Fontignie}, {Fox}, {Franceschini},
  {Frerking}, {Fulton}, {Garcia}, {Gastaud}, {Gear}, {Glenn}, {Goizel},
  {Griffin}, {Grundy}, {Guest}, {Guillemet}, {Hargrave}, {Harwit}, {Hastings},
  {Hatziminaoglou}, {Herman}, {Hinde}, {Hristov}, {Huang}, {Imhof}, {Isaak},
  {Israelsson}, {Ivison}, {Jennings}, {Kiernan}, {King}, {Lange}, {Latter},
  {Laurent}, {Laurent}, {Leeks}, {Lellouch}, {Levenson}, {Li}, {Li},
  {Lilienthal}, {Lim}, {Liu}, {Lu}, {Madden}, {Mainetti}, {Marliani}, {McKay},
  {Mercier}, {Molinari}, {Morris}, {Moseley}, {Mulder}, {Mur}, {Naylor},
  {Nguyen}, {O'Halloran}, {Oliver}, {Olofsson}, {Olofsson}, {Orfei}, {Page},
  {Pain}, {Panuzzo}, {Papageorgiou}, {Parks}, {Parr-Burman}, {Pearce},
  {Pearson}, {P{\'e}rez-Fournon}, {Pinsard}, {Pisano}, {Podosek}, {Pohlen},
  {Polehampton}, {Pouliquen}, {Rigopoulou}, {Rizzo}, {Roseboom}, {Roussel},
  {Rowan-Robinson}, {Rownd}, {Saraceno}, {Sauvage}, {Savage}, {Savini},
  {Sawyer}, {Scharmberg}, {Schmitt}, {Schneider}, {Schulz}, {Schwartz},
  {Shafer}, {Shupe}, {Sibthorpe}, {Sidher}, {Smith}, {Smith}, {Smith},
  {Spencer}, {Stobie}, {Sudiwala}, {Sukhatme}, {Surace}, {Stevens}, {Swinyard},
  {Trichas}, {Tourette}, {Triou}, {Tseng}, {Tucker}, {Turner}, {Vaccari},
  {Valtchanov}, {Vigroux}, {Virique}, {Voellmer}, {Walker}, {Ward}, {Waskett},
  {Weilert}, {Wesson}, {White}, {Whitehouse}, {Wilson}, {Winter}, {Woodcraft},
  {Wright}, {Xu}, {Zavagno}, {Zemcov}, {Zhang}, \& {Zonca}}]{Griffin10}
{Griffin}, M.~J., {Abergel}, A., {Abreu}, A., {et~al.} 2010, \aap, 518, L3

\bibitem[{{Gusdorf} {et~al.}(2012){Gusdorf}, {Anderl}, {G{\"u}sten}, {Stutzki},
  {H{\"u}bers}, {Hartogh}, {Heyminck}, \& {Okada}}]{Gusdorf12}
{Gusdorf}, A., {Anderl}, S., {G{\"u}sten}, R., {et~al.} 2012, \aap, 542, L19

\bibitem[{{Gusdorf} {et~al.}(2015){Gusdorf}, {Riquelme}, {Anderl},
  {Eisl{\"o}ffel}, {Codella}, {G{\'o}mez-Ruiz}, {Graf}, {Kristensen},
  {Leurini}, {Parise}, {Requena-Torres}, {Ricken}, \& {G{\"u}sten}}]{Gusdorf15}
{Gusdorf}, A., {Riquelme}, D., {Anderl}, S., {et~al.} 2015, \aap, 575, A98

\bibitem[{{Habart} {et~al.}(2010){Habart}, {Dartois}, {Abergel}, {Baluteau},
  {Naylor}, {Polehampton}, {Joblin}, {Ade}, {Anderson}, {Andr{\'e}}, {Arab},
  {Bernard}, {Blagrave}, {Bontemps}, {Boulanger}, {Cohen}, {Compiegne}, {Cox},
  {Davis}, {Emery}, {Fulton}, {Gry}, {Huang}, {Jones}, {Kirk}, {Lagache},
  {Lim}, {Madden}, {Makiwa}, {Martin}, {Miville-Desch{\^e}nes}, {Molinari},
  {Moseley}, {Motte}, {Okumura}, {Pinheiro Gon{\c c}alves}, {Rodon}, {Russeil},
  {Saraceno}, {Sidher}, {Spencer}, {Swinyard}, {Ward-Thompson}, {White}, \&
  {Zavagno}}]{Habart10}
{Habart}, E., {Dartois}, E., {Abergel}, A., {et~al.} 2010, \aap, 518, L116

\bibitem[{{Hailey-Dunsheath} {et~al.}(2012){Hailey-Dunsheath}, {Sturm},
  {Fischer}, {Sternberg}, {Graci{\'a}-Carpio}, {Davies},
  {Gonz{\'a}lez-Alfonso}, {Mark}, {Poglitsch}, {Contursi}, {Genzel}, {Lutz},
  {Tacconi}, {Veilleux}, {Verma}, \& {de Jong}}]{Hailey-Dunsheath12}
{Hailey-Dunsheath}, S., {Sturm}, E., {Fischer}, J., {et~al.} 2012, \apj, 755,
  57

\bibitem[{{Hartmann}(1999)}]{Hartmann99}
{Hartmann}, D.~H. 1999, Proceedings of the National Academy of Science, 96,
  4752

\bibitem[{{Higdon} {et~al.}(2004){Higdon}, {Devost}, {Higdon}, {Brandl},
  {Houck}, {Hall}, {Barry}, {Charmandaris}, {Smith}, {Sloan}, \&
  {Green}}]{Higdon04}
{Higdon}, S.~J.~U., {Devost}, D., {Higdon}, J.~L., {et~al.} 2004, \pasp, 116,
  975

\bibitem[{{Hollenbach} {et~al.}(1989){Hollenbach}, {Chernoff}, \&
  {McKee}}]{Hollenbach89}
{Hollenbach}, D.~J., {Chernoff}, D.~F., \& {McKee}, C.~F. 1989, in ESA Special
  Publication, Vol. 290, Infrared Spectroscopy in Astronomy, ed.
  E.~{B{\"o}hm-Vitense}

\bibitem[{{Indebetouw} {et~al.}(2013){Indebetouw}, {Brogan}, {Chen}, {Leroy},
  {Johnson}, {Muller}, {Madden}, {Cormier}, {Galliano}, {Hughes}, {Hunter},
  {Kawamura}, {Kepley}, {Lebouteiller}, {Meixner}, {Oliveira}, {Onishi}, \&
  {Vasyunina}}]{Indebetouw13}
{Indebetouw}, R., {Brogan}, C., {Chen}, C.-H.~R., {et~al.} 2013, \apj, 774, 73

\bibitem[{{Indebetouw} {et~al.}(2009){Indebetouw}, {de Messi{\`e}res},
  {Madden}, {Engelbracht}, {Smith}, {Meixner}, {Brandl}, {Smith}, {Boulanger},
  {Galliano}, {Gordon}, {Hora}, {Sewilo}, {Tielens}, {Werner}, \&
  {Wolfire}}]{Indebetouw09}
{Indebetouw}, R., {de Messi{\`e}res}, G.~E., {Madden}, S., {et~al.} 2009, \apj,
  694, 84

\bibitem[{{Indriolo} {et~al.}(2017){Indriolo}, {Bergin}, {Goicoechea},
  {Cernicharo}, {Gerin}, {Gusdorf}, {Lis}, \& {Schilke}}]{Indriolo17}
{Indriolo}, N., {Bergin}, E.~A., {Goicoechea}, J.~R., {et~al.} 2017, \apj, 836,
  117

\bibitem[{{Indriolo} \& {McCall}(2012)}]{Indriolo12a}
{Indriolo}, N. \& {McCall}, B.~J. 2012, \apj, 745, 91

\bibitem[{{Indriolo} {et~al.}(2015){Indriolo}, {Neufeld}, {Gerin}, {Schilke},
  {Benz}, {Winkel}, {Menten}, {Chambers}, {Black}, {Bruderer}, {Falgarone},
  {Godard}, {Goicoechea}, {Gupta}, {Lis}, {Ossenkopf}, {Persson},
  {Sonnentrucker}, {van der Tak}, {van Dishoeck}, {Wolfire}, \&
  {Wyrowski}}]{Indriolo15}
{Indriolo}, N., {Neufeld}, D.~A., {Gerin}, M., {et~al.} 2015, \apj, 800, 40

\bibitem[{{Israel} {et~al.}(2014){Israel}, {G{\"u}sten}, {Meijerink}, {Loenen},
  {Requena-Torres}, {Stutzki}, {van der Werf}, {Harris}, {Kramer},
  {Martin-Pintado}, \& {Weiss}}]{Israel14}
{Israel}, F.~P., {G{\"u}sten}, R., {Meijerink}, R., {et~al.} 2014, \aap, 562,
  A96

\bibitem[{{Joblin} {et~al.}(2018){Joblin}, {Bron}, {Pinto}, {Pilleri}, {Le
  Petit}, {Gerin}, {Le Bourlot}, {Fuente}, {Berne}, {Goicoechea}, {Habart},
  {K{\"o}hler}, {Teyssier}, {Nagy}, {Montillaud}, {Vastel}, {Cernicharo},
  {R{\"o}llig}, {Ossenkopf-Okada}, \& {Bergin}}]{Joblin18}
{Joblin}, C., {Bron}, E., {Pinto}, C., {et~al.} 2018, \aap, 615, A129

\bibitem[{{Johansson} {et~al.}(1998){Johansson}, {Greve}, {Booth}, {Boulanger},
  {Garay}, {de Graauw}, {Israel}, {Kutner}, {Lequeux}, {Murphy}, {Nyman}, \&
  {Rubio}}]{Johansson98}
{Johansson}, L.~E.~B., {Greve}, A., {Booth}, R.~S., {et~al.} 1998, \aap, 331,
  857

\bibitem[{{Kamenetzky} {et~al.}(2012){Kamenetzky}, {Glenn}, {Rangwala},
  {Maloney}, {Bradford}, {Wilson}, {Bendo}, {Baes}, {Boselli}, {Cooray},
  {Isaak}, {Lebouteiller}, {Madden}, {Panuzzo}, {Schirm}, {Spinoglio}, \&
  {Wu}}]{Kamenetzky12}
{Kamenetzky}, J., {Glenn}, J., {Rangwala}, N., {et~al.} 2012, \apj, 753, 70

\bibitem[{{Kamenetzky} {et~al.}(2017){Kamenetzky}, {Rangwala}, \&
  {Glenn}}]{Kamenetzky17}
{Kamenetzky}, J., {Rangwala}, N., \& {Glenn}, J. 2017, \mnras, 471, 2917

\bibitem[{{Kennicutt} \& {Evans}(2012)}]{Kennicutt12}
{Kennicutt}, R.~C. \& {Evans}, N.~J. 2012, \araa, 50, 531

\bibitem[{{Kennicutt} \& {Hodge}(1986)}]{Kennicutt86}
{Kennicutt}, Jr., R.~C. \& {Hodge}, P.~W. 1986, \apj, 306, 130

\bibitem[{{Kim} {et~al.}(1998){Kim}, {Staveley-Smith}, {Dopita}, {Freeman},
  {Sault}, {Kesteven}, \& {McConnell}}]{Kim98}
{Kim}, S., {Staveley-Smith}, L., {Dopita}, M.~A., {et~al.} 1998, \apj, 503, 674

\bibitem[{{K{\"o}hler} {et~al.}(2014){K{\"o}hler}, {Habart}, {Arab},
  {Bernard-Salas}, {Ayasso}, {Abergel}, {Zavagno}, {Polehampton}, {van der
  Wiel}, {Naylor}, {Makiwa}, {Dassas}, {Joblin}, {Pilleri}, {Bern{\'e}},
  {Fuente}, {Gerin}, {Goicoechea}, \& {Teyssier}}]{Kohler14}
{K{\"o}hler}, M., {Habart}, E., {Arab}, H., {et~al.} 2014, \aap, 569, A109

\bibitem[{{Larson} {et~al.}(2015){Larson}, {Evans}, {Green}, \&
  {Yang}}]{Larson15}
{Larson}, R.~L., {Evans}, II, N.~J., {Green}, J.~D., \& {Yang}, Y.-L. 2015,
  \apj, 806, 70

\bibitem[{{Le Bourlot} {et~al.}(2012){Le Bourlot}, {Le Petit}, {Pinto},
  {Roueff}, \& {Roy}}]{LeBourlot12}
{Le Bourlot}, J., {Le Petit}, F., {Pinto}, C., {Roueff}, E., \& {Roy}, F. 2012,
  \aap, 541, A76

\bibitem[{{Le Bourlot} {et~al.}(1999){Le Bourlot}, {Pineau des For{\^e}ts}, \&
  {Flower}}]{LeBourlot99}
{Le Bourlot}, J., {Pineau des For{\^e}ts}, G., \& {Flower}, D.~R. 1999, \mnras,
  305, 802

\bibitem[{{Le Petit} {et~al.}(2006){Le Petit}, {Nehm{\'e}}, {Le Bourlot}, \&
  {Roueff}}]{LePetit06}
{Le Petit}, F., {Nehm{\'e}}, C., {Le Bourlot}, J., \& {Roueff}, E. 2006, \apjs,
  164, 506

\bibitem[{{Lebouteiller} {et~al.}(2010){Lebouteiller}, {Bernard-Salas},
  {Sloan}, \& {Barry}}]{Lebouteiller10}
{Lebouteiller}, V., {Bernard-Salas}, J., {Sloan}, G.~C., \& {Barry}, D.~J.
  2010, \pasp, 122, 231

\bibitem[{{Lebouteiller} {et~al.}(2012){Lebouteiller}, {Cormier}, {Madden},
  {Galliano}, {Indebetouw}, {Abel}, {Sauvage}, {Hony}, {Contursi}, {Poglitsch},
  {R{\'e}my}, {Sturm}, \& {Wu}}]{Lebouteiller12}
{Lebouteiller}, V., {Cormier}, D., {Madden}, S.~C., {et~al.} 2012, \aap, 548,
  A91

\bibitem[{{Lee} {et~al.}(2016){Lee}, {Madden}, {Lebouteiller}, {Gusdorf},
  {Godard}, {Wu}, {Galametz}, {Cormier}, {Le Petit}, {Roueff}, {Bron},
  {Carlson}, {Chevance}, {Fukui}, {Galliano}, {Hony}, {Hughes}, {Indebetouw},
  {Israel}, {Kawamura}, {Le Bourlot}, {Lesaffre}, {Meixner}, {Muller}, {Nayak},
  {Onishi}, {Roman-Duval}, \& {Sewi{\l}o}}]{Lee16}
{Lee}, M.-Y., {Madden}, S.~C., {Lebouteiller}, V., {et~al.} 2016, \aap, 596,
  A85

\bibitem[{{Lesaffre} {et~al.}(2013){Lesaffre}, {Pineau des For{\^e}ts},
  {Godard}, {Guillard}, {Boulanger}, \& {Falgarone}}]{Lesaffre13}
{Lesaffre}, P., {Pineau des For{\^e}ts}, G., {Godard}, B., {et~al.} 2013, \aap,
  550, A106

\bibitem[{{Luks} \& {Rohlfs}(1992)}]{Luks92}
{Luks}, T. \& {Rohlfs}, K. 1992, \aap, 263, 41

\bibitem[{{Maggi} {et~al.}(2016){Maggi}, {Haberl}, {Kavanagh}, {Sasaki},
  {Bozzetto}, {Filipovi{\'c}}, {Vasilopoulos}, {Pietsch}, {Points}, {Chu},
  {Dickel}, {Ehle}, {Williams}, \& {Greiner}}]{Maggi16}
{Maggi}, P., {Haberl}, F., {Kavanagh}, P.~J., {et~al.} 2016, \aap, 585, A162

\bibitem[{{Makiwa} {et~al.}(2013){Makiwa}, {Naylor}, {Ferlet}, {Salji},
  {Swinyard}, {Polehampton}, \& {van der Wiel}}]{Makiwa13}
{Makiwa}, G., {Naylor}, D.~A., {Ferlet}, M., {et~al.} 2013, \ao, 52, 3864

\bibitem[{{Mashian} {et~al.}(2015){Mashian}, {Sturm}, {Sternberg}, {Janssen},
  {Hailey-Dunsheath}, {Fischer}, {Contursi}, {Gonz{\'a}lez-Alfonso},
  {Graci{\'a}-Carpio}, {Poglitsch}, {Veilleux}, {Davies}, {Genzel}, {Lutz},
  {Tacconi}, {Verma}, {Wei{\ss}}, {Polisensky}, \& {Nikola}}]{Mashian15}
{Mashian}, N., {Sturm}, E., {Sternberg}, A., {et~al.} 2015, \apj, 802, 81

\bibitem[{{Mathis} {et~al.}(1983){Mathis}, {Mezger}, \& {Panagia}}]{Mathis83}
{Mathis}, J.~S., {Mezger}, P.~G., \& {Panagia}, N. 1983, \aap, 128, 212

\bibitem[{{Meijerink} {et~al.}(2013){Meijerink}, {Kristensen}, {Wei{\ss}}, {van
  der Werf}, {Walter}, {Spaans}, {Loenen}, {Fischer}, {Israel}, {Isaak},
  {Papadopoulos}, {Aalto}, {Armus}, {Charmandaris}, {Dasyra}, {Diaz-Santos},
  {Evans}, {Gao}, {Gonz{\'a}lez-Alfonso}, {G{\"u}sten}, {Henkel}, {Kramer},
  {Lord}, {Mart{\'{\i}}n-Pintado}, {Naylor}, {Sanders}, {Smith}, {Spinoglio},
  {Stacey}, {Veilleux}, \& {Wiedner}}]{Meijerink13}
{Meijerink}, R., {Kristensen}, L.~E., {Wei{\ss}}, A., {et~al.} 2013, \apjl,
  762, L16

\bibitem[{{Meixner} {et~al.}(2006){Meixner}, {Gordon}, {Indebetouw}, {Hora},
  {Whitney}, {Blum}, {Reach}, {Bernard}, {Meade}, {Babler}, {Engelbracht},
  {For}, {Misselt}, {Vijh}, {Leitherer}, {Cohen}, {Churchwell}, {Boulanger},
  {Frogel}, {Fukui}, {Gallagher}, {Gorjian}, {Harris}, {Kelly}, {Kawamura},
  {Kim}, {Latter}, {Madden}, {Markwick-Kemper}, {Mizuno}, {Mizuno}, {Mould},
  {Nota}, {Oey}, {Olsen}, {Onishi}, {Paladini}, {Panagia}, {Perez-Gonzalez},
  {Shibai}, {Sato}, {Smith}, {Staveley-Smith}, {Tielens}, {Ueta}, {van Dyk},
  {Volk}, {Werner}, \& {Zaritsky}}]{Meixner06}
{Meixner}, M., {Gordon}, K.~D., {Indebetouw}, R., {et~al.} 2006, \aj, 132, 2268

\bibitem[{{Minamidani} {et~al.}(2008){Minamidani}, {Mizuno}, {Mizuno},
  {Kawamura}, {Onishi}, {Hasegawa}, {Tatematsu}, {Ikeda}, {Moriguchi},
  {Yamaguchi}, {Ott}, {Wong}, {Muller}, {Pineda}, {Hughes}, {Staveley-Smith},
  {Klein}, {Mizuno}, {Nikoli{\'c}}, {Booth}, {Heikkil{\"a}}, {Nyman}, {Lerner},
  {Garay}, {Kim}, {Fujishita}, {Kawase}, {Rubio}, \& {Fukui}}]{Minamidani08}
{Minamidani}, T., {Mizuno}, N., {Mizuno}, Y., {et~al.} 2008, \apjs, 175, 485

\bibitem[{{Nestingen-Palm} {et~al.}(2017){Nestingen-Palm}, {Stanimirovi{\'c}},
  {Gonz{\'a}lez-Casanova}, {Babler}, {Jameson}, \&
  {Bolatto}}]{Nestingen-Palm17}
{Nestingen-Palm}, D., {Stanimirovi{\'c}}, S., {Gonz{\'a}lez-Casanova}, D.~F.,
  {et~al.} 2017, \apj, 845, 53

\bibitem[{{Newville} {et~al.}(2014){Newville}, {Stensitzki}, {Allen}, \&
  {Ingargiola}}]{Newville14}
{Newville}, M., {Stensitzki}, T., {Allen}, D.~B., \& {Ingargiola}, A. 2014,
  {LMFIT: Non-Linear Least-Square Minimization and Curve-Fitting for Python}

\bibitem[{{Okada} {et~al.}(2019){Okada}, {G{\"u}sten}, {Requena-Torres},
  {R{\"o}llig}, {Stutzki}, {Graf}, \& {Hughes}}]{Okada19}
{Okada}, Y., {G{\"u}sten}, R., {Requena-Torres}, M.~A., {et~al.} 2019, \aap,
  621, A62

\bibitem[{{Ott}(2010)}]{Ott10}
{Ott}, S. 2010, in Astronomical Society of the Pacific Conference Series, Vol.
  434, Astronomical Data Analysis Software and Systems XIX, ed. Y.~{Mizumoto},
  K.-I. {Morita}, \& M.~{Ohishi}, 139

\bibitem[{{Papadopoulos} {et~al.}(2014){Papadopoulos}, {Zhang}, {Xilouris},
  {Weiss}, {van der Werf}, {Israel}, {Greve}, {Isaak}, \&
  {Gao}}]{Papadopoulos14}
{Papadopoulos}, P.~P., {Zhang}, Z.-Y., {Xilouris}, E.~M., {et~al.} 2014, \apj,
  788, 153

\bibitem[{{Pellegrini} {et~al.}(2010){Pellegrini}, {Baldwin}, \&
  {Ferland}}]{Pellegrini10}
{Pellegrini}, E.~W., {Baldwin}, J.~A., \& {Ferland}, G.~J. 2010, \apjs, 191,
  160

\bibitem[{{Pellegrini} {et~al.}(2013){Pellegrini}, {Smith}, {Wolfire},
  {Draine}, {Crocker}, {Croxall}, {van der Werf}, {Dale}, {Rigopoulou},
  {Wilson}, {Schinnerer}, {Groves}, {Kreckel}, {Sandstrom}, {Armus},
  {Calzetti}, {Murphy}, {Walter}, {Koda}, {Bayet}, {Beirao}, {Bolatto},
  {Bradford}, {Brinks}, {Hunt}, {Kennicutt}, {Knapen}, {Leroy}, {Rosolowsky},
  {Vigroux}, \& {Hopwood}}]{Pellegrini13}
{Pellegrini}, E.~W., {Smith}, J.~D., {Wolfire}, M.~G., {et~al.} 2013, \apjl,
  779, L19

\bibitem[{{P{\'e}rez-Beaupuits} {et~al.}(2010){P{\'e}rez-Beaupuits}, {Spaans},
  {Hogerheijde}, {G{\"u}sten}, {Baryshev}, \& {Boland}}]{JP10}
{P{\'e}rez-Beaupuits}, J.~P., {Spaans}, M., {Hogerheijde}, M.~R., {et~al.}
  2010, \aap, 510, A87

\bibitem[{{Pilbratt} {et~al.}(2010){Pilbratt}, {Riedinger}, {Passvogel},
  {Crone}, {Doyle}, {Gageur}, {Heras}, {Jewell}, {Metcalfe}, {Ott}, \&
  {Schmidt}}]{Pilbratt10}
{Pilbratt}, G.~L., {Riedinger}, J.~R., {Passvogel}, T., {et~al.} 2010, \aap,
  518, L1

\bibitem[{{Pineda} {et~al.}(2012){Pineda}, {Mizuno}, {R{\"o}llig}, {Stutzki},
  {Kramer}, {Klein}, {Rubio}, {Kawamura}, {Minamidani}, {Benz}, {Burton},
  {Fukui}, {Koo}, \& {Onishi}}]{Pineda12}
{Pineda}, J.~L., {Mizuno}, N., {R{\"o}llig}, M., {et~al.} 2012, \aap, 544, A84

\bibitem[{{Poglitsch} {et~al.}(2010){Poglitsch}, {Waelkens}, {Geis},
  {Feuchtgruber}, {Vandenbussche}, {Rodriguez}, {Krause}, {Renotte}, {van
  Hoof}, {Saraceno}, {Cepa}, {Kerschbaum}, {Agn{\`e}se}, {Ali}, {Altieri},
  {Andreani}, {Augueres}, {Balog}, {Barl}, {Bauer}, {Belbachir}, {Benedettini},
  {Billot}, {Boulade}, {Bischof}, {Blommaert}, {Callut}, {Cara}, {Cerulli},
  {Cesarsky}, {Contursi}, {Creten}, {De Meester}, {Doublier}, {Doumayrou},
  {Duband}, {Exter}, {Genzel}, {Gillis}, {Gr{\"o}zinger}, {Henning},
  {Herreros}, {Huygen}, {Inguscio}, {Jakob}, {Jamar}, {Jean}, {de Jong},
  {Katterloher}, {Kiss}, {Klaas}, {Lemke}, {Lutz}, {Madden}, {Marquet},
  {Martignac}, {Mazy}, {Merken}, {Montfort}, {Morbidelli}, {M{\"u}ller},
  {Nielbock}, {Okumura}, {Orfei}, {Ottensamer}, {Pezzuto}, {Popesso},
  {Putzeys}, {Regibo}, {Reveret}, {Royer}, {Sauvage}, {Schreiber}, {Stegmaier},
  {Schmitt}, {Schubert}, {Sturm}, {Thiel}, {Tofani}, {Vavrek}, {Wetzstein},
  {Wieprecht}, \& {Wiezorrek}}]{Poglitsch10}
{Poglitsch}, A., {Waelkens}, C., {Geis}, N., {et~al.} 2010, \aap, 518, L2

\bibitem[{{Pon} {et~al.}(2016){Pon}, {Kaufman}, {Johnstone}, {Caselli},
  {Fontani}, {Butler}, {Jim{\'e}nez-Serra}, {Palau}, \& {Tan}}]{Pon16}
{Pon}, A., {Kaufman}, M.~J., {Johnstone}, D., {et~al.} 2016, \apj, 827, 107

\bibitem[{{Rangwala} {et~al.}(2011){Rangwala}, {Maloney}, {Glenn}, {Wilson},
  {Rykala}, {Isaak}, {Baes}, {Bendo}, {Boselli}, {Bradford}, {Clements},
  {Cooray}, {Fulton}, {Imhof}, {Kamenetzky}, {Madden}, {Mentuch}, {Sacchi},
  {Sauvage}, {Schirm}, {Smith}, {Spinoglio}, \& {Wolfire}}]{Rangwala11}
{Rangwala}, N., {Maloney}, P.~R., {Glenn}, J., {et~al.} 2011, \apj, 743, 94

\bibitem[{{R{\"o}llig} {et~al.}(2006){R{\"o}llig}, {Ossenkopf}, {Jeyakumar},
  {Stutzki}, \& {Sternberg}}]{Rollig06}
{R{\"o}llig}, M., {Ossenkopf}, V., {Jeyakumar}, S., {Stutzki}, J., \&
  {Sternberg}, A. 2006, \aap, 451, 917

\bibitem[{{Rosenberg} {et~al.}(2014){Rosenberg}, {Kazandjian}, {van der Werf},
  {Israel}, {Meijerink}, {Wei{\ss}}, {Requena-Torres}, \&
  {G{\"u}sten}}]{Rosenberg14a}
{Rosenberg}, M.~J.~F., {Kazandjian}, M.~V., {van der Werf}, P.~P., {et~al.}
  2014, \aap, 564, A126

\bibitem[{{Rosenberg} {et~al.}(2015){Rosenberg}, {van der Werf}, {Aalto},
  {Armus}, {Charmandaris}, {D{\'{\i}}az-Santos}, {Evans}, {Fischer}, {Gao},
  {Gonz{\'a}lez-Alfonso}, {Greve}, {Harris}, {Henkel}, {Israel}, {Isaak},
  {Kramer}, {Meijerink}, {Naylor}, {Sanders}, {Smith}, {Spaans}, {Spinoglio},
  {Stacey}, {Veenendaal}, {Veilleux}, {Walter}, {Wei{\ss}}, {Wiedner}, {van der
  Wiel}, \& {Xilouris}}]{Rosenberg15}
{Rosenberg}, M.~J.~F., {van der Werf}, P.~P., {Aalto}, S., {et~al.} 2015, \apj,
  801, 72

\bibitem[{{Russell} \& {Dopita}(1992)}]{Russell92}
{Russell}, S.~C. \& {Dopita}, M.~A. 1992, \apj, 384, 508

\bibitem[{{Schaefer}(2008)}]{Schaefer08}
{Schaefer}, B.~E. 2008, \aj, 135, 112

\bibitem[{{Schirm} {et~al.}(2014){Schirm}, {Wilson}, {Parkin}, {Kamenetzky},
  {Glenn}, {Rangwala}, {Spinoglio}, {Pereira-Santaella}, {Baes}, {Barlow},
  {Clements}, {Cooray}, {De Looze}, {Karczewski}, {Madden}, {R{\'e}my-Ruyer},
  \& {Wu}}]{Schirm14}
{Schirm}, M.~R.~P., {Wilson}, C.~D., {Parkin}, T.~J., {et~al.} 2014, \apj, 781,
  101

\bibitem[{{Selman} {et~al.}(1999){Selman}, {Melnick}, {Bosch}, \&
  {Terlevich}}]{Selman99}
{Selman}, F., {Melnick}, J., {Bosch}, G., \& {Terlevich}, R. 1999, \aap, 341,
  98

\bibitem[{{Selman} \& {Melnick}(2013)}]{Selman13}
{Selman}, F.~J. \& {Melnick}, J. 2013, \aap, 552, A94

\bibitem[{{Smith} {et~al.}(2007){Smith}, {Armus}, {Dale}, {Roussel}, {Sheth},
  {Buckalew}, {Jarrett}, {Helou}, \& {Kennicutt}}]{Smith07}
{Smith}, J.~D.~T., {Armus}, L., {Dale}, D.~A., {et~al.} 2007, \pasp, 119, 1133

\bibitem[{{Sternberg} \& {Dalgarno}(1989)}]{Sternberg89}
{Sternberg}, A. \& {Dalgarno}, A. 1989, \apj, 338, 197

\bibitem[{{Stock} {et~al.}(2015){Stock}, {Wolfire}, {Peeters}, {Tielens},
  {Vandenbussche}, {Boersma}, \& {Cami}}]{Stock15}
{Stock}, D.~J., {Wolfire}, M.~G., {Peeters}, E., {et~al.} 2015, \aap, 579, A67

\bibitem[{{Stoerzer} {et~al.}(1996){Stoerzer}, {Stutzki}, \&
  {Sternberg}}]{Stoerzer96}
{Stoerzer}, H., {Stutzki}, J., \& {Sternberg}, A. 1996, \aap, 310, 592

\bibitem[{{Swinyard} {et~al.}(2014){Swinyard}, {Polehampton}, {Hopwood},
  {Valtchanov}, {Lu}, {Fulton}, {Benielli}, {Imhof}, {Marchili}, {Baluteau},
  {Bendo}, {Ferlet}, {Griffin}, {Lim}, {Makiwa}, {Naylor}, {Orton},
  {Papageorgiou}, {Pearson}, {Schulz}, {Sidher}, {Spencer}, {Wiel}, \&
  {Wu}}]{Swinyard14}
{Swinyard}, B.~M., {Polehampton}, E.~T., {Hopwood}, R., {et~al.} 2014, \mnras,
  440, 3658

\bibitem[{{Tielens} \& {Hollenbach}(1985)}]{Tielens85a}
{Tielens}, A.~G.~G.~M. \& {Hollenbach}, D. 1985, \apj, 291, 722

\bibitem[{{Townsley} {et~al.}(2006{\natexlab{a}}){Townsley}, {Broos},
  {Feigelson}, {Brandl}, {Chu}, {Garmire}, \& {Pavlov}}]{Townsley06a}
{Townsley}, L.~K., {Broos}, P.~S., {Feigelson}, E.~D., {et~al.}
  2006{\natexlab{a}}, \aj, 131, 2140

\bibitem[{{Townsley} {et~al.}(2006{\natexlab{b}}){Townsley}, {Broos},
  {Feigelson}, {Garmire}, \& {Getman}}]{Townsley06b}
{Townsley}, L.~K., {Broos}, P.~S., {Feigelson}, E.~D., {Garmire}, G.~P., \&
  {Getman}, K.~V. 2006{\natexlab{b}}, \aj, 131, 2164

\bibitem[{{van der Tak} {et~al.}(2007){van der Tak}, {Black}, {Sch{\"o}ier},
  {Jansen}, \& {van Dishoeck}}]{vanderTak07}
{van der Tak}, F.~F.~S., {Black}, J.~H., {Sch{\"o}ier}, F.~L., {Jansen}, D.~J.,
  \& {van Dishoeck}, E.~F. 2007, \aap, 468, 627

\bibitem[{{van der Werf} {et~al.}(2010){van der Werf}, {Isaak}, {Meijerink},
  {Spaans}, {Rykala}, {Fulton}, {Loenen}, {Walter}, {Wei{\ss}}, {Armus},
  {Fischer}, {Israel}, {Harris}, {Veilleux}, {Henkel}, {Savini}, {Lord},
  {Smith}, {Gonz{\'a}lez-Alfonso}, {Naylor}, {Aalto}, {Charmandaris}, {Dasyra},
  {Evans}, {Gao}, {Greve}, {G{\"u}sten}, {Kramer}, {Mart{\'{\i}}n-Pintado},
  {Mazzarella}, {Papadopoulos}, {Sanders}, {Spinoglio}, {Stacey}, {Vlahakis},
  {Wiedner}, \& {Xilouris}}]{vanderWerf10}
{van der Werf}, P.~P., {Isaak}, K.~G., {Meijerink}, R., {et~al.} 2010, \aap,
  518, L42

\bibitem[{{Weingartner} \& {Draine}(2001)}]{Weingartner01}
{Weingartner}, J.~C. \& {Draine}, B.~T. 2001, \apjs, 134, 263

\bibitem[{{Weingartner} {et~al.}(2006){Weingartner}, {Draine}, \&
  {Barr}}]{Weingartner06}
{Weingartner}, J.~C., {Draine}, B.~T., \& {Barr}, D.~K. 2006, \apj, 645, 1188

\bibitem[{{Wong} {et~al.}(2011){Wong}, {Hughes}, {Ott}, {Muller}, {Pineda},
  {Bernard}, {Chu}, {Fukui}, {Gruendl}, {Henkel}, {Kawamura}, {Klein},
  {Looney}, {Maddison}, {Mizuno}, {Paradis}, {Seale}, \& {Welty}}]{Wong11}
{Wong}, T., {Hughes}, A., {Ott}, J., {et~al.} 2011, \apjs, 197, 16

\bibitem[{{Wu} {et~al.}(2018){Wu}, {Bron}, {Onaka}, {Le Petit}, {Galliano},
  {Languignon}, {Nakamura}, \& {Okada}}]{RWu18}
{Wu}, R., {Bron}, E., {Onaka}, T., {et~al.} 2018, ArXiv e-prints
  [\eprint[arXiv]{1801.01643}]

\bibitem[{{Wu} {et~al.}(2015){Wu}, {Madden}, {Galliano}, {Wilson},
  {Kamenetzky}, {Lee}, {Schirm}, {Hony}, {Lebouteiller}, {Spinoglio},
  {Cormier}, {Glenn}, {Maloney}, {Pereira-Santaella}, {R{\'e}my-Ruyer}, {Baes},
  {Boselli}, {Bournaud}, {De Looze}, {Hughes}, {Panuzzo}, \&
  {Rangwala}}]{RWu15}
{Wu}, R., {Madden}, S.~C., {Galliano}, F., {et~al.} 2015, \aap, 575, A88

\end{thebibliography}

\begin{appendix} 

\section{FTS data} 
\label{s:appendix1}

In Fig. \ref{f:appendix1}, we present the FTS CO, \CI, and \NII integrated intensity images of 30 Doradus. 
All images are at 42$''$ resolution ($\sim$10 pc at the LMC distance) with a pixel size of 30$''$. 
In each image, the spectra of individual pixels are overlaid in red (detections with $S/N_{\rm s} > 5$) and blue (non-detections with $S/N_{\rm s} \leq 5$). 
These spectra are plotted with the same $x$- and $y$-axis ranges (in GHz and 10$^{-18}$ W m$^{-2}$ Hz$^{-1}$ sr$^{-1}$), 
which can be found in the \textit{top left} corner of the image with an example spectrum. 
The example spectrum is from the pixel that was observed with the central SLWC3 and SSWD4 detectors 
of the first jiggle position of the Obs. ID = 1342219550 (yellow and orange crosses). 

\begin{figure*}
\centering 
\includegraphics[scale=0.22]{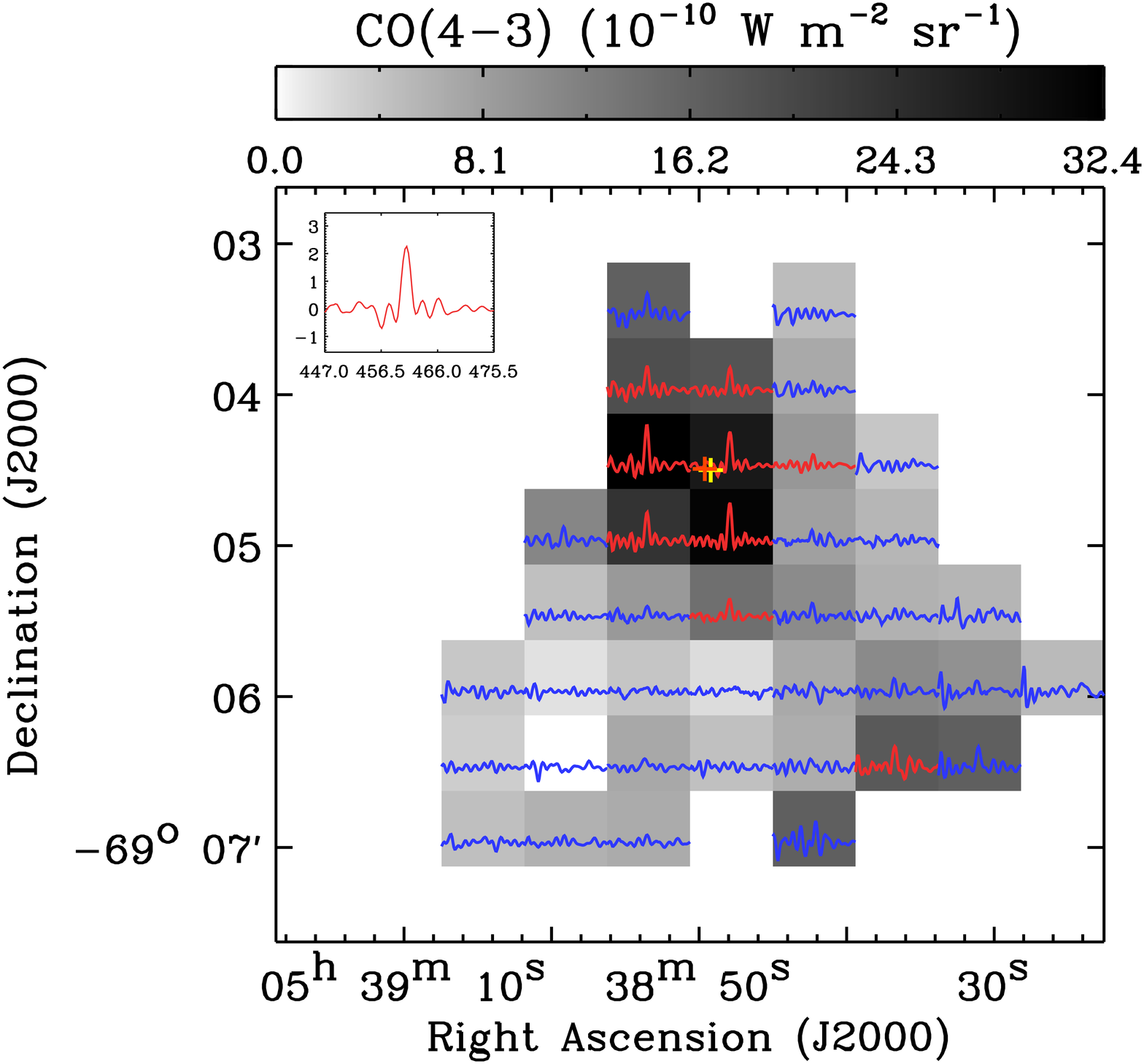} 
\includegraphics[scale=0.22]{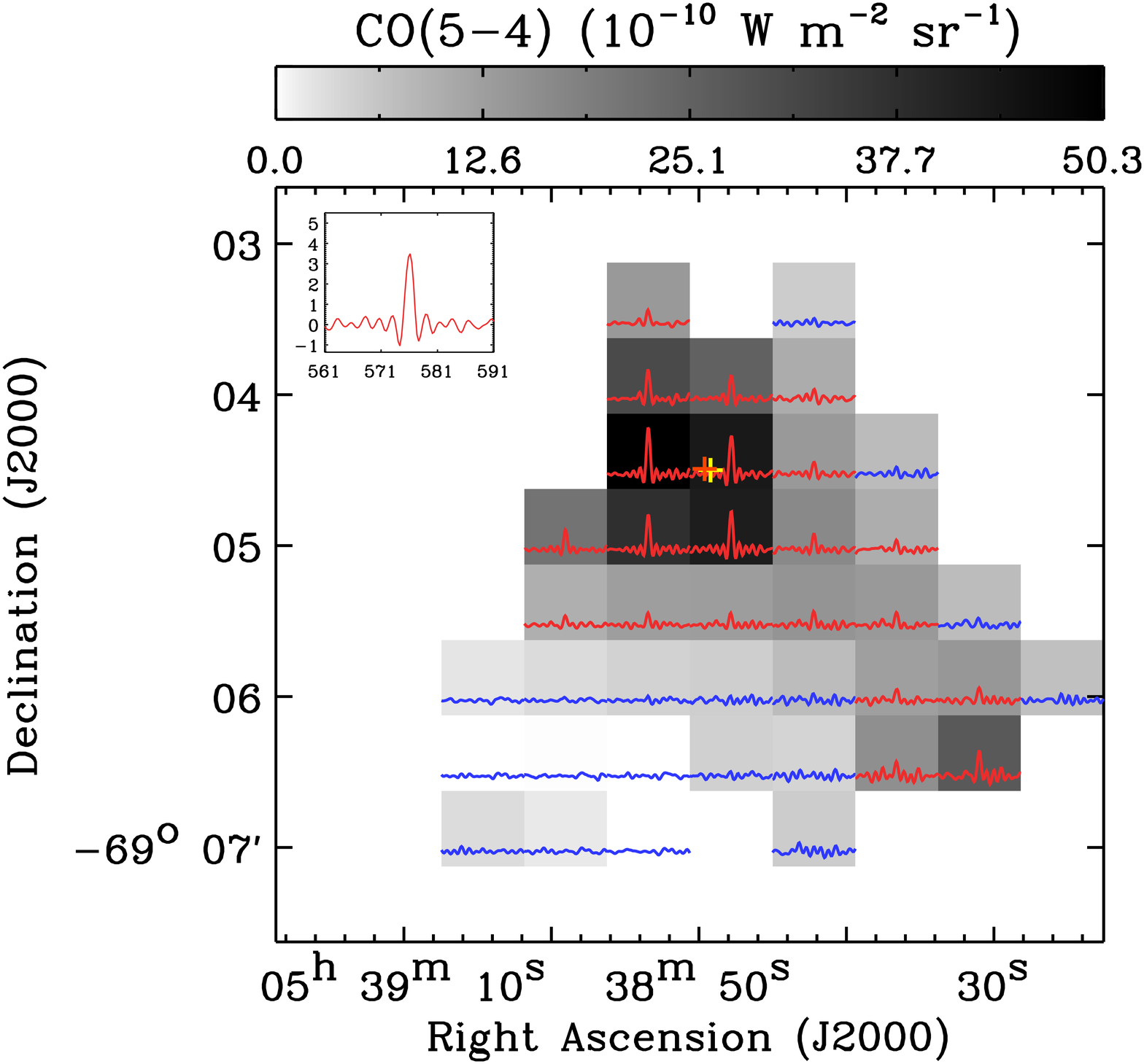}
\includegraphics[scale=0.22]{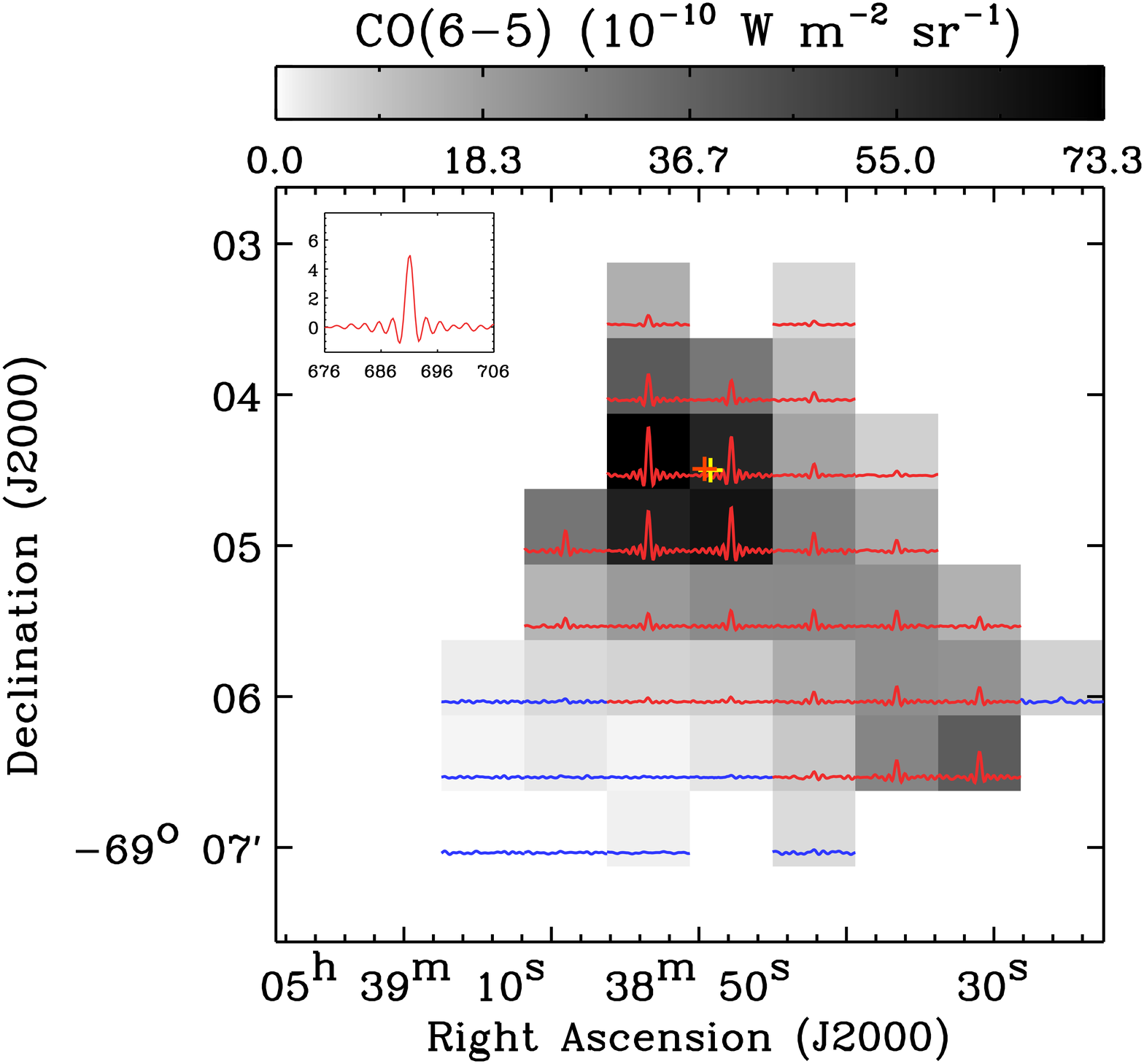}
\includegraphics[scale=0.22]{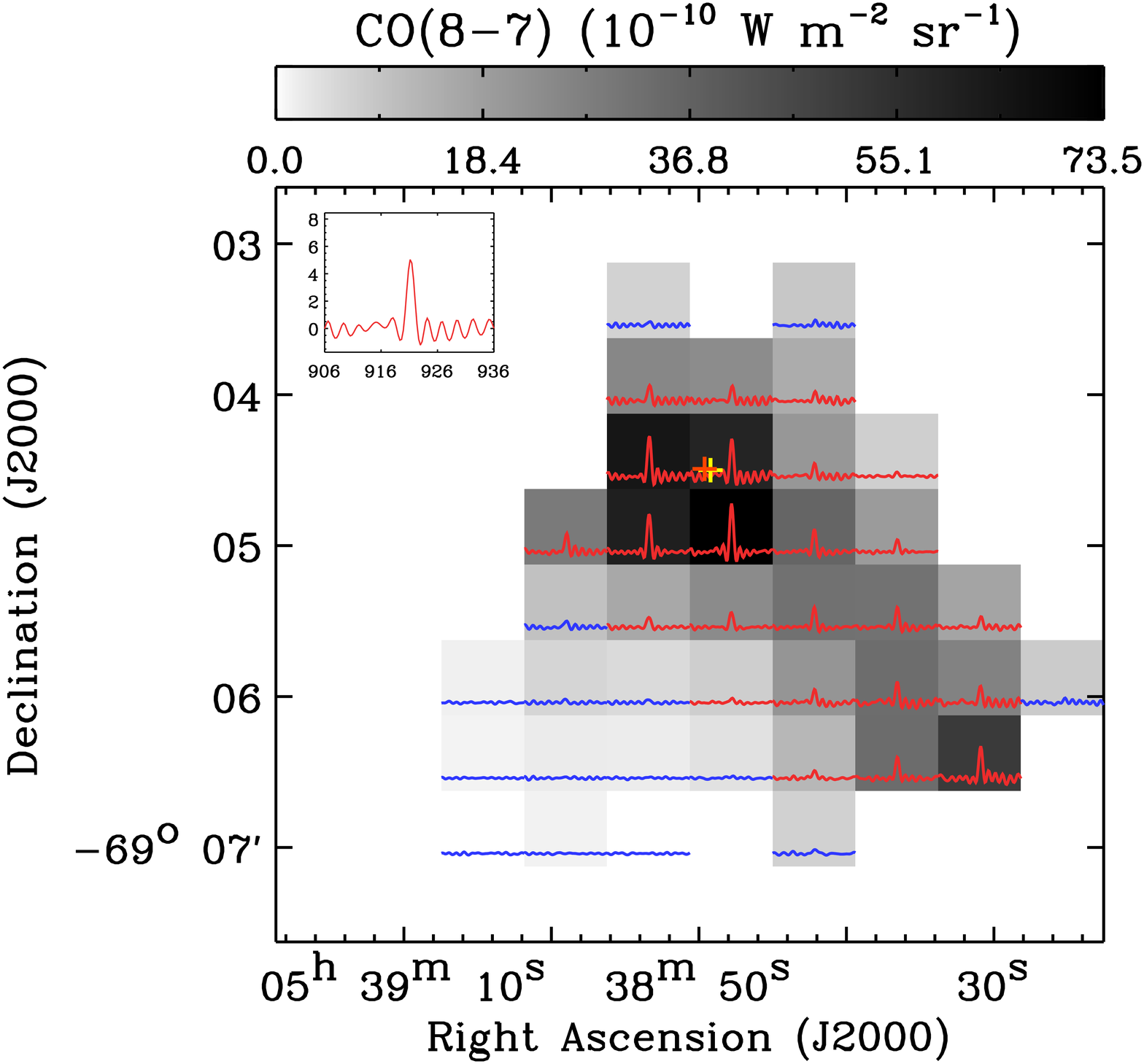}
\includegraphics[scale=0.22]{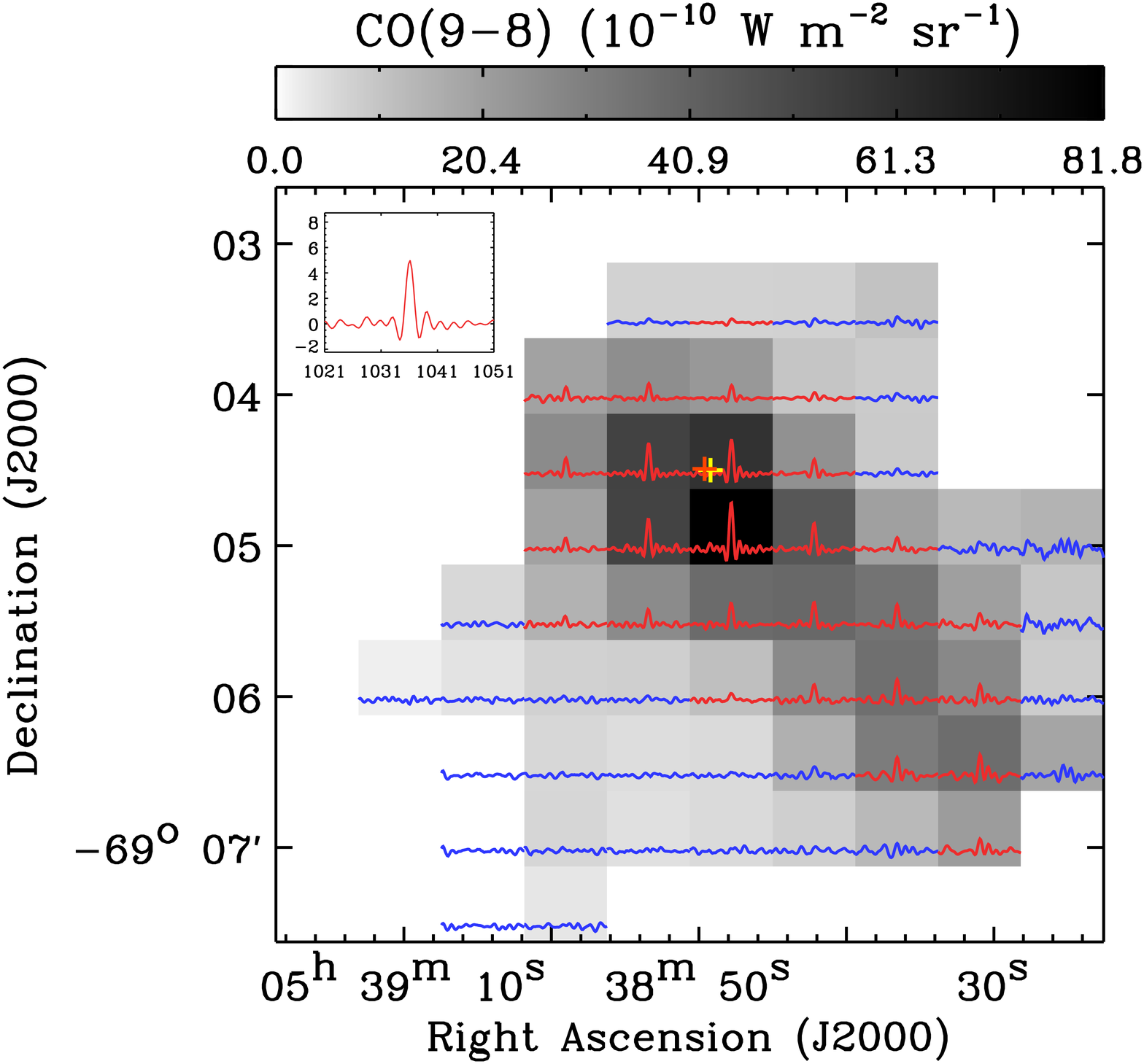}
\includegraphics[scale=0.22]{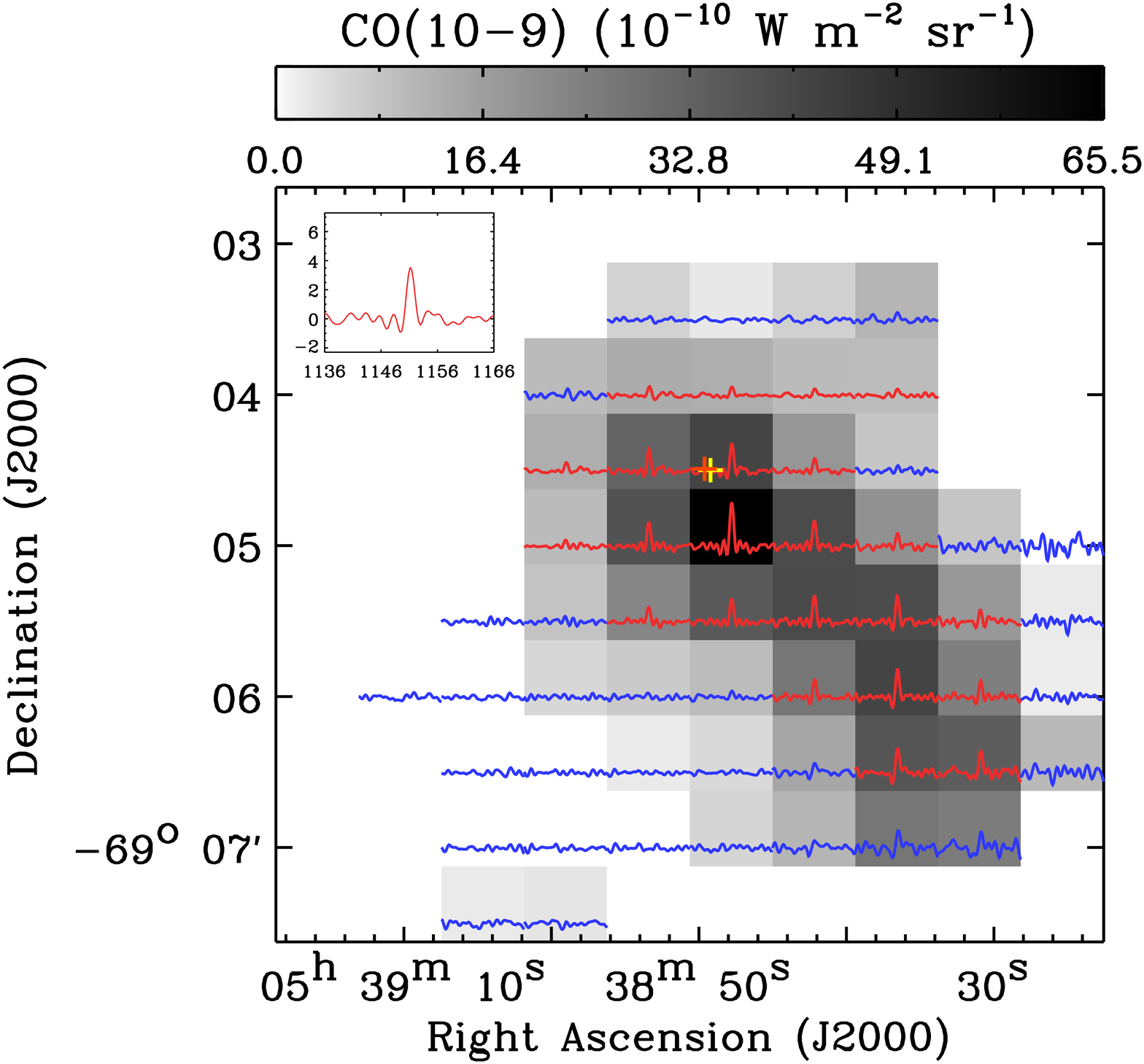}
\caption{\label{f:appendix1} FTS CO, \CI, and \NII integrated intensity images of 30 Doradus. 
See Appendix \ref{s:appendix1} for details on these figures.}
\end{figure*}

\begin{figure*}
\centering
\ContinuedFloat
\includegraphics[scale=0.22]{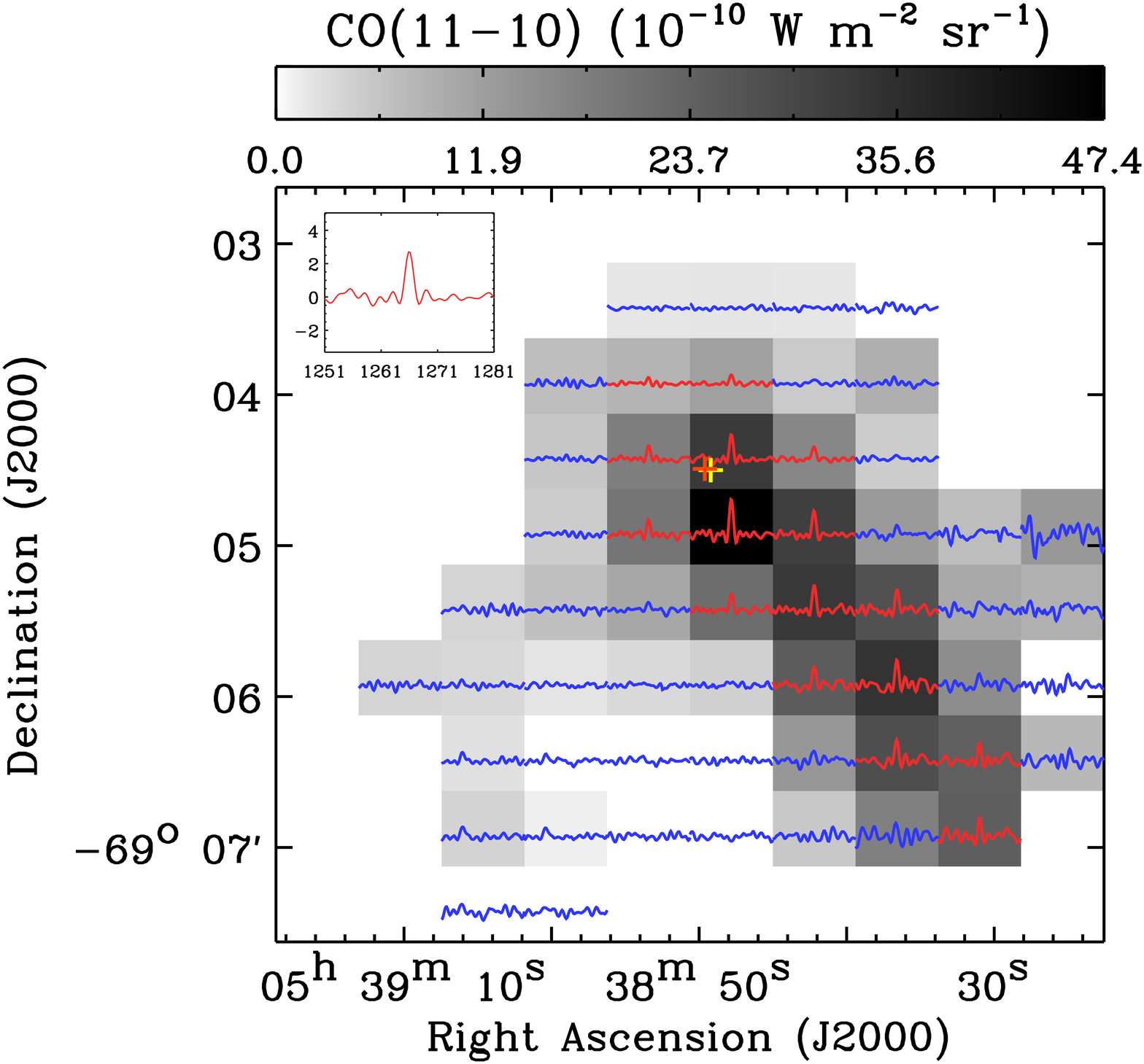} 
\includegraphics[scale=0.22]{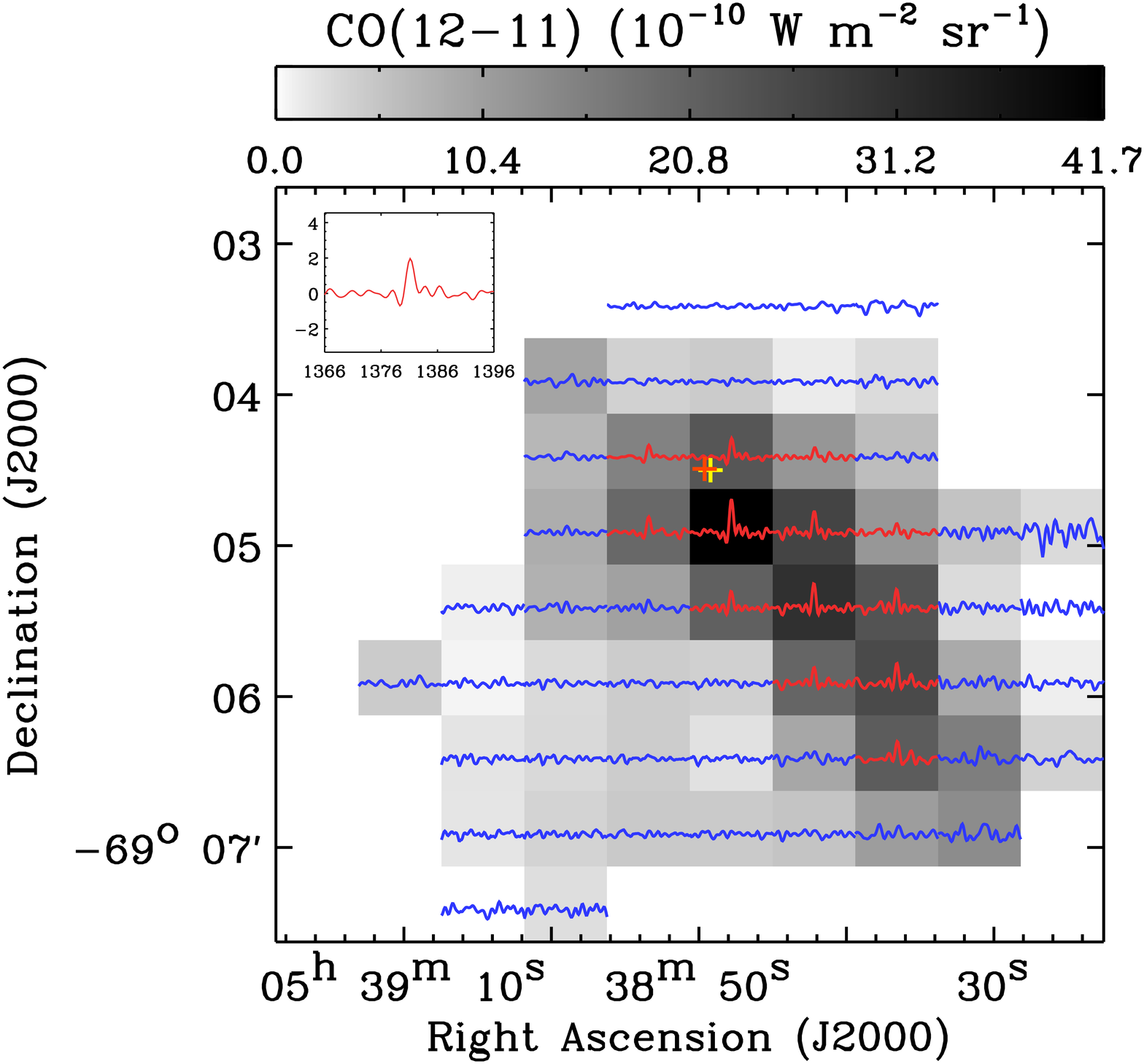}
\includegraphics[scale=0.22]{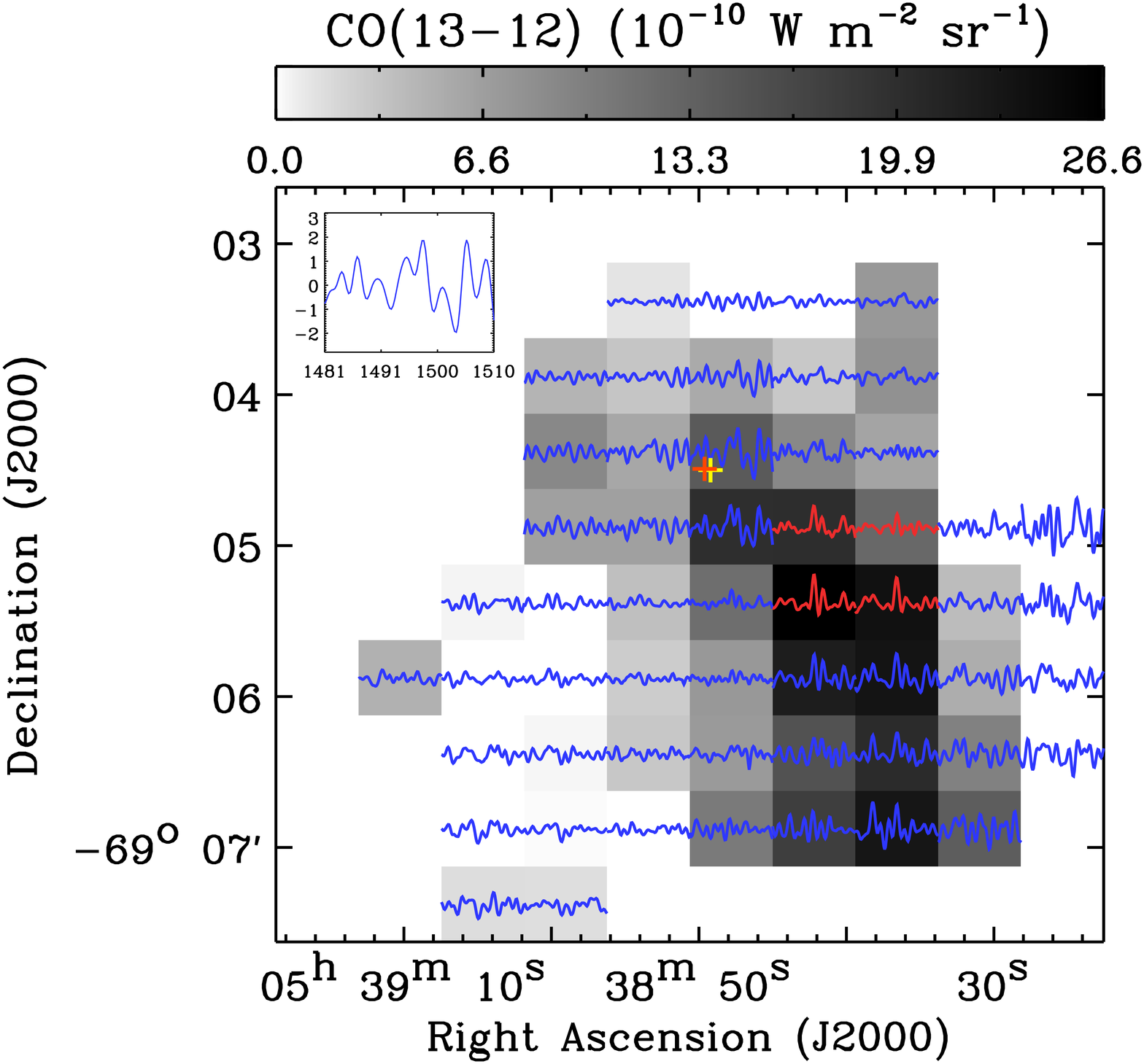}
\includegraphics[scale=0.22]{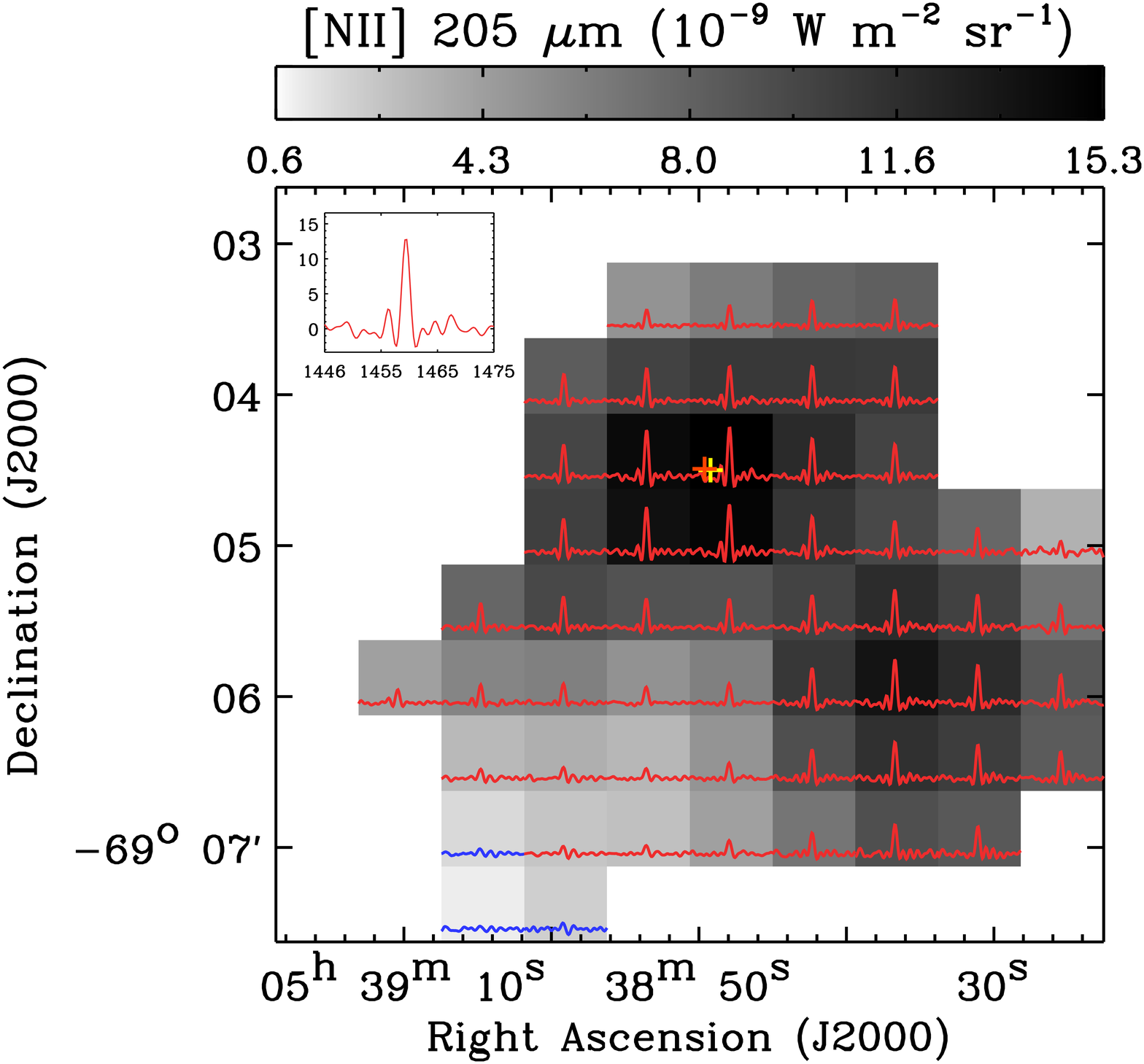}
\includegraphics[scale=0.22]{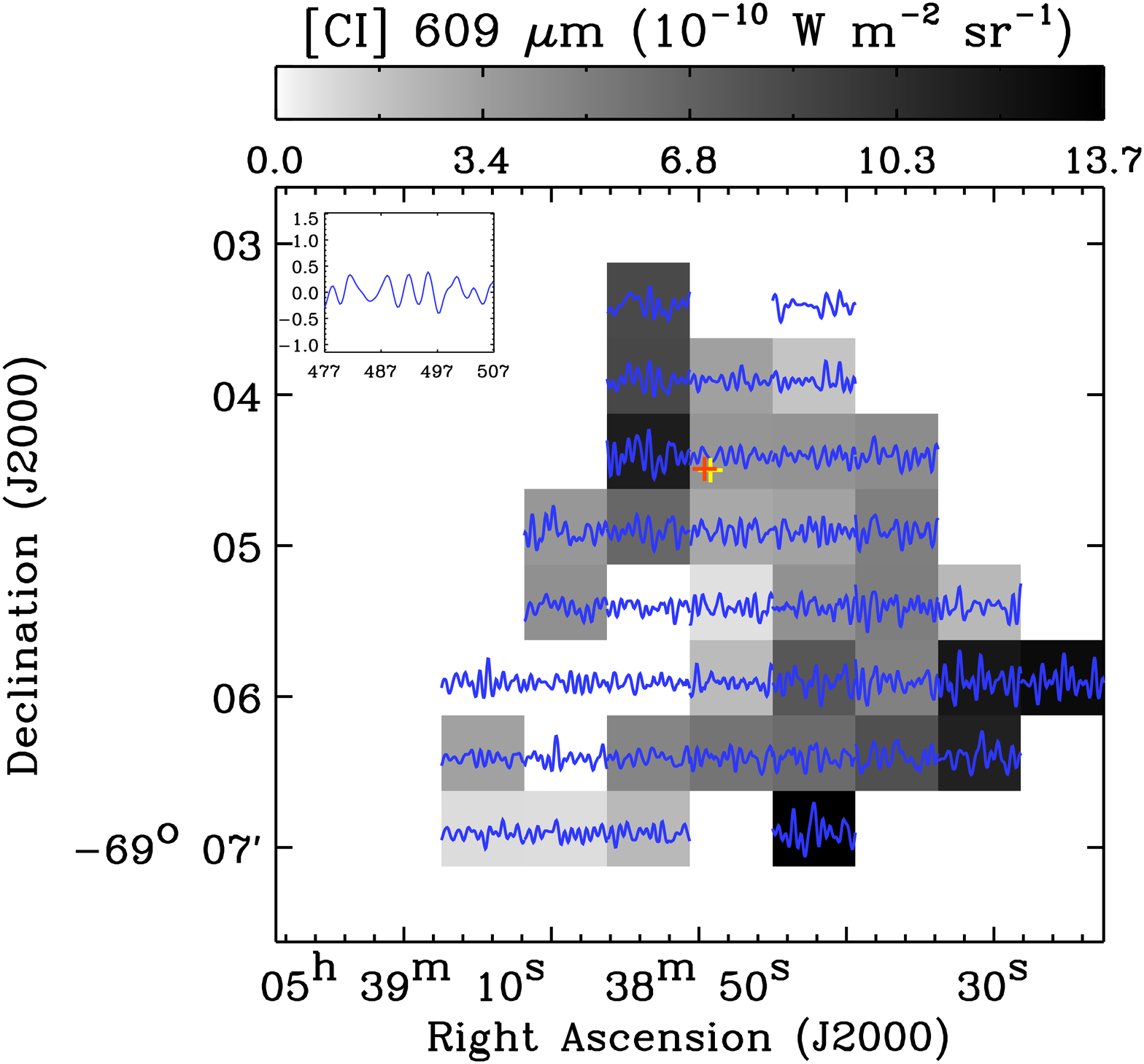}
\includegraphics[scale=0.22]{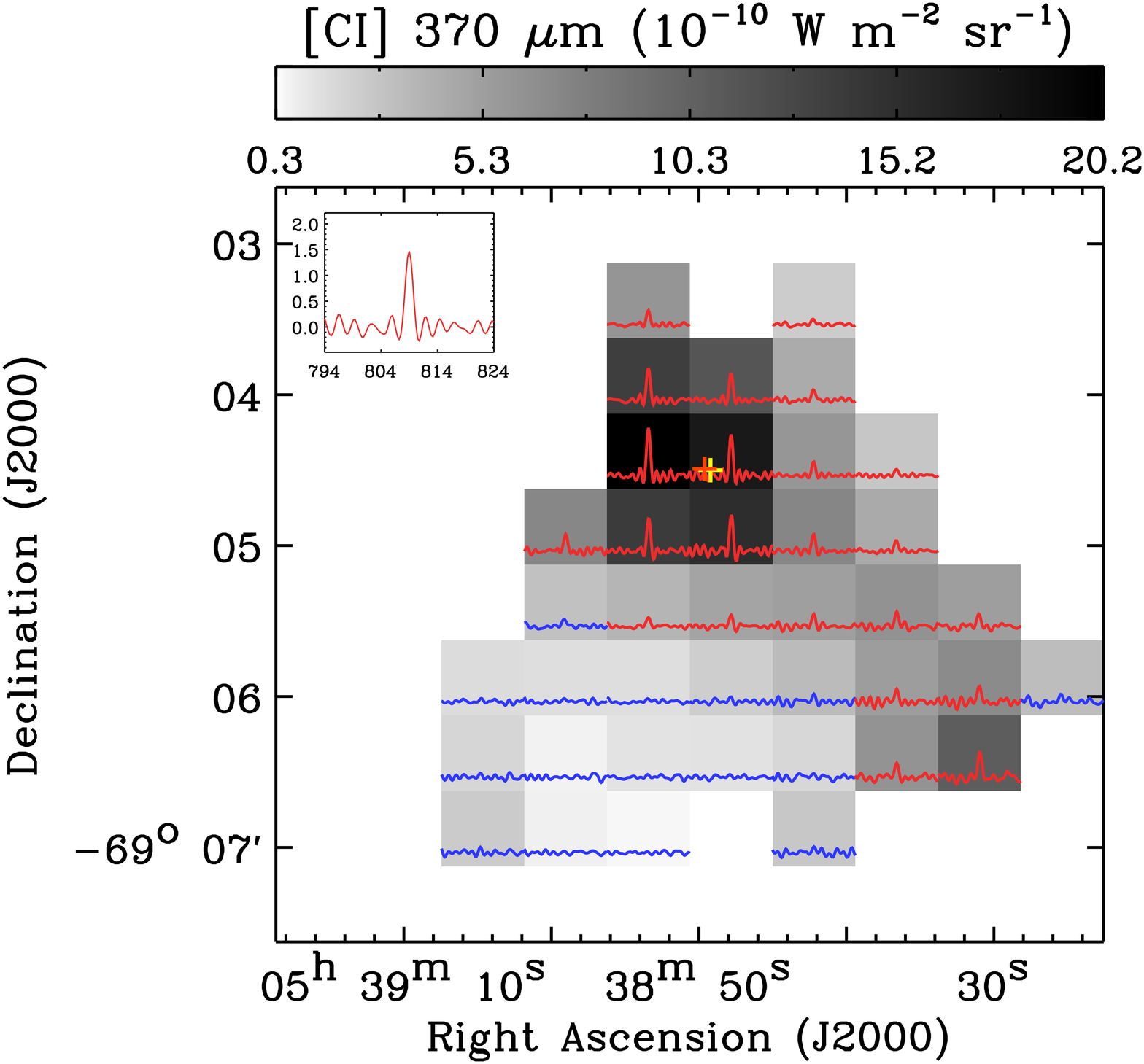}
\caption[]{(continued)}
\end{figure*}

\section{Stellar UV radiation field}
\label{s:appendix2}

To derive the stellar UV radiation field $G_{\rm stars}$ of 30 Doradus, we essentially followed \cite{Chevance16} and provide here a summary of our derivation. 
First, we cross-matched the catalogs of massive stars (W-R and OB-type stars) published by \cite{Crowther98}, \cite{Selman99}, and \cite{Doran13} 
and extracted the temperatures of $\sim$1.3 $\times$ 10$^{4}$ sources. 
The catalog by \cite{Doran13} was used as our main reference when possible, 
since it likely provides more reliable estimates of stellar properties based on spectroscopic observations, 
and the other two catalogs were used to complement it. 
We then integrated a blackbody from 912 \AA ~to 2400 \AA ~for each star to be consistent with the definition of $G_{\rm UV}$ in the Meudon PDR model 
and calculated $G_{\rm stars}$ for each 30$''$-size pixel of a 5$'$ $\times$ 5$'$ image (matching the coverage and pixel size of the FTS maps) 
by assuming that all the stars lie on the same plane as R136 and summing the UV fluxes of the stars. 
The derived $G_{\rm stars}$ on the plane of R136 is presented in Fig. \ref{f:UV_sources}
and can be considered as the maximum incident UV radiation field we expect, 
since no absorption was taken into account. 
While absorption by dust would have an important impact on the derivation of $G_{\rm stars}$, 
estimating absorption is currently not straightforward due to the lack of information on the location of absorbers.  


Compared to \cite{Chevance16}, the only difference is our usage of the catalog by \cite{Doran13}. 
This results in $\sim$50\% stronger UV radiation field on average, 
suggesting that a factor of two or so uncertainty could arise from the selection of stellar catalogs. 
Another source of uncertainty is our assumption of the stellar distribution, 
and we examined this issue by deriving $G_{\rm stars}$ with a random three-dimensional distribution of the stars.   
We then found that the assumption of the stars on the same plane tends to underestimate $G_{\rm stars}$ by up to $\sim$30\%. 

\section{PDR contribution to FIR emission}
\label{s:appendix3} 

FIR emission can originate not only from PDRs (neutral gas), but also from \HII regions (ionized gas).
To be properly compared to the predictions from the Meudon PDR model, 
our $L_{\rm FIR}$ derived from dust SED modeling (Sect. \ref{s:dust_properties}) 
then needs to be corrected for the contribution from the ionized medium. 
To do so, we used the PAH and \OIII 88 $\mu$m images from \cite{Chevance16} 
and performed the following steps (essentially what \citealt{Chevance16} did, but on 42$''$ scales). 
First, we assumed that PAH and \OIII 88 $\mu$m emission trace PDRs and \HII regions respectively 
and adopted a linear relation $L_{\rm FIR}$ = $\alpha L_{\rm PAH} + \beta L_{\rm \OIII}$. 
The coefficients ($\alpha$, $\beta$) = (4.6, 8.8) were then derived by fitting a multiple regression model to all available pixels, 
and the PDR-only component of $L_{\rm FIR}$ was estimated as $L_{\rm FIR}^{\rm PDR}$ = $L_{\rm FIR} - \beta L_{\rm \OIII}$. 
Note that our analysis is based on the assumption that the PAH-to-dust mass ratio does not change in the PDRs and drops to zero in the \HII regions.  

This $L_{\rm FIR}^{\rm PDR}$ is what we used as an input for PDR modeling in Sect. \ref{s:UV_photons}, 
and we assigned 30\% of $L_{\rm FIR}^{\rm PDR}$ as 1$\sigma$ uncertainties, considering the simple empirical relation we adopted. 
The resulting $L_{\rm FIR}^{\rm PDR}$ shows a good agreement with \cite{Chevance16}, 
e.g., the PDR contribution to $L_{\rm FIR}$ ranges from $\sim$40\% to $\sim$80\% across the 30 Doradus region 
($\sim$30\% to $\sim$90\% in \citealt{Chevance16}). 

\section{Comparison with \cite{Chevance16}}
\label{s:appendix4}

As described in Sect. \ref{s:30dor_pdr}, \cite{Chevance16} recently studied the properties of PDRs in 30 Doradus on 12$''$ scales 
by performing Meudon PDR modeling of \CII 158 $\mu$m, \OI 145 $\mu$m, and FIR emission. 
Since we employed essentially the same datasets and modeling approach for our analyses (Sect. \ref{s:PDR_results1}), 
it is important to double-check that our results are indeed consistent with those in \cite{Chevance16}. 

\begin{figure*}
\centering
\includegraphics[scale=0.65]{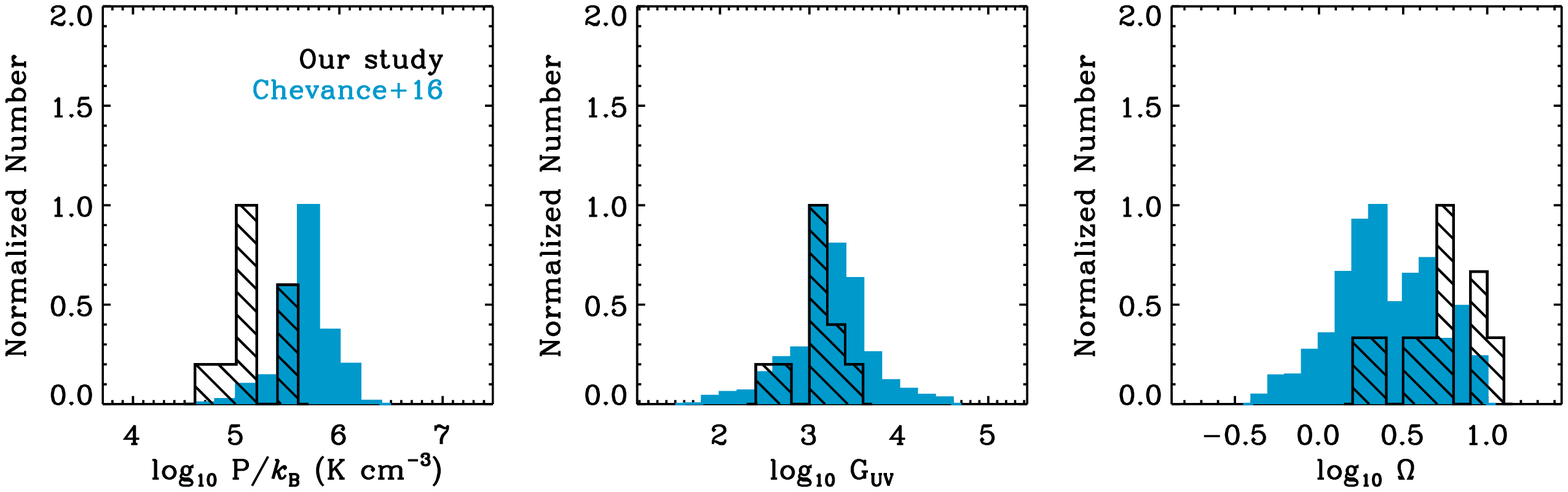} 
\caption{\label{f:appendix4} Comparison with \cite{Chevance16}. 
The normalized histograms of the PDR parameters in our study (42$''$ scales) and \cite{Chevance16} (12$''$ scales) 
are shown in black (hatched) and blue (solid) respectively
($P/k_{\rm B}$, $G_{\rm UV}$, and $\Omega$ on the \textit{right}, \textit{middle}, and \textit{left panels}).}
\end{figure*}

As for the spatial distributions of the PDR parameters (e.g., peak locations of $P$, $G_{\rm UV}$, and $\Omega$), 
we found that our results are consistent with \cite{Chevance16}. 
However, a comparison of the absolute values showed a large discrepancy, as can be seen in Fig. \ref{f:appendix4}. 
Specifically, we noticed that our $P$ and $G_{\rm UV}$ distributions primarily trace the lower part of the \cite{Chevance16} histograms, 
while the opposite is the case for $\Omega$.  

This large discrepancy in the absolute values of the PDR parameters could result from several differences between the two studies, 
e.g., resolution (42$''$ vs. 12$''$), spatial coverage (our maps are smaller), and slightly inconsistent PDR models 
(version 1.5.2 with $A_{V}$ $\sim$ 2 mag vs. version 1.6.0 with $A_{V}$ = 3 mag). 
Among these possibilities, we probed the impact of different resolutions by performing Meundon PDR modeling of 
\CII 158 $\mu$m, \OI 145 $\mu$m, and FIR emission for one pixel (chosen as the one where the $G_{\rm UV}$ distribution in \cite{Chevance16} peaks), 
but on 12$''$ scales. 
For this purpose, we used the same model grid with $A_{V}$ = 2 mag, as well as the same fitting method, as in Sect. \ref{s:PDR_results1} 
and constrained $P/k_{\rm B}$ = 7 $\times$ 10$^{5}$ K cm$^{-3}$, $G_{\rm UV}$ = 3 $\times$ 10$^{4}$, and $\Omega$ = 0.7. 
These results are in good agreement with what \cite{Chevance16} estimated 
($P/k_{\rm B}$ = 10$^{6}$ K cm$^{-3}$, $G_{\rm UV}$ = 3 $\times$ 10$^{4}$, and $\Omega$ = 0.5), 
essentially suggesting that the discrepancy between our study and \cite{Chevance16} 
mostly results from the difference in the angular resolution. 
As \cite{Chevance16} pointed out as well, utilizing lower resolution data skews the PDR solutions toward lower $P$ and $G_{\rm UV}$, 
since the intensities measured on larger scales tend to be more dominated by the emission from diffuse regions with less UV illumination.

\section{Comparison with \cite{Okada19}}
\label{s:appendix5}

Recently, \cite{Okada19} presented an independent study on CO 
(CO $J$=2--1, 3--2, 4--3, and 6--5, as well as $^{13}$CO $J$=3--2; observed with APEX)  
and fine-structure lines 
(\CII 158 $\mu$m, \CI 609 $\mu$m and 370 $\mu$m, and \OI 63 $\mu$m and 145 $\mu$m; observed with APEX and \textit{SOFIA}) in 30 Doradus. 
Among the observed transitions, CO, $^{13}$CO, \CII and \CI were mapped over an area that is comparable to our FTS coverage, 
while \OI was obtained for selected positions only.  
This study based on high spatial ($\sim$6$''$ to $\sim$30$''$) and spectral ($\sim$1 km s$^{-1}$) resolution data 
is complementary to our work, and we provide here a summary of it. 

First of all, their high resolution data clearly demonstrate the complexity of the neutral ISM in 30 Doradus. 
For example, the authors found that CO, $^{13}$CO, and \CI spectra are similar,  
while \CII 158 $\mu$m shows a wider linewidth and/or additional velocity components. 
In addition, \OI spectra match CO spectra at some locations, 
but they are more similar to \CII 158 $\mu$m profiles at other locations. 
In terms of spatial distribution, \CII 158 $\mu$m and CO $J$=4--3 show relatively similar structures, 
except for several mismatching peaks.  

The complexity of the neutral ISM in 30 Doradus was also manifested in KOSMA-$\tau$ PDR modeling by \cite{Okada19}. 
The KOSMA-$\tau$ model calculates the thermal and chemical structures of a PDR as the Meudon PDR model does, 
but with a different geometry of the medium. 
Specifically, while the Meudon PDR model considers a plane-parallel slab of gas and dust, 
the KOSMA-$\tau$ model assumes an ensemble of clumps with a power-law mass spectrum $dN/dM$ $\propto$ $M^{-\alpha}$ 
($\alpha$ = 1.8 was used in \citealt{Okada19}) 
to take into account the clumpiness of the ISM. 
Line and continuum intensities are then estimated by adding the contribution from each clump for a model grid of three parameters, 
total mass ($M_{\rm tot}$), average gas density ($\overline{n}$), and UV radiation field ($G'_{\rm UV}$). 
For more details on the KOSMA-$\tau$ PDR model, we refer to \cite{Stoerzer96} and \cite{Rollig06}. 
While we cannot make a direct comparison with \cite{Okada19} due to systematic differences in PDR modeling 
(e.g., input parameters and modeling approach), 
our results are essentially consistent in a way that
both our study and \cite{Okada19} show that one ISM component is not sufficient to analyze 30 Doradus. 
For example, \cite{Okada19} modeled their observed transitions along with dust continuum emission for 30Dor-10 at  
($\alpha$, $\delta$)$_{\rm J2000}$ = (05$^{\rm h}$38$^{\rm m}$48.8$^{\rm s}$, $-69^{\circ}$04$'$42.1$''$) 
and found that CO and \CI are relatively well reproduced, while \CII and \OI are overestimated by a factor of a few.  
This best-fit KOSMA-$\tau$ model, however, predicts the CO SLED to be flat already around at $J$=6--5, 
suggesting that the gas is not warm enough ($J_{\rm p}$ = 9--8 in our FTS observations). 
The constrained UV radiation field ($\sim$200 Draine fields) is indeed weaker than 
what we estimated for the low- and high-$P$ PDR components 
($\sim$3 $\times$ 10$^{3}$ Draine fields).
In addition, the large beam filling factor of $\gtrsim$ 1 is not consistent with 
what the ALMA CO(2--1) observations suggest (\citealt{Indebetouw13}).  
All in all, both our study and \cite{Okada19} highlight that many high resolution tracers of gas and dust are required  
to build a consistent picture of the multi-phase, multi-component ISM in complex regions like 30 Doradus.


\end{appendix}

\end{document}